\documentclass[preprint,aps,pra,superscriptaddress,showpacs,floatfix]{revtex4-1} 
\usepackage{hyperref}

\usepackage{graphicx,subfigure}

\usepackage{times}
\usepackage {euscript}
\usepackage{amsmath}
\usepackage{amsthm}
\usepackage{epsf}
\usepackage{epsfig}

\usepackage[left=2cm,right=2cm,top=2cm,bottom=2cm]{geometry}
\usepackage{bm}
\usepackage[utf8]{inputenc}
\renewcommand{\vec}[1]{{\mbox{\boldmath$#1$}}}

\usepackage{color}
\definecolor{BLUE}{rgb}{0.0,0.0,1.0}

\usepackage{soul}
\soulregister\cite7
\soulregister\ref7

\begin{document}  
\title{Strong relativistic effect on (quasi)resonance electron-capture processes in non-symmetrical heavy ion-atom collisions
%Single- and double-electron transfer in low- and intermediate-energy Th$^{90+}$-Ru$^{42+}$ collisions
}  
%  
% \date{}  
%  
\author{Y.S.~Kozhedub}
\email{y.kozhedub@spbu.ru}
\affiliation {Department of Physics, St.~Petersburg State University, 
Universitetskaya 7/9, 199034 St.~Petersburg, Russia}

\author{A.I.~Bondarev}
\affiliation {Department of Physics, St.~Petersburg State University, 
Universitetskaya 7/9, 199034 St.~Petersburg, Russia}
\affiliation {Center for Advanced Studies, Peter the Great St.~Petersburg Polytechnic University, Polytechnicheskaya 29, 195251 St.~Petersburg, Russia}

\author{X.~Cai}
\affiliation {Institute of Modern Physics, Chinese Academy of Sciences, 
Nanchang rd. 509, 730000 Lanzhou, China}

\author{X.~Ma}
\affiliation {Institute of Modern Physics, Chinese Academy of Sciences, 
Nanchang rd. 509, 730000 Lanzhou, China}

\author{G.~Plunien}
\affiliation {Institut f\"ur Theoretische Physik, Technische Universit\"at Dresden,  
Mommsenstra{\ss}e 13, D-01062 Dresden, Germany}

\author{V.M.~Shabaev}
\affiliation {Department of Physics, St.~Petersburg State University, 
Universitetskaya 7/9, 199034 St.~Petersburg, Russia}

\author{C.~Shao}
\affiliation {Institute of Modern Physics, Chinese Academy of Sciences, 
Nanchang rd. 509, 730000 Lanzhou, China}

\author{I.I.~Tupitsyn}
\affiliation {Department of Physics, St.~Petersburg State University, 
Universitetskaya 7/9, 199034 St.~Petersburg, Russia}

\author{B.~Yang}
\affiliation {Institute of Modern Physics, Chinese Academy of Sciences, 
Nanchang rd. 509, 730000 Lanzhou, China}

\author{D.~Yu}
\affiliation {Institute of Modern Physics, Chinese Academy of Sciences, 
Nanchang rd. 509, 730000 Lanzhou, China}

% \author[SPSU]{A.I.~Bondarev}
% \author[GSI]{A.~Gumberidze}
% \author[GSI]{S.~Hagmann}
% \author[GSI]{C.~Kozhuharov}
% \author[SPSU]{I.A.~Maltsev}
% \author[SPSU]{V.M.~Shabaev}
% \author[GSI,HI_Jena,FSU_Jena]{Th.~St\"ohlker}
% \author[SPSU]{I.I.~Tupitsyn}

%  
%%%%%%%%%%%%%%%%%%%%%%%%%%%%%%%%%%%%%%%%%%%%%%%%%%%%%%%%%%%%%%%%%%%%%  
\begin{abstract}  
(Quasi)resonance electron-capture processes in non-symmetrical $(Z_{\rm P} \simeq 2Z_{\rm T})$ heavy ion-atom collisions are studied employing a semiclassical atomic Dirac-Fock-Sturm orbital coupled-channel approach within an independent-particle model.  
%A special attention has been paid to study a role of the relativistic effects. 
Systematic calculations of the electron-capture cross sections of the target $K$-shell electrons to the $L$ subshells of the projectile have been carried out for the collisions of bare thorium $(Z_{\rm P}=90)$ and zinc $(Z_{\rm P}=30)$ nuclei with hydrogenlike ions $(Z_{\rm T}=36$-$47)$ and $(Z_{\rm T}=12$-$15)$, correspondingly. 
Strong relativistic effects, crucial for the case of Th$^{90+}$-Ru$^{43+}(1s)$ collisions in the low-energy regime, are found. 
Various one- and two-electron capture processes occurring in course of the collisions 
of the two-electron system Th$^{90+}$-Ru$^{42+}(1s^2)$ have been investigated in details in the wide range of collision energies $0.5$-$50$~MeV/u. The impact parameter dependencies of the double electron-capture processes are also presented. 
Our study demonstrates a very significant role of the relativistic  
effects for the processes, which becomes crucial in the low-energy regime. 
\end{abstract}

\maketitle  
%  
%%%%%%%%%%%%%%%%%%%%%%%%%%%%%%%%%%%%%%%%%%%%%%%%%%%%%%%%%%%%%%%%%%%%%%%%%%%%%%%% 
%  

%%%%%%%%%%%%%%%%%%%%%%%%%%%%%%%%%%%%%%%%%%%%%%%%%%%%%%%%%%%%%%%%%%%%%%%%%%%%%%%% 
\section{Introduction}  
Heavy-ion collisions play a very important role in investigations of the 
relativistic quantum dynamics of electrons in presence of strong electromagnetic 
fields~\cite{Eichler:1995}. Moreover, if the total charge of the colliding   
nuclei is larger than the critical one, $Z_{\rm total} = Z_1 + Z_2 > 173$, such 
collisions can provide a unique tool for tests of quantum electrodynamics (QED) in the  
supercritical regime~\cite{Greiner:1985,Maltsev:PRL:2019}.    
Experimental investigations aimed at comprehensive study of various 
processes in low-energy heavy ion-atom collisions including
combined nuclear charges greater than the critical one are planned in the 
nearest future. The realization of the 
FAIR (Germany)~(see, e.g., Refs.~\cite{Gumberidze:NIMB:2009,Lestinsky:EPJST:2016}) ,
NICA (Russia)~(see, e.g., Ref.~\cite{Ter-Akopian:IJMPE:2015})  ,
HIAF (China)~(see, e.g., Ref.~\cite{Ma:nimb:2017})  and 
FISIC (France)~(see, e.g., Ref.~\cite{FISIC}) projects will open novel and unique 
opportunities with a large discovery potential for studying various effects in
these collisions. The corresponding theoretical calculations which would 
be able to describe in details relativistic quantum   
dynamics of electrons and radiation processes in these collisions are urgently required.

A special interest should be attracted to (quasi)resonance processes, which can be very sensitive to the relativistic and QED effects. Among them are (quasi)resonance electron-capture (EC) processes at the low- and intermediate collision energies. An example of such processes is the K-K charge transfer in heavy (quasi)symmetrical ion-atom collisions. Thus, the impact-parameter dependence of the probability of charge transfer in the low-energy collisions of bare and hydrogenlike uranium is very sensitive to the relativistic effects~\cite{Tupitsyn:pra:2010,Maltsev:ps:2013}. Meanwhile, if one considers the total cross section of the process, the relativistic effects would be not so significant. 
In the present paper, we study EC processes in heavy ion-atom collisions, where the ground state of the target is in a resonance with the $n=2$ states of the projectile. We started our investigation with consideration of one-electron systems: collisions of bare zinc or thorium nuclei with hydrogenlike targets, where the (quasi)resonance condition is fulfilled. It was found that the relativistic effects are crucial for the Th$^{90+}$-Ru$^{43+}(1s)$ collisions in the low-energy regime. Then the two-electron collisions of the  Th$^{90+}$-Ru$^{42+}(1s^2)$ were investigated in details. Evaluations of various single-electron-capture (SEC) and double-electron-capture (DEC) processes for a wide collision energy region ranging from $0.5$ to $50$ MeV/u were performed. Special attention was paid to investigation of the relativistic effects.
A semiclassical atomic Dirac-Fock-Sturm orbital coupled-channel method within an independent particle model~\cite{Kozhedub:pra:2014,Kozhedub:ps:2013,Tupitsyn:pra:2010} was used in our calculations. 
Large basis sets were employed to reach reasonable convergence of the cross sections. We also studied stability of the obtained results with respect to the used model parameters such as a screening potential and a type of colliding ion trajectories. 
%The non-relativistic C$^{4+}(1s)^2$-He$(1s)^2$ collision system, which was quite well studied theoretically as well as experimentally, was used as a test system for our calculation.     

The paper is organized as follows. In the next
section we briefly outline the method used in
the calculations. 
In Sec.~\ref{sec:Results}, the results of calculations are presented and discussed.
Subsections~\ref{sec:results:one-electron} and ~\ref{sec:results:many-electron} are devoted to one-electron and many-electron systems, respectively.
A brief conclusion is given in the final Sec.~\ref{sec:Conclusion}. 
Atomic units (a.u.) ($\hbar=e=m_e=1$)  are used throughout the paper unless otherwise stated. 
%%%%%%%%%%%%%%%%%%%%%%%%%%%%%%%%%%%%%%%%%%%%%%%%%%%%%%%%%%%%%%%%%%%%%  
\section{Theory}  
\label{sec:Theory}  
%%%%%%%%%%%%%%%%%%%%%%%%%%%%%%%%%%%%%%%%%%%%%%%%%%%%%%%%%%%%%%%%%%%%%  
%  
 
Here we briefly present the formalism used, for a   
complete description see   
Refs.~\cite{Kozhedub:pra:2014,Kozhedub:ps:2013,Tupitsyn:pra:2010}.  
Using the semiclassical approximation, where the atomic nuclei move along the   
classical trajectories and are considered as sources of a time-dependent   
external potential, we have to solve the time-dependent many-particle Dirac   
equation for the electrons involved in the process.   
We employ a method based on an independent particle model,   
where the many-electron Hamiltonian $\hat H$ is approximated by a sum of   
effective single-electron Hamiltonians, $\hat  
H^{\rm eff}=\sum \hat h^{\rm eff}$, reducing the electronic many-particle   
problem to a set of single-particle Dirac equations for all ($N$) electrons of 
the colliding system:  
\begin{equation}    
i \frac{\partial \psi_i(\vec{r},t)}{\partial t} = \hat h^{\rm eff}(\vec{r},t) \,    
\psi_i(\vec{r},t), \quad i=1,\ldots,N,    
\label{eq:Dirac}    
\end{equation}    
subject to the initial 
conditions: 
\begin{equation}    
\lim_{t\rightarrow -\infty}(\psi_i(\vec{r},t) - \psi_{i}^{0}(\vec{r},t))=0    
 , \quad i=1,\ldots,N.    
\label{eq:}    
\end{equation}    
% The many-electron wave function is given by a Slater determinant made-up from 
% the single-particle wave functions.     
%%%%%%%%%%%%%%%%%%%%%%    
As the effective single-electron Hamiltonian $\hat h^{\rm eff}$ we use the   
two-center Dirac-Kohn-Sham Hamiltonian:    
\begin{equation}    
\hat h^{\rm eff} =c (\vec{\alpha} \cdot \vec{p}) +    
\beta \, c^2 +     
V_{\rm nucl}^{A}(\vec{r}_A) + V_{\rm nucl}^{B}(\vec{r}_B) +    
V_{\rm C}[\rho] + V_{\rm xc}[\rho]    
\,,    
\end{equation}    
where $c$ is the speed of light and $\vec{\alpha}$, $\beta$ are the    
Dirac matrices.     
Index $\alpha = A,B $ indicates the centers, 
$\vec{r}_{\alpha}=\vec{r}-\vec{R}_{\alpha}$, 
$\vec{R}_{\alpha}$ is the radius-vector of the centers (nucleus),
$V_{\rm nucl}^{\alpha}(\vec{r}_\alpha)$ and    
$V_{\rm C}[\rho] = \int \, d^3\vec{r^{\prime}} \,     
\frac{\rho(\vec{r}^{\prime})}{|\vec{r}-\vec{r}^{\prime}|}$    
are the electron-nucleus and the electron-electron Coulomb    
interaction potentials, respectively, 
and
$\rho(\vec{r})$ is the electron density of the system.     
The exchange-correlation    
potential $V_{\rm xc}[\rho]$ is taken in the Perdew-Zunger    
parametrization~\cite{Perdew:prb:1981}.     
$V_{\rm C}$ and $V_{\rm xc}$ together provide the electron screening potential.
  
The effective single-particle equations~(\ref{eq:Dirac}) are solved by means of
the coupled-channel approach with atomic-like Dirac-Sturm-Fock orbitals,   
localized at the ions (atoms)~\cite{Tupitsyn:pra:2010,Tupitsyn:pra:2012}. The many-particle   
probabilities are calculated in terms of the single-particle amplitudes 
employing   
the formalism of inclusive 
probabilities~\cite{Ludde:jpb:1985,Kurpick:cpc:1993}, 
which allows one to describe the many-electron collision dynamics.  

\section{Results of the calculations and discussion}  
% \section{Results}  
\label{sec:Results}  
%%%%%%%%%%%%%%%%%%%%%%%%%%%%%%%%%%%%%%%%%%%%%%%%%%%%%%%%%%%%%%%%%%%%%%%%%%%%%%%% 
% %  
\clearpage
The (quasi)resonance EC from the ground target state to the projectile $n=2$ states in heavy ion-atom collisions is considered. The resonance condition for the processes is realized when the energies of the states are close to one another or $Z_{\rm T} \simeq Z_{\rm P}/2$, equivalently ($Z_{\rm T}$ and $Z_{\rm P}$ are the target and projectile nuclear charges, respectively).
%%%%%%%%%%%%%%%%%%%%%%%%%%%%%%%%%%%
\subsection{One-electron systems}  
\label{sec:results:one-electron}  
We start our consideration with one-electron systems (collisions of a bare nucleus with hydrogenlike ions) and evaluate cross sections of EC to the projectile $n=2$ substates.

\begin{figure}
\begin{center}
\includegraphics[width=0.9\textwidth]{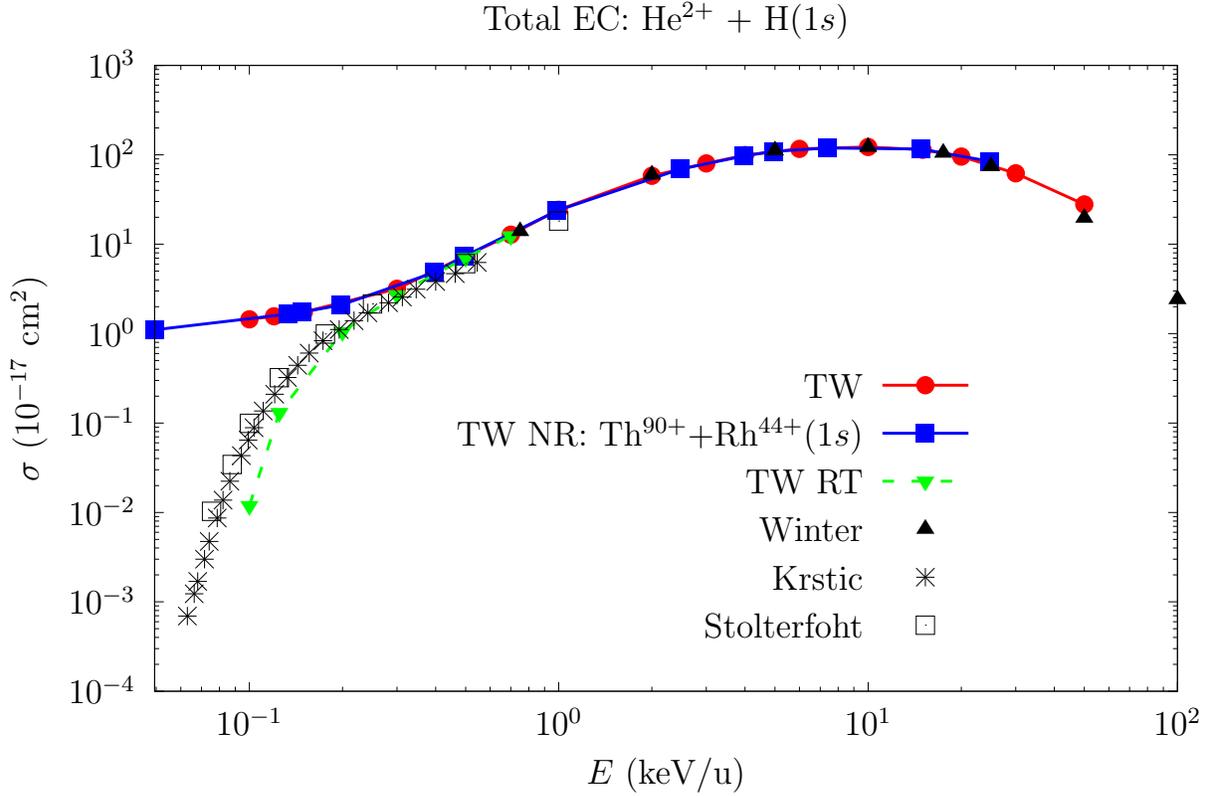}
\end{center}
\caption{\label{fig:He_total_EC}
Total EC cross sections for the He$^{2+}$-H$(1s)$ collision as functions of the impact energy.
The present results (TW) are compared with 
the coupled-channel calculations of Winter~\cite{Winter:pra:2007}, 
the hidden crossing coupled-channel approach  by  Krsti\'{c}~\cite{Krstic:JPB:2004},
the electron-nuclear dynamic approach by Stolterfoht {\it et al.}~\cite{Stolterfoht:PRL:2007}. 
"TW" (This Work) data are obtained for the straight-line trajectories of colliding nuclei, 
"TW NR: Th$^{90+}$+Rh$^{44+}(1s)$" data are obtained for the Th$^{90+}$+Rh$^{44+}(1s)$ collision within the non-relativistic (NR) limit, for the straight-line nuclear collision  trajectories and are presented using the "scaling factor", 
"TW RT" data are obtained for the Rutherford trajectories of colliding nuclei.
}
\end{figure}

In order to test our approach, it was applied to the non-relativistic well studied collision He$^{2+}$-H$(1s)$ (see, e.g., Refs.~\cite{Krstic:JPB:2004,Stolterfoht:PRL:2007,Winter:pra:2007,Minami:JPB:2008,Faulkner:PSCF:2019} and references therein).
In figure~\ref{fig:He_total_EC} we present the energy dependence of the total EC cross section in comparison with the results of other calculations. 
At energies higher than $1$~keV/u our data (TW) agree well with the most recent coupled-channel calculations of Winter~\cite{Winter:pra:2007}. There are some deviation in the high collision energy region results due to the representation of atomic (ion) states for moving nuclei used in our calculations, which becomes not good enough. It deals with an absence of the translation factor in our basis set, which starts to be important at collision energies $E>20$~keV/u. 
In slow collisions ($E<0.7$~keV/u) the description of 
internuclear motion affects the EC process. The results of calculations for the straight-line (indicated by "TW") and Rutherford (indicated by "TW RT") trajectories of colliding nuclei are presented in Fig.~\ref{fig:He_total_EC}, where an essential distinction of the results is observed. Such strong effect results from a resonance type of the process under consideration, which is very sensitive to various perturbations. As an example, for the He$^{2+}$-H$(1s)$ collision a strong isotope effect on the EC was observed and investigated by Stolterfoht {\it et al.}~\cite{Stolterfoht:PRL:2007}. 
Our results obtained for the Rutherford trajectories are generally in a good agreement  with the hidden crossing coupled-channel approach by Krsti\'{c}~\cite{Krstic:JPB:2004} and the electron-nuclear dynamic approach by Stolterfoht {\it et al.}~\cite{Stolterfoht:PRL:2007}, which take into account the effect of electron-nuclei interaction on the internuclear motion. 

To test conformity of our evaluation for heavy systems we performed calculations 
for the Th$^{90+}$+Rh$^{44+}(1s)$ collision in the non-relativistic limit 
($c\rightarrow\infty$), by using the same numerical routine, but multiplying the standard value of the speed of light by the factor $1000$. The system is similar to 
the He$^{2+}$-H$(1s)$ one with the ``scaling factor'' $f=45$.
According to the scaling law~\cite{Sharma:pra:1976}, valid for the non-relativistic approximation, the straight-line trajectories of colliding nuclei and the Coulomb nuclear potentials, for the He$^{2+}$-H$(1s)$ and Th$^{90+}$+Rh$^{44+}(1s)$ collisions we should obtain the same values of the EC cross section within the scaling rules:
the collision energy values are divided by $f^2$  and the values of the cross sections  are multiplied by the factor $f^2$. The corresponding results of calculations for the  Th$^{90+}$+Rh$^{44+}(1s)$ collision are shown in Fig.\ref{fig:He_total_EC} (indicated as "TW NR: Th$^{90+}$+Rh$^{44+}(1s)$") and good agreement is obtained with the He$^{2+}$-H$(1s)$ ones. 
%%%%%%%%%%%%%%%%%%%%%%%%%%%%%%%%%%%%%%%%%%%%%%%%%%%%%%%%%%%%%%%%%%%%%  
%%%%%%%%%%%%%%%%%%%%%%%%%%%%%%%%%%%%%%%%%%%%%%%%%%%%%%%%%%%%%%%%%%%%%  

\begin{figure}
\begin{center}
\subfigure[\quad $2s$]{%
\includegraphics[width=0.45\textwidth]{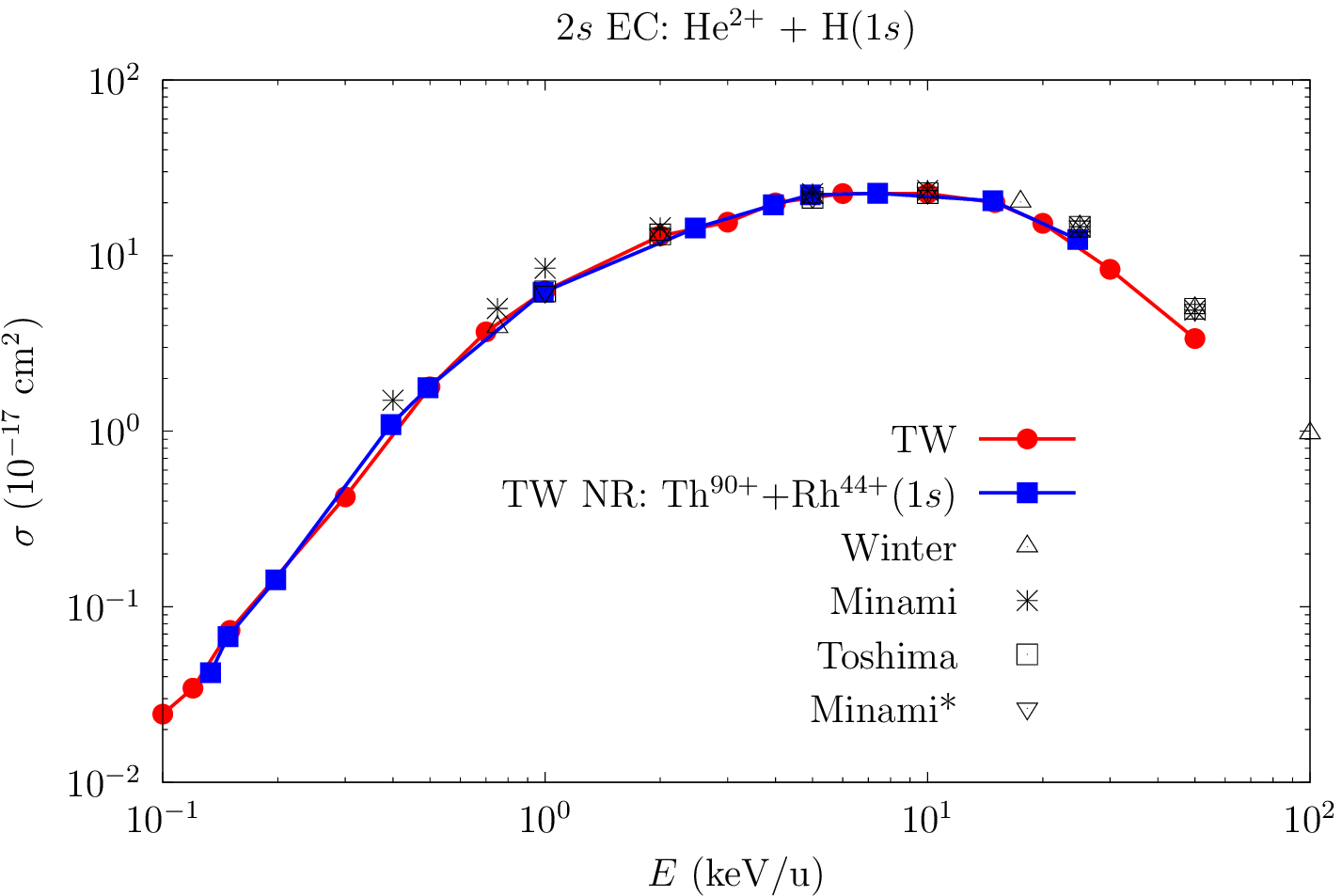}
}
\subfigure[\quad $2p$]{%
\includegraphics[width=0.45\textwidth]{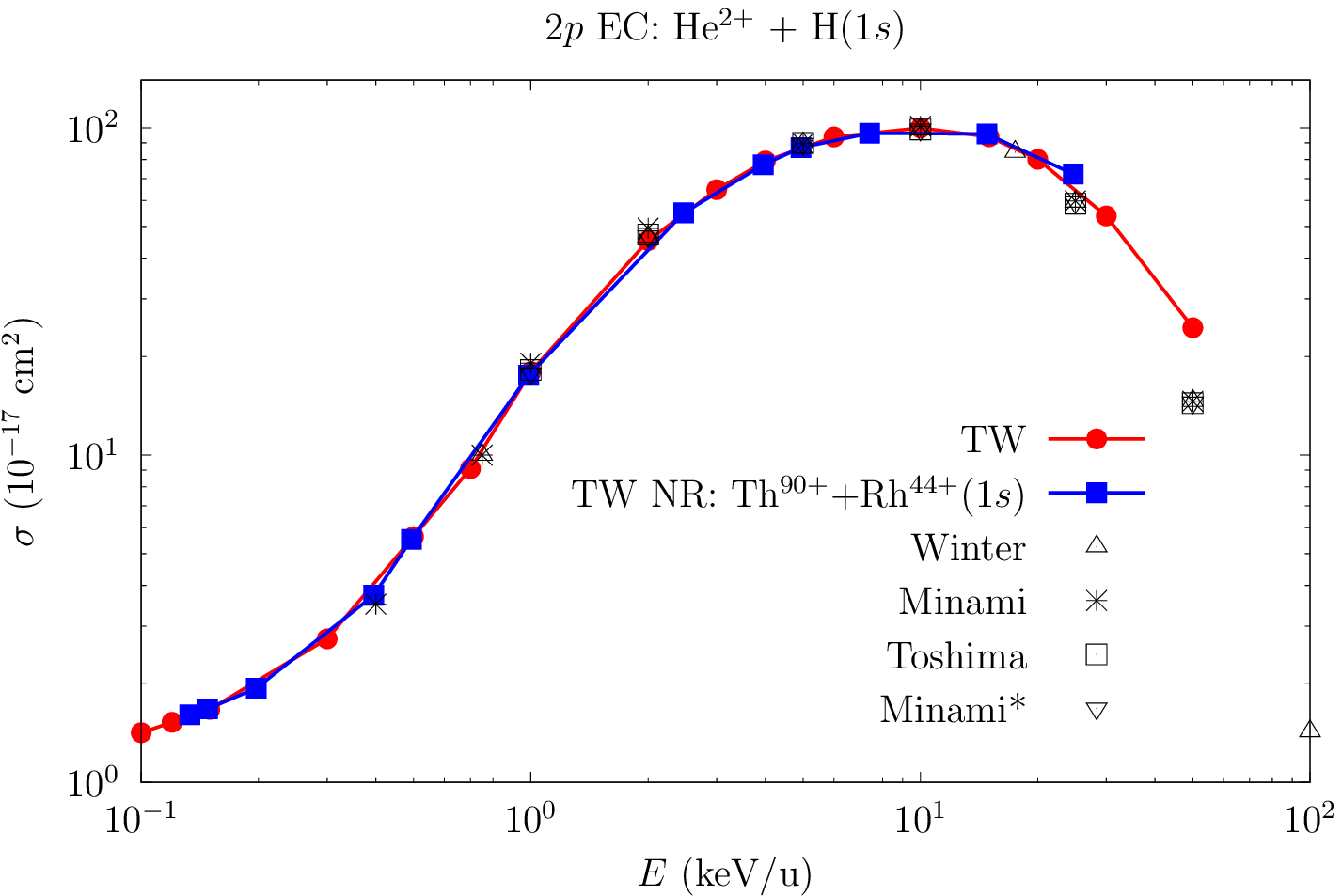}
}\\
\end{center}
\caption{\label{fig:He_state_EC}
State-selective EC cross sections for the He$^{2+}$-H$(1s)$ collision as functions of the impact energy. 
The present (TW) results are compared with 
the coupled-channel calculations of Winter~\cite{Winter:pra:2007}, 
the lattice time-dependent Schr{\"o}dinger equation approach by Minami {\it et al.}~\cite{Minami:JPB:2008},
the Gaussian basis approach by Toshima and Tawara~\cite{Toshima:1995},
the coupled-channel calculations of Minami {\it et al.}~\cite{Minami:JPB:2008} based on 
and  the  even-tempered  basis  approach  of  Kuang  and  Lin~\cite{Kuang:JPB:1996a,Kuang:JPB:1996b} (Minami*).
"TW" (this work) data are obtained for the straight-line trajectories of colliding nuclei,
"TW NR: Th$^{90+}$+Rh$^{44+}(1s)$" data are obtained for the Th$^{90+}$+Rh$^{44+}(1s)$ collision within the non-relativistic (NR) limit, for the straight-line nuclear collision  trajectories and are presented using the "scaling factor".
}
\end{figure}

In Fig.~\ref{fig:He_state_EC} we present the energy dependence of the EC cross section to the $2s$ and $2p$ states in comparison with available results of other calculations. Generally good agreement is obtained with 
the coupled-channel calculations of Winter~\cite{Winter:pra:2007}, 
the lattice time-dependent Schr{\"o}dinger equation approach by Minami {\it et al.}~\cite{Minami:JPB:2008},
the Gaussian basis approach by Toshima and Tawara~\cite{Toshima:1995},
the coupled-channel calculations of Minami {\it et al.}~\cite{Minami:JPB:2008} based on 
and  the  even-tempered  basis  approach  of  Kuang  and  Lin~\cite{Kuang:JPB:1996a,Kuang:JPB:1996b} (Minami*). 
The scaled non-relativistic data for heavy system Th$^{90+}$+Rh$^{44+}(1s)$ are also presented and agree well with the He$^{2+}$-H$(1s)$ ones.

The calculations are performed using the basis set consisting of the positive-energy 
Dirac-Sturm orbitals $1s$-$8s$, $2p$-$8p$, $3d$-$8d$, $4f$-$8f$ and $1s$-$5s$, $2p$-$5p$, $3d$-$5d$ (in the standard non-relativistic notation) at the projectile and target, respectively, supplemented with Sturm orbitals corresponding to the negative-energy Dirac spectrum. It should be noted that the constructed basis satisfies the dual kinetic balance conditions~\cite{Shabaev:prl:2004} and does not contain the so-called “spurious” states~\cite{Tupitsyn:OS:2004}.
We also note that evaluation of the cross sections at small collision energies demands much more accurate integrating over the impact parameter due to more complicated impact parameter dependency of the probabilities, see subsection~\ref{sec:results:many-electron}. 

%%%%%%%%%%%%%%%%%%%%%%%%%%%%%%%%%%%%%%%%%%%%%%%%%%%%%%%%%%%%%%%%%%%%%  
%%%%%%%%%%%%%%%%%%%%%%%%%%%%%%%%%%%%%%%%%%%%%%%%%%%%%%%%%%%%%%%%%%%%%  

\begin{figure}
\begin{center}
\subfigure[\quad $Z=36$]{%
\includegraphics[width=0.45\textwidth]{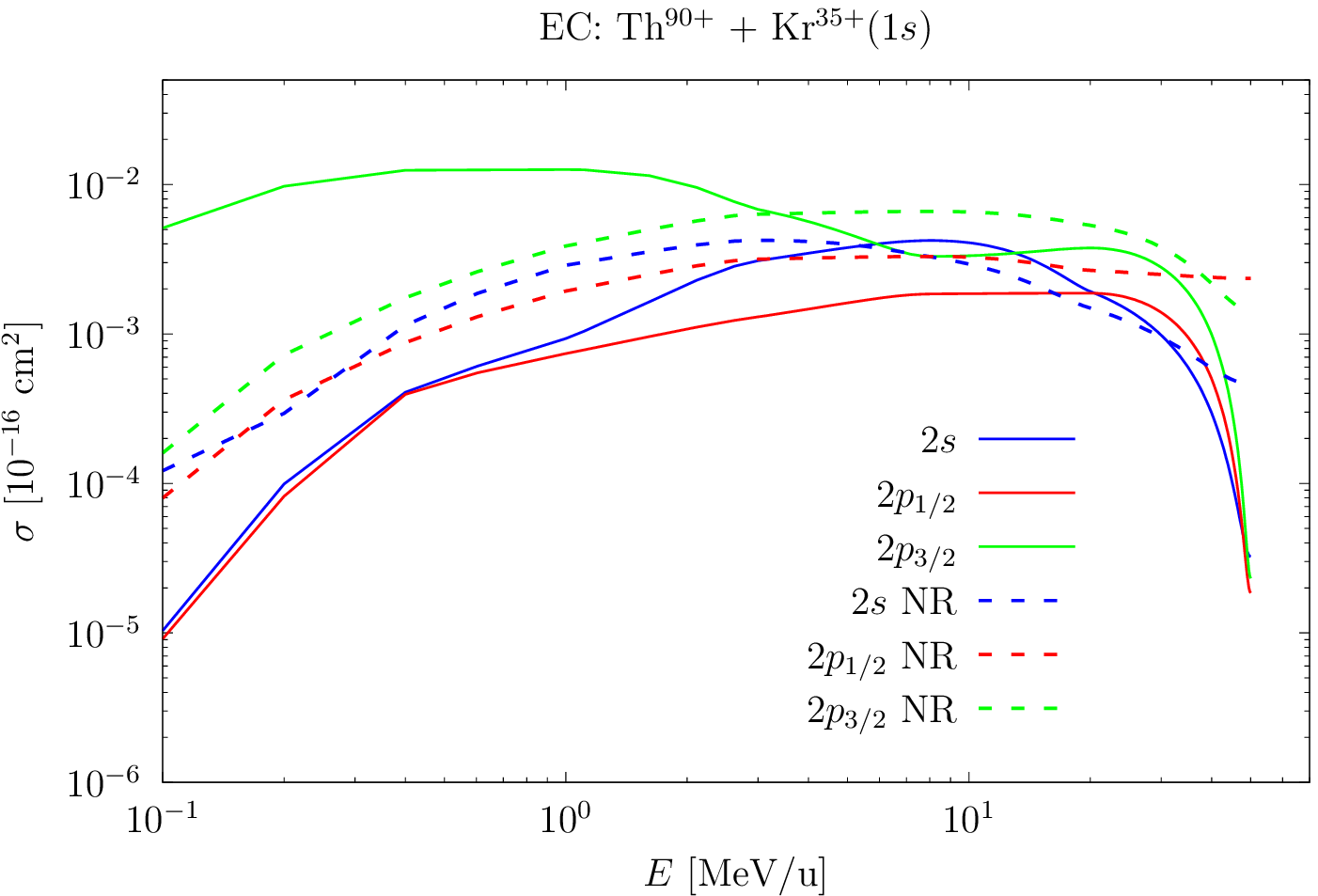}
}
\subfigure[\quad $Z=40$]{%
\includegraphics[width=0.45\textwidth]{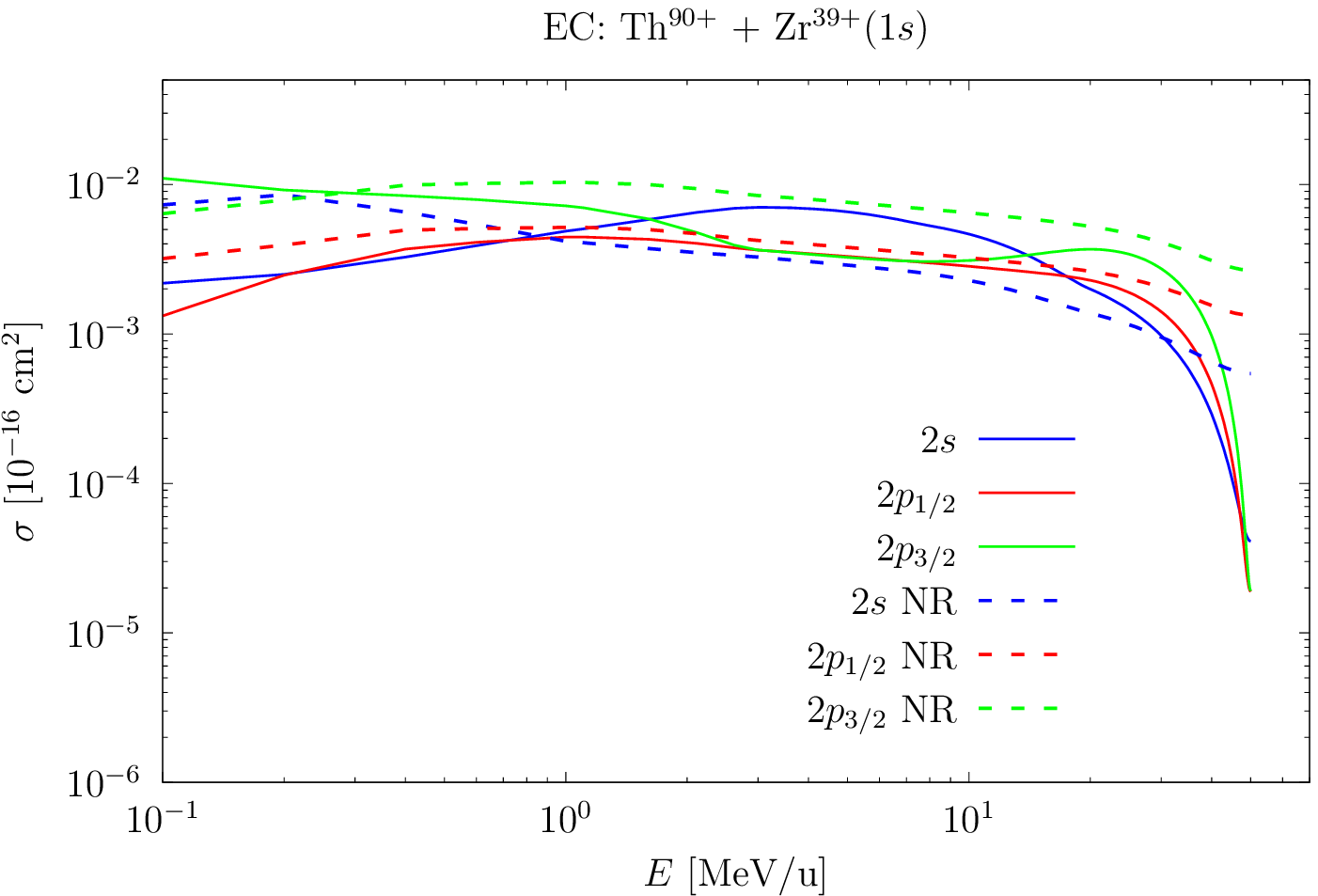}
}\\
\subfigure[\quad $Z=44$]{%
\includegraphics[width=0.45\textwidth]{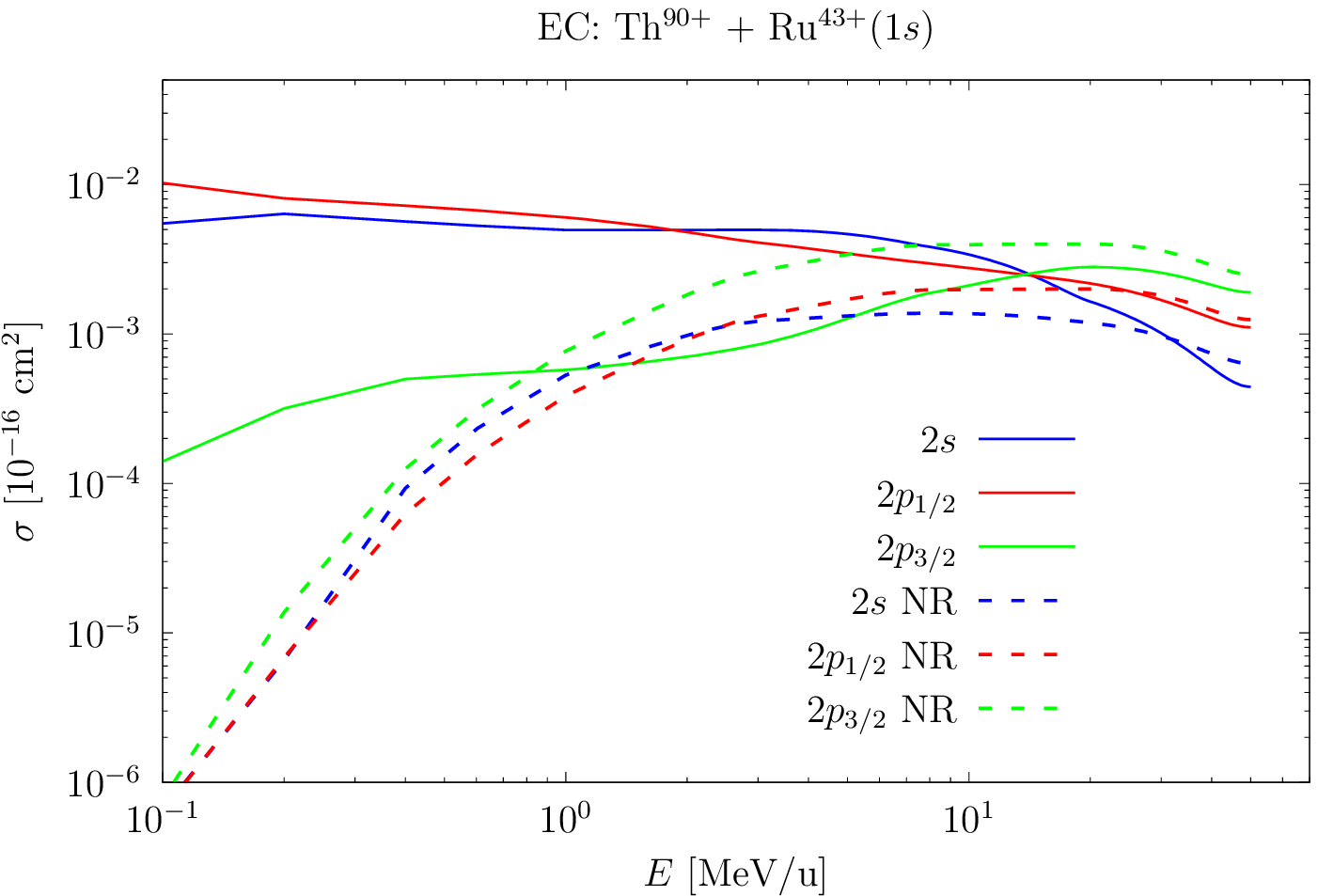}
}
%%%%
\subfigure[\quad $Z=47$]{%
\includegraphics[width=0.45\textwidth]{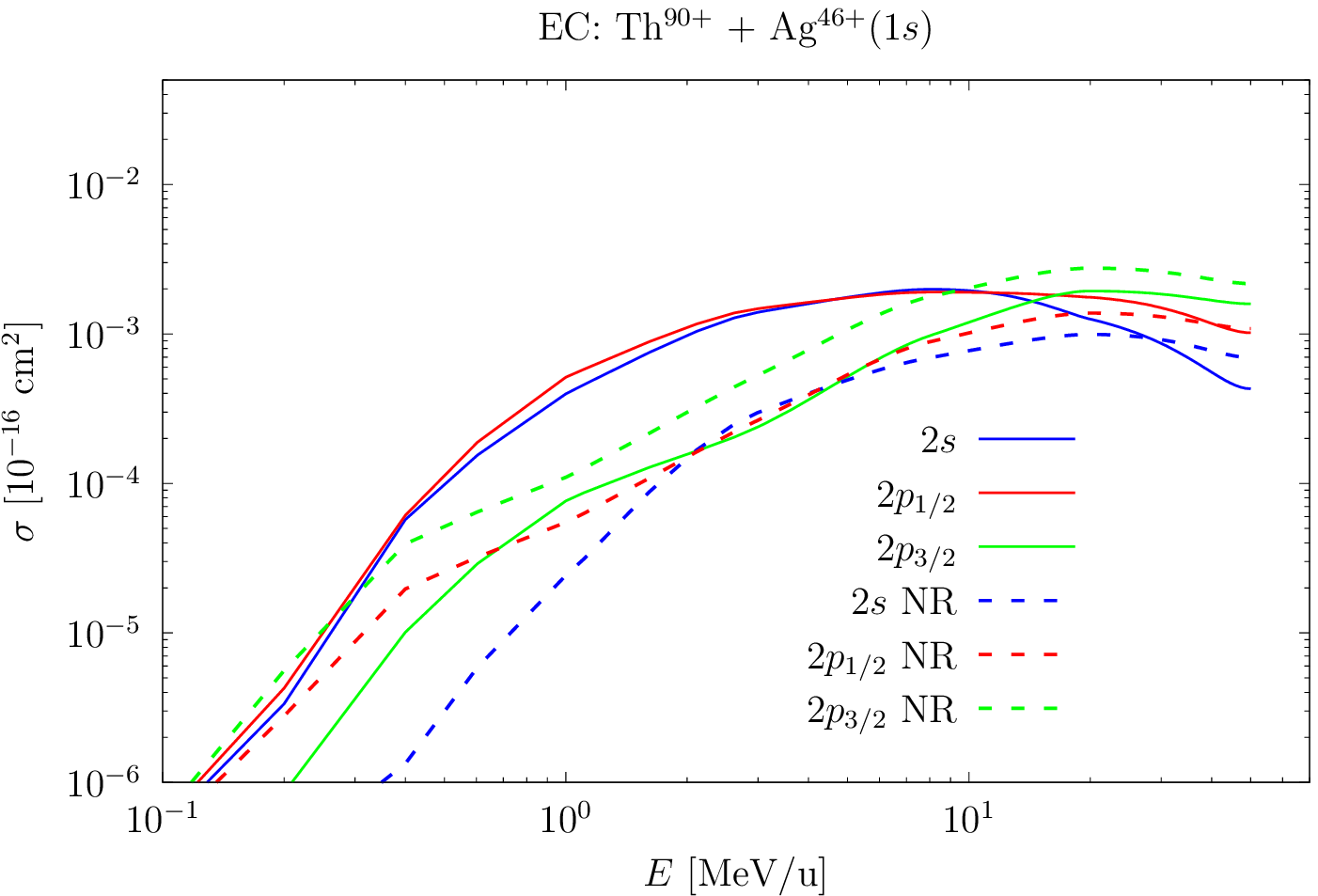}
}
%%%%
%
\end{center}
\caption{\label{fig:Th_EC}
State selective electron capture cross sections for the one-electron Th$^{90+}$-A$^{(Z-1)+}(1s)$ collisions as functions of the impact energy.}
\end{figure}

\begin{figure}
\begin{center}
\subfigure[\quad $Z=36$]{%
\includegraphics[width=0.40\textwidth]{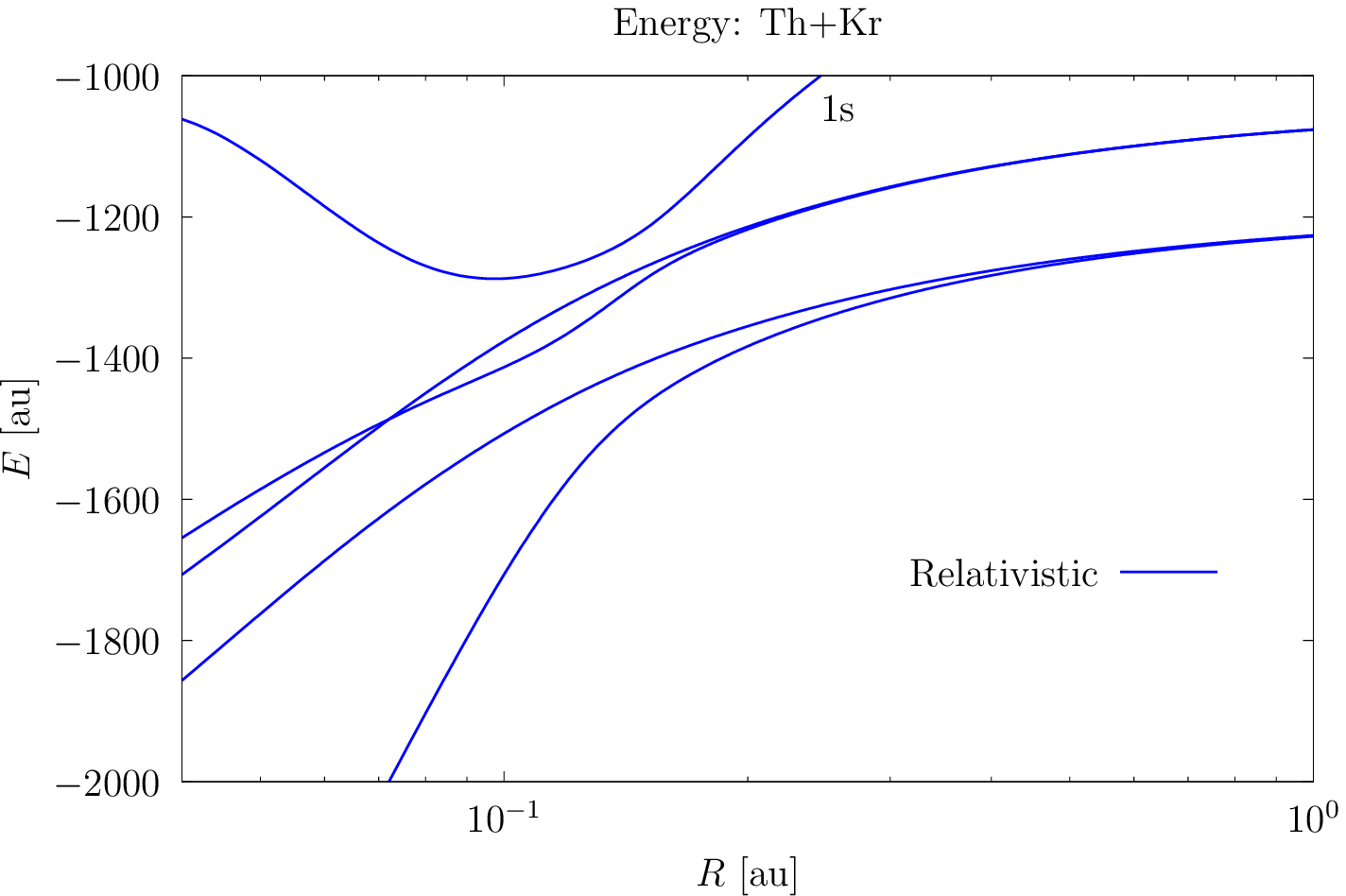}
}
\subfigure[\quad $Z=36$, non-relativistic]{%
\includegraphics[width=0.40\textwidth]{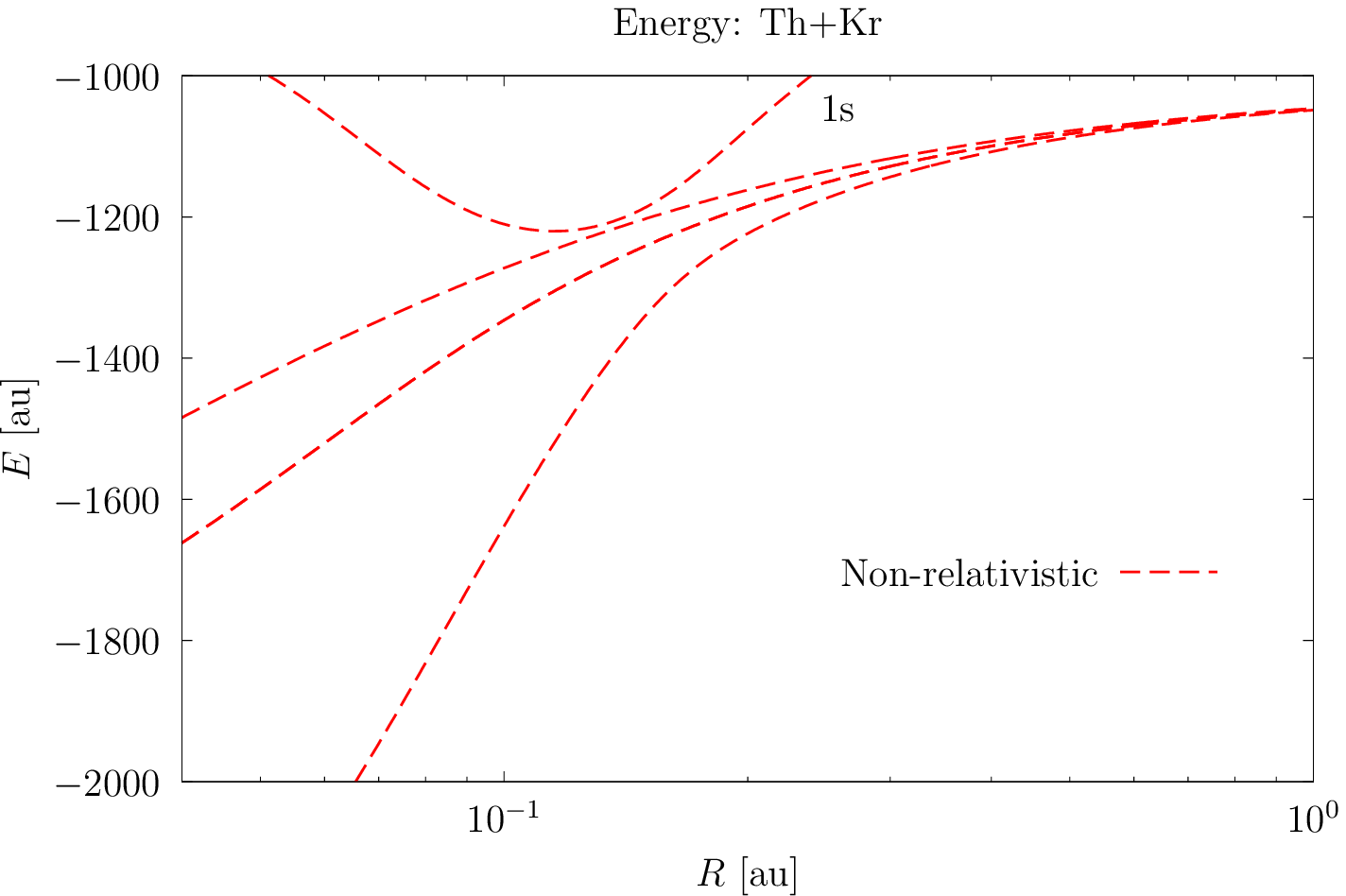}
}\\
%%%%
\subfigure[\quad $Z=40$]{%
\includegraphics[width=0.4\textwidth]{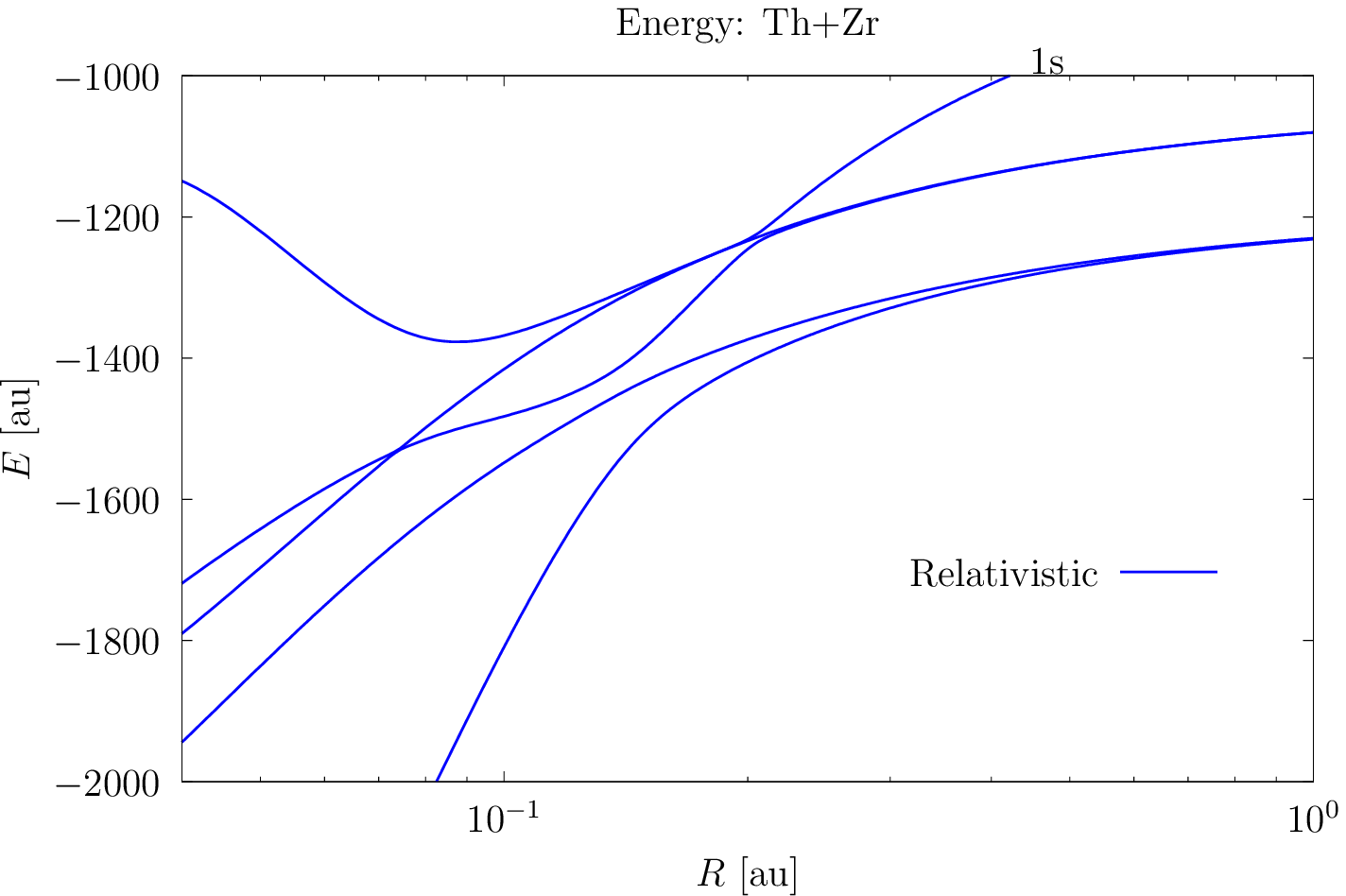}
}
\subfigure[\quad $Z=40$, non-relativistic]{%
\includegraphics[width=0.4\textwidth]{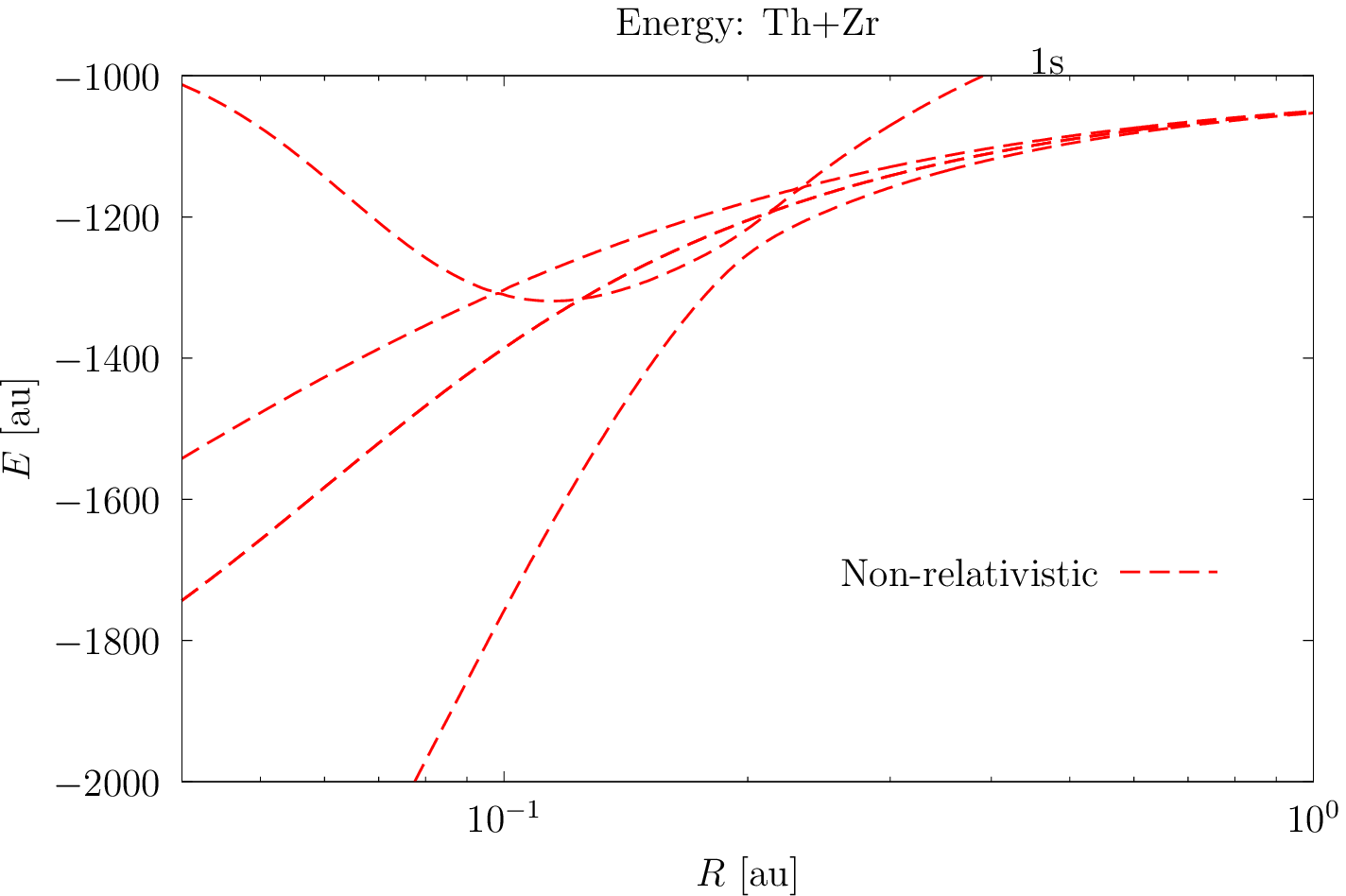}
}\\
%%%%
\subfigure[\quad $Z=44$]{%
\includegraphics[width=0.4\textwidth]{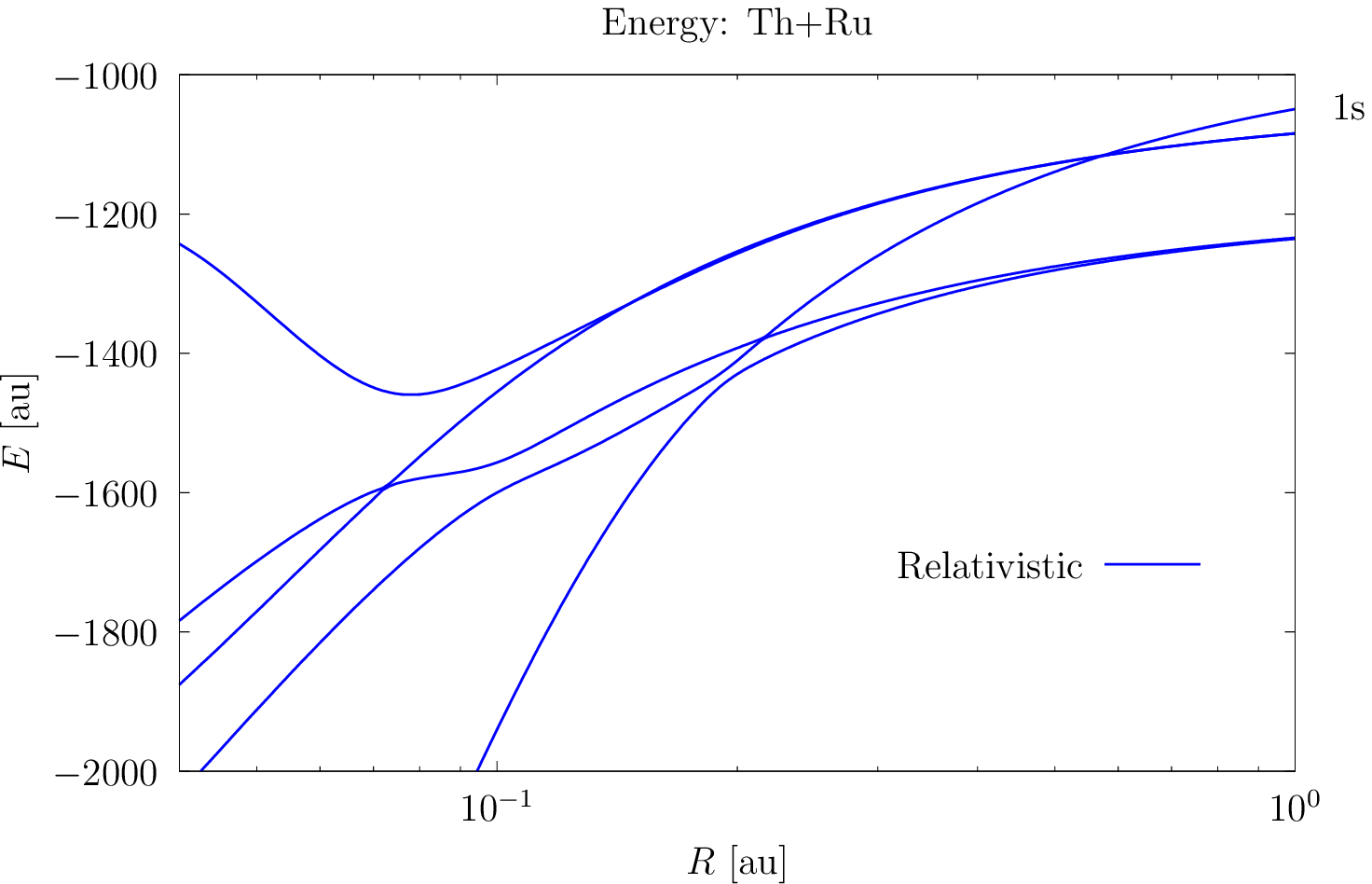}
}
\subfigure[\quad $Z=44$, non-relativistic]{%
\includegraphics[width=0.4\textwidth]{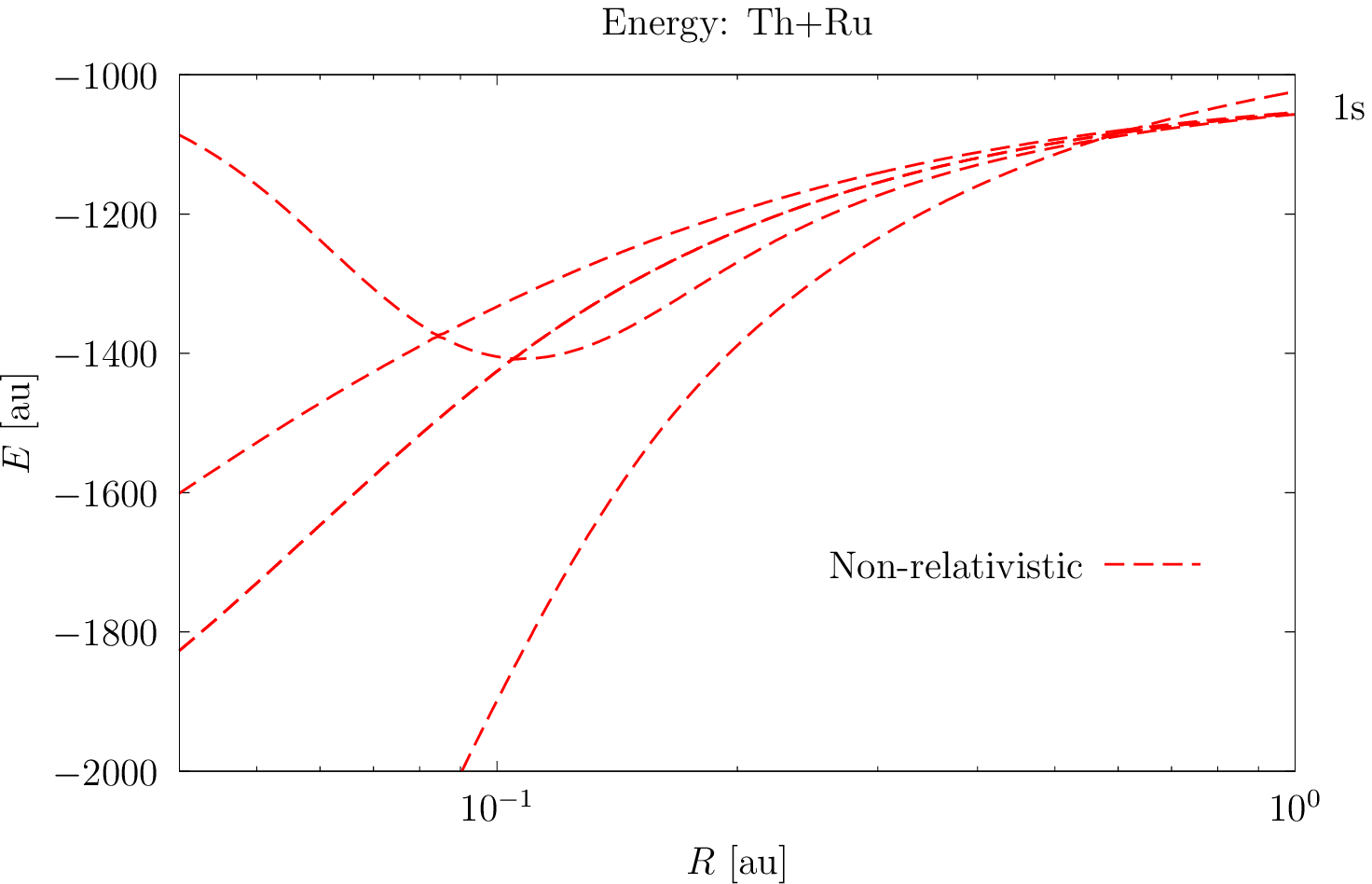}
}\\
%%%%
\subfigure[\quad $Z=47$]{%
\includegraphics[width=0.4\textwidth]{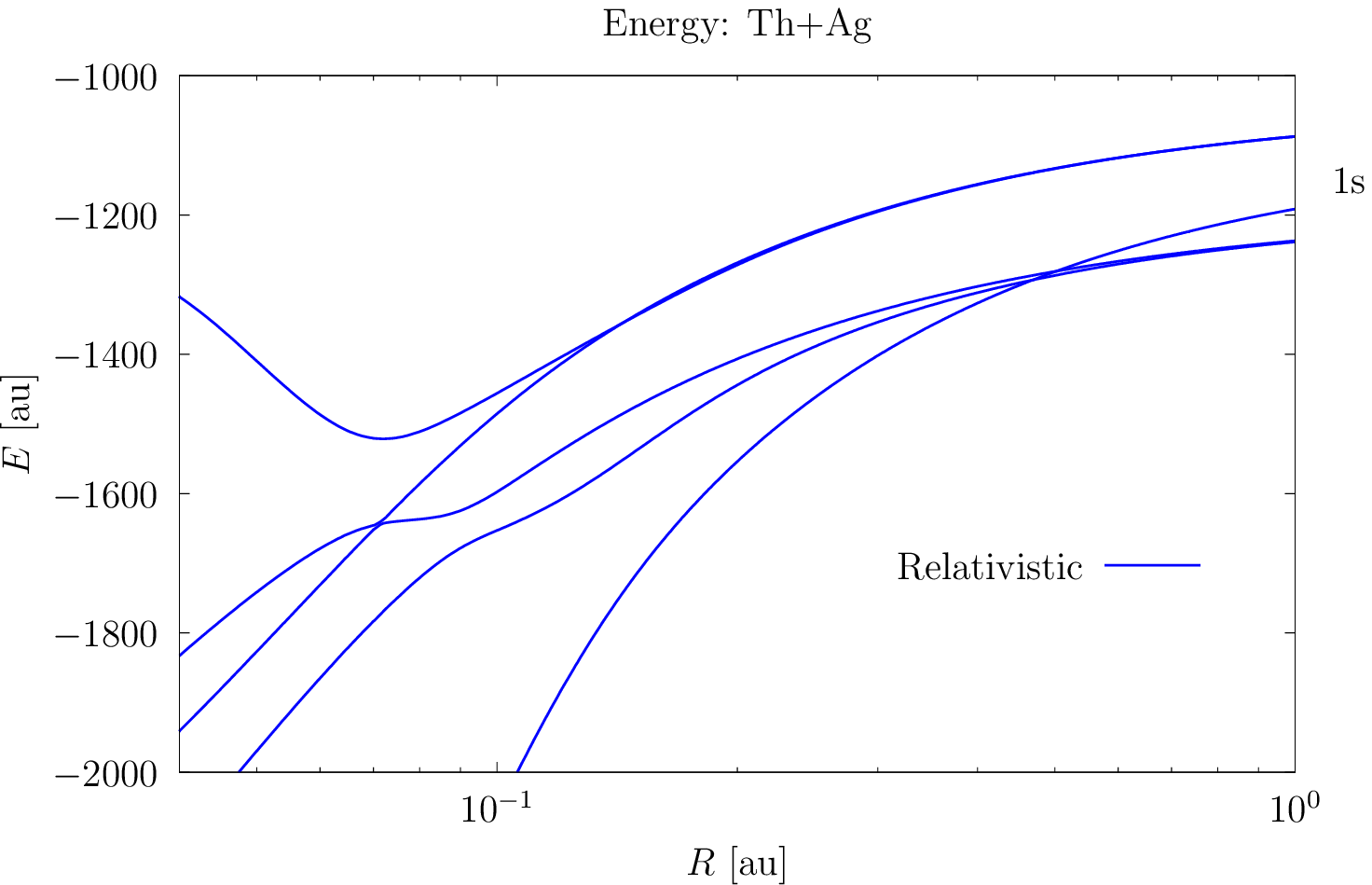}
}
\subfigure[\quad $Z=47$, non-relativistic]{%
\includegraphics[width=0.4\textwidth]{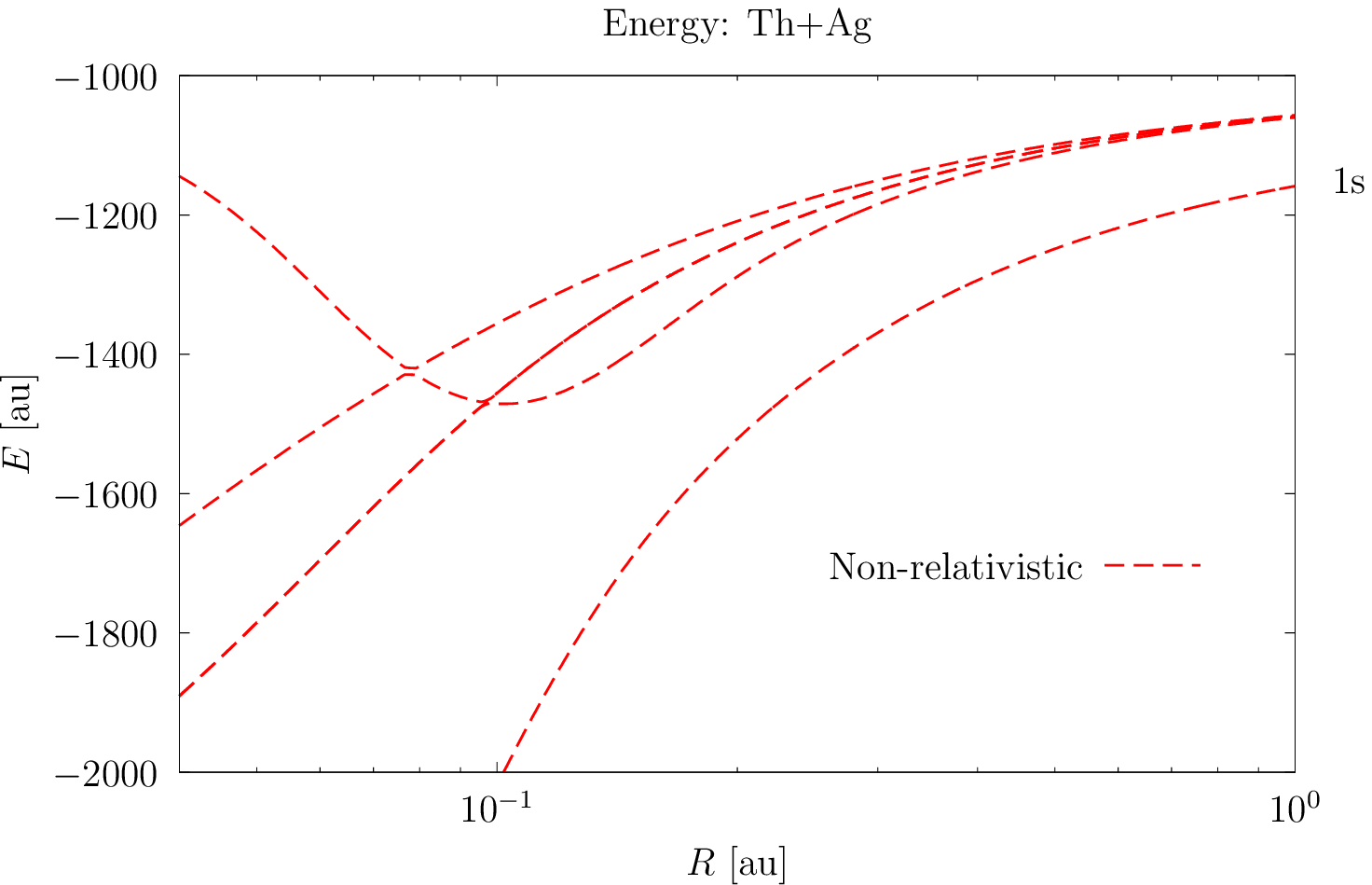}
}
%%%%
%
\end{center}
\caption{\label{fig:energy}
Correlation diagrams for the Th$^{90+}$-A$^{(Z-1)+}(1s)$ systems. The energy level (at large internuclear distances) of the $1s$ target state is indicated by "$1s$".}
\end{figure}

%%%%%%%%%%%%%%%%%%%%%%%%%%%%%%%%%%%%%%%%%%%%%%%%%%%%%%%%%%%%%%%%%%%%%  
The results of EC cross section calculations for the collisions of bare thorium ($Z=90$) nucleus with hydrogenlike targets: krypton ($Z=36$), zirconium ($Z=40$), ruthenium ($Z=44$) and silver ($Z=47$) are presented in Fig.~\ref{fig:Th_EC} as functions of the collision energy. 
%%%%%%%%%%%%%%%%%%%%%%%%%%%%%%%%%%%%%%%%%%%%%%%%%%%%%%%%%%%%%%%%%%%%%  
%%%%%%%%%%%%%%%%%%%%%%%%%%%%%%%%%%%%%%%%%%%%%%%%%%%%%%%%%%%%%%%%%%%%%  
The data, obtained in the non-relativistic limit are also displayed in the figure for comparison. Depending on the target one could see totally different behavior of the EC cross section (the relativistic and non-relativistic as well) in the low-energy region. In order to adept at the data we present correlation diagrams (dependencies of the quasimolecular energy levels on the internuclear distance) in Fig.~\ref{fig:energy}.
%%%%%%%%%%%%%%%%%%%%%%%%%%%%%%%%%%%%%%%%%%%%%%%%%%%%%%%%%%%%%%%%%%%%%  
%%%%%%%%%%%%%%%%%%%%%%%%%%%%%%%%%%%%%%%%%%%%%%%%%%%%%%%%%%%%%%%%%%%%%  
According to the adiabatic picture (relevant in the low-energy regime), in course of the collision the electron moves along a quasimolecular energy level and can transit to another ones only at the level (quasi)crossings (see, for example, Ref.~\cite{Eichler:2005}). 
Let us start our discussion of the obtained results with the simplest case of the Th$^{90+}$-Ag$^{46+}(1s)$ collision, Figs.~\ref{fig:energy}(g)-(h). 
One could see that the initially occupied target $1s$ level does not cross (for the non-relativistic case) or crosses other levels only at large internuclear distances (for the relativistic case), where level couplings and, hence, transition probabilities are negligible. That is why the EC cross sections go to zero at the low collision energies, Fig.~\ref{fig:Th_EC}(d). In the case of Th$^{90+}$-Zr$^{39+}(1s)$ collision, the initially target $1s$ level crosses the initially projectile $n=2$ levels with strong coupling and disturbing ones, Figs.~\ref{fig:energy}(c)-(d), resulting in large values of the EC cross sections in the low-energy regime in the relativistic as well as the non-relativistic calculations, Fig.~\ref{fig:Th_EC}(b). 
There are strong differences in the relativistic and non-relativistic correlation diagrams in the case of Th+Ru, Figs.~\ref{fig:energy}(e)-(f). The presence and absence of the crossings (with strong coupling) results in totally different behavior of the EC cross section at the low energies for the relativistic and non-relativistic regimes, Fig.~\ref{fig:Th_EC}(c). An intermediate variant corresponds to the Th$^{90+}$-Kr$^{35+}(1s)$ collision, Fig.~\ref{fig:Th_EC}(a), and correlation diagrams, Figs.~\ref{fig:energy}(a)-(b). 

\begin{figure}
\begin{center}
\subfigure[\quad $Z=12$]{%
\includegraphics[width=0.45\textwidth]{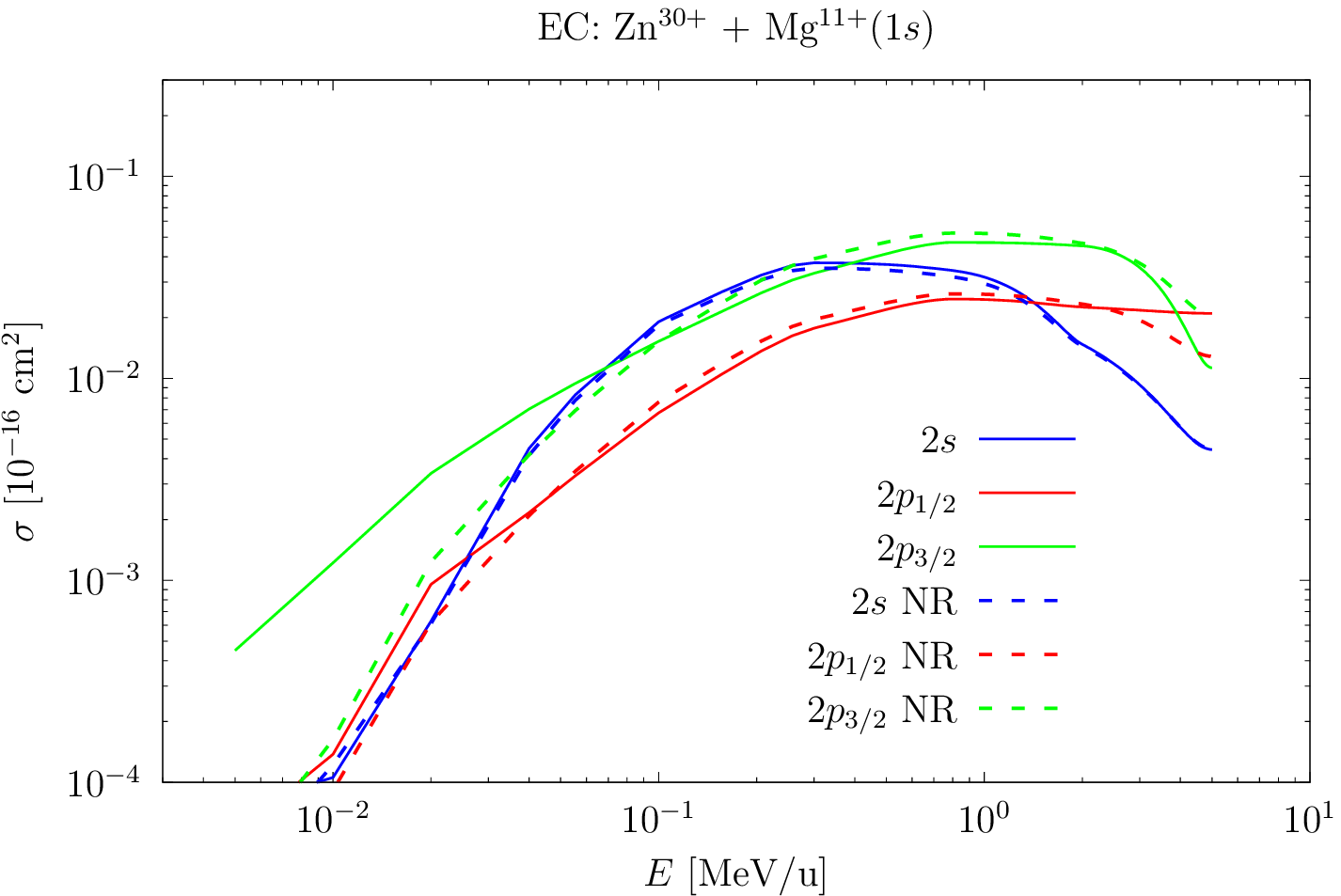}
}
\subfigure[\quad $Z=13$]{%
\includegraphics[width=0.45\textwidth]{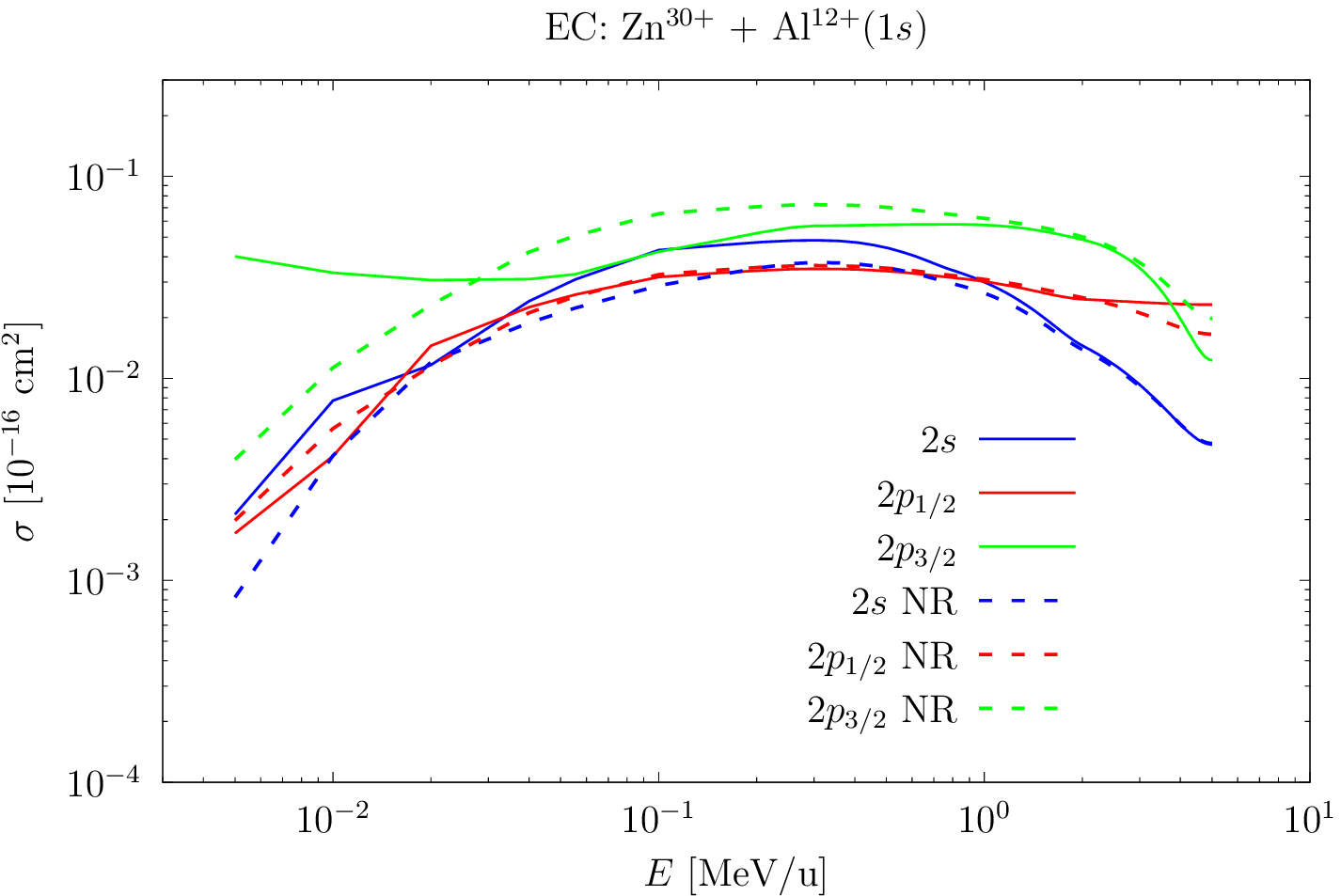}
}\\
\subfigure[\quad $Z=14$]{%
\includegraphics[width=0.45\textwidth]{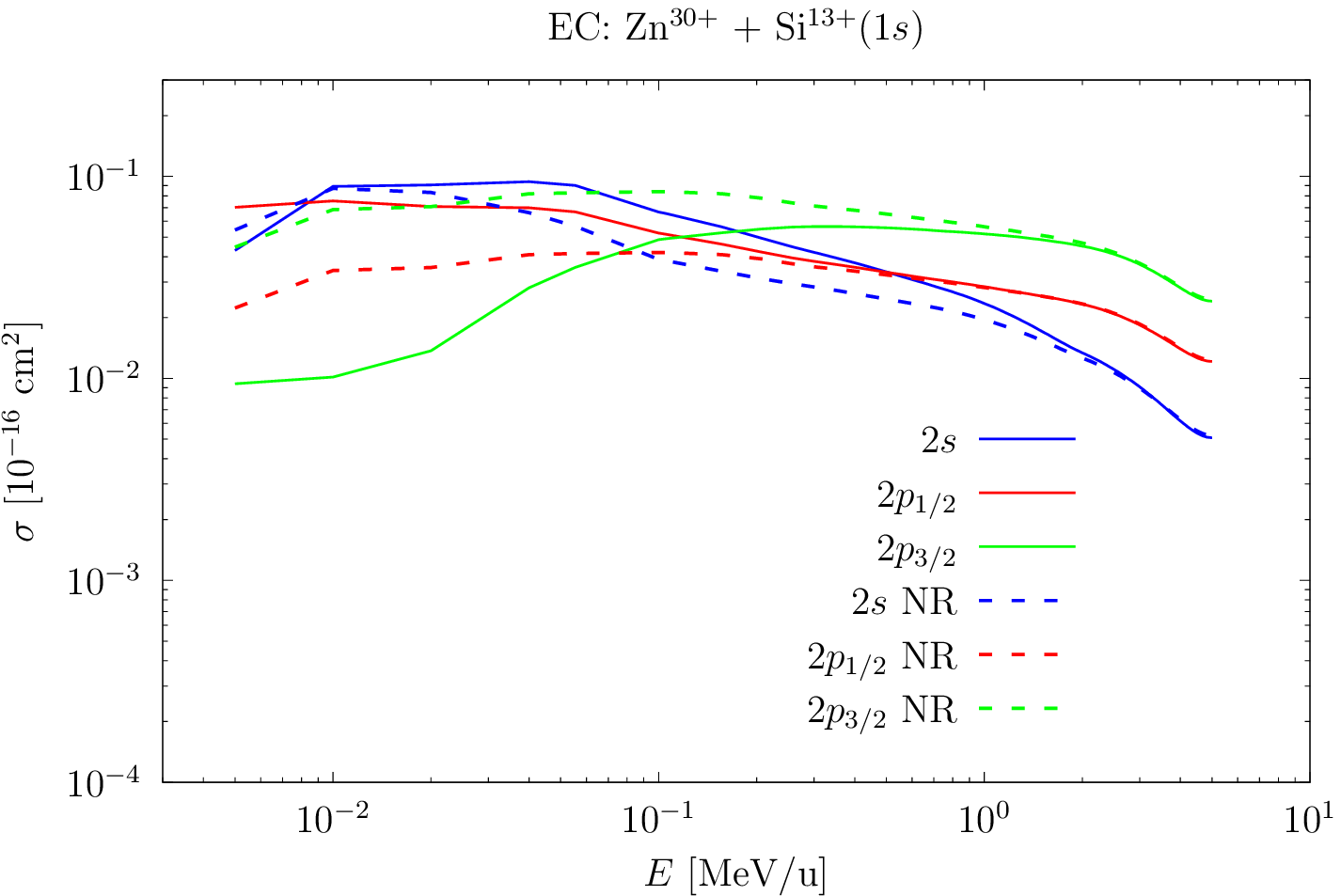}
}
%%%%
\subfigure[\quad $Z=15$]{%
\includegraphics[width=0.45\textwidth]{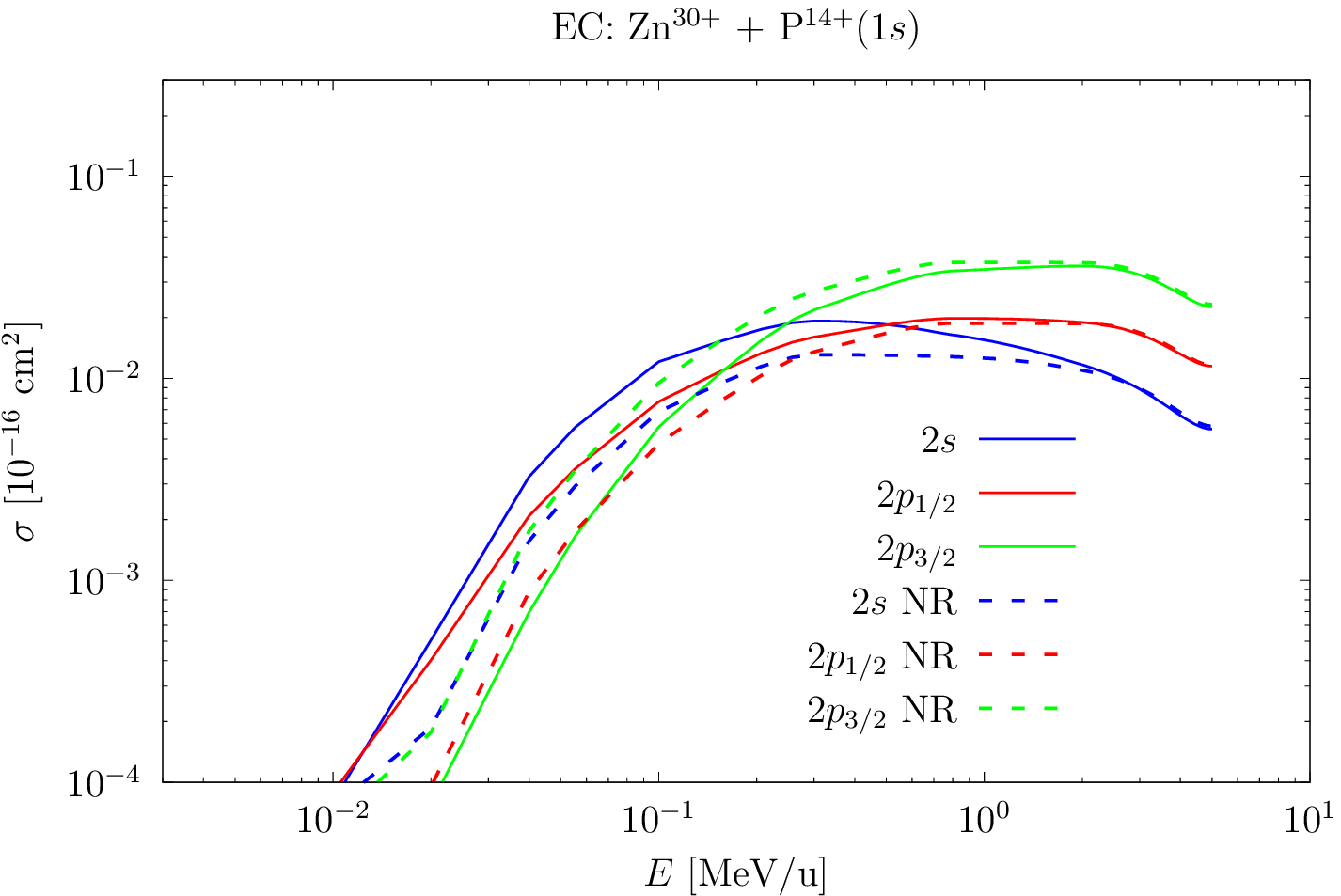}
}
%%%%
%
\end{center}
\caption{\label{fig:Zn_EC}
State selective electron capture cross sections for the one-electron Zn$^{30+}$-A$^{(Z-1)+}(1s)$ collisions as functions of the impact energy.}
\end{figure}

\begin{figure}
\begin{center}
\subfigure[\quad $Z=12$]{%
\includegraphics[width=0.4\textwidth]{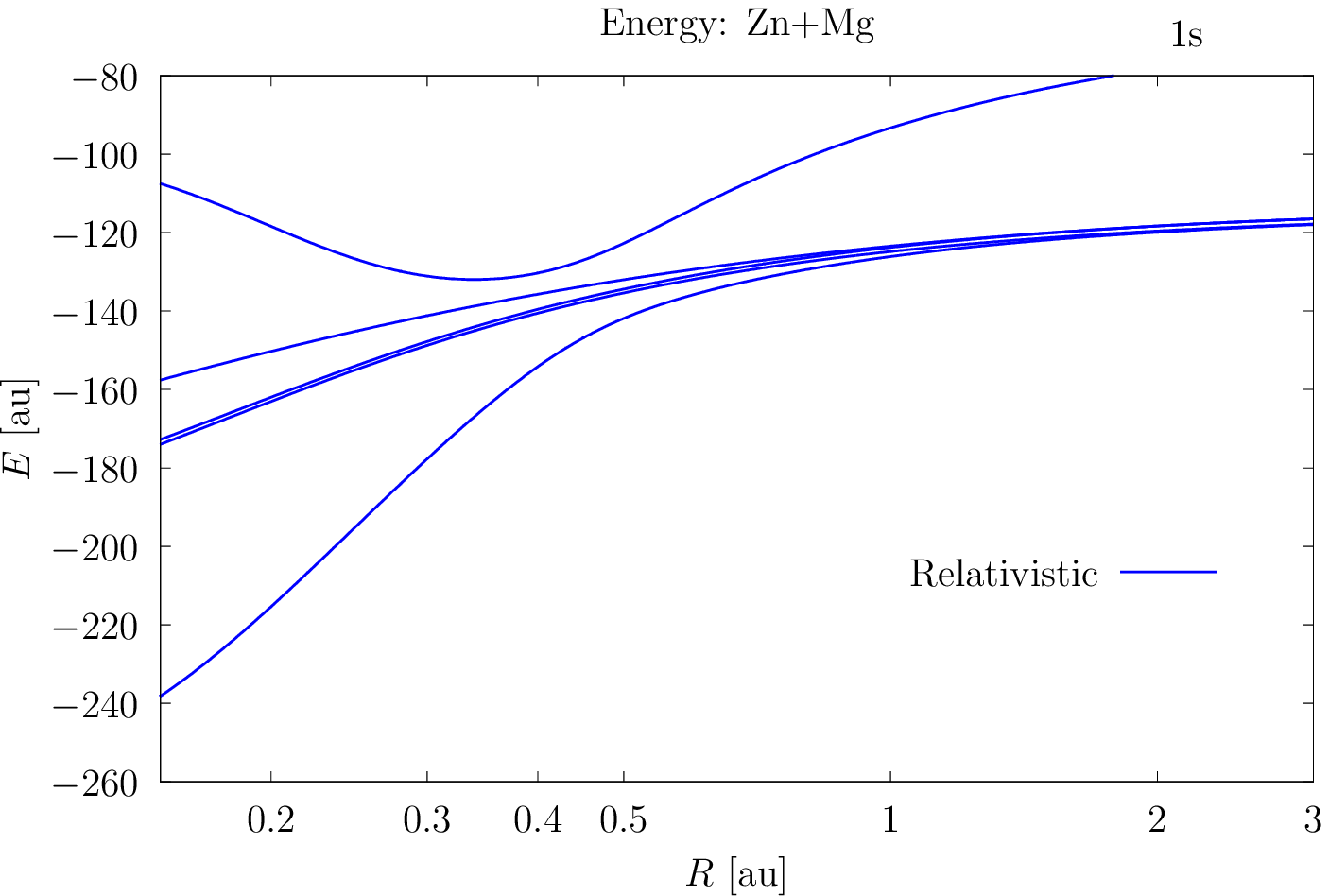}
}
\subfigure[\quad $Z=12$, non-relativistic]{%
\includegraphics[width=0.4\textwidth]{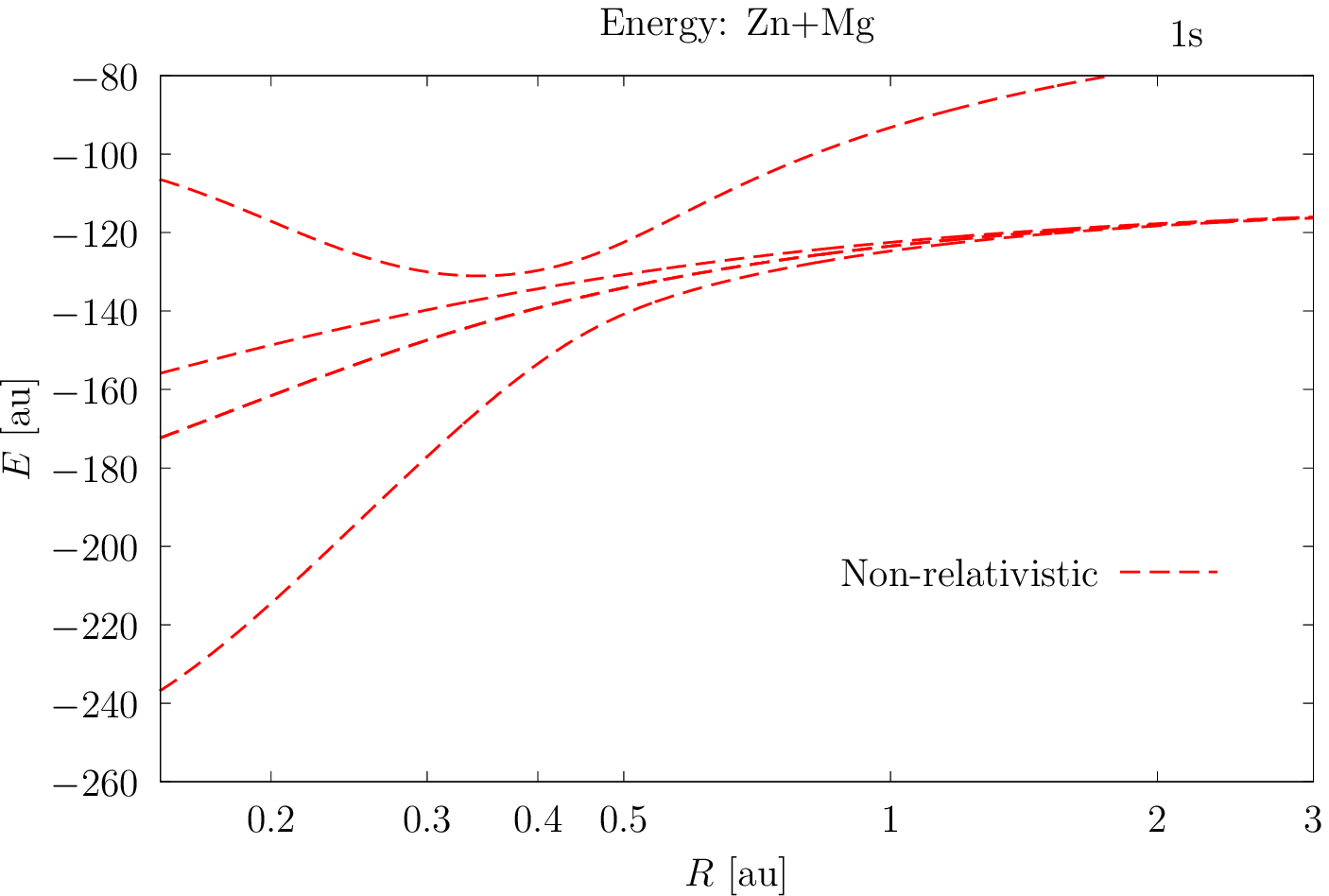}
}\\
%%%%
\subfigure[\quad $Z=13$]{%
\includegraphics[width=0.4\textwidth]{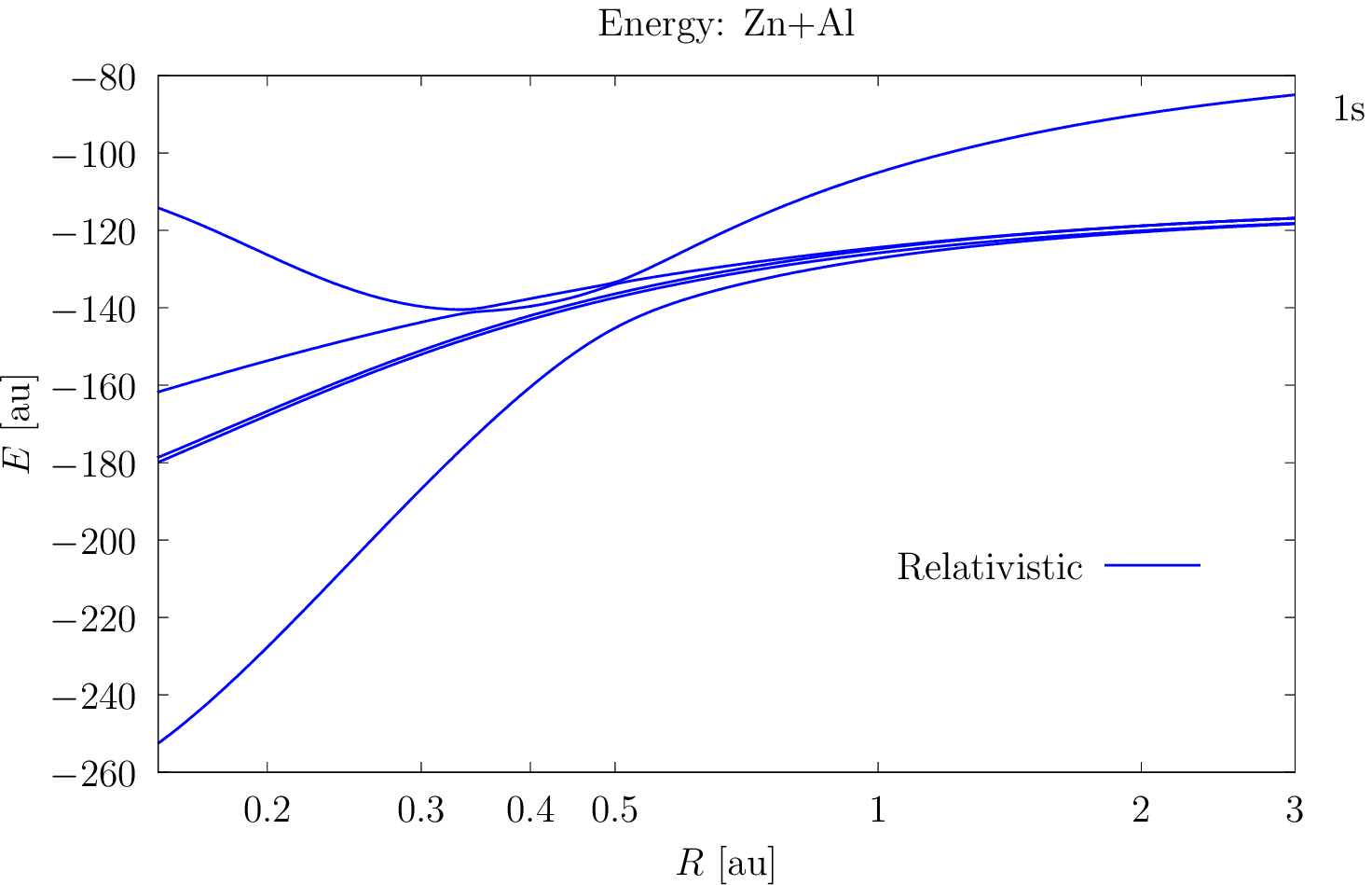}
}
\subfigure[\quad $Z=13$, non-relativistic]{%
\includegraphics[width=0.4\textwidth]{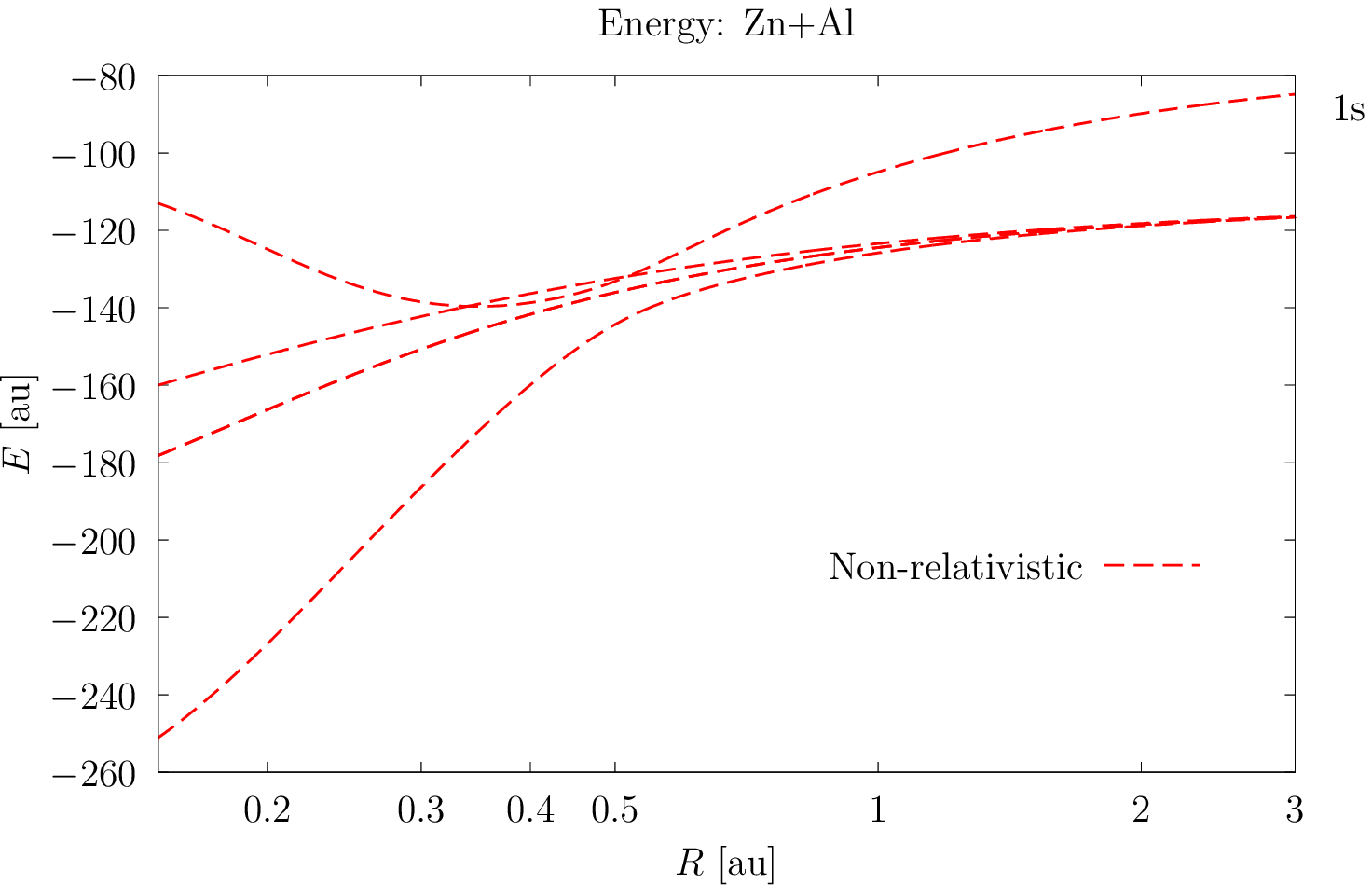}
}\\
%%%%
\subfigure[\quad $Z=14$]{%
\includegraphics[width=0.4\textwidth]{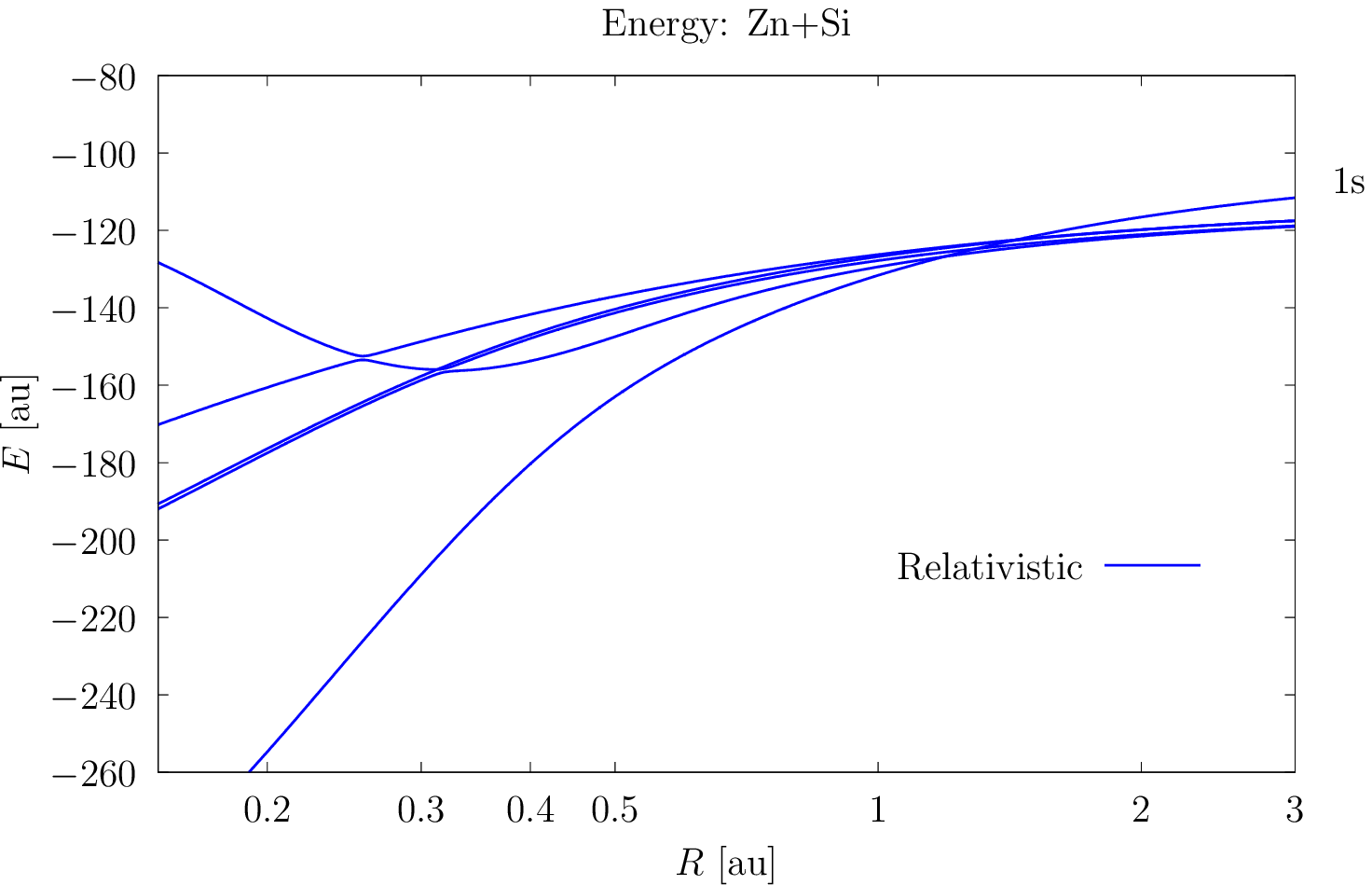}
}
\subfigure[\quad $Z=14$, non-relativistic]{%
\includegraphics[width=0.4\textwidth]{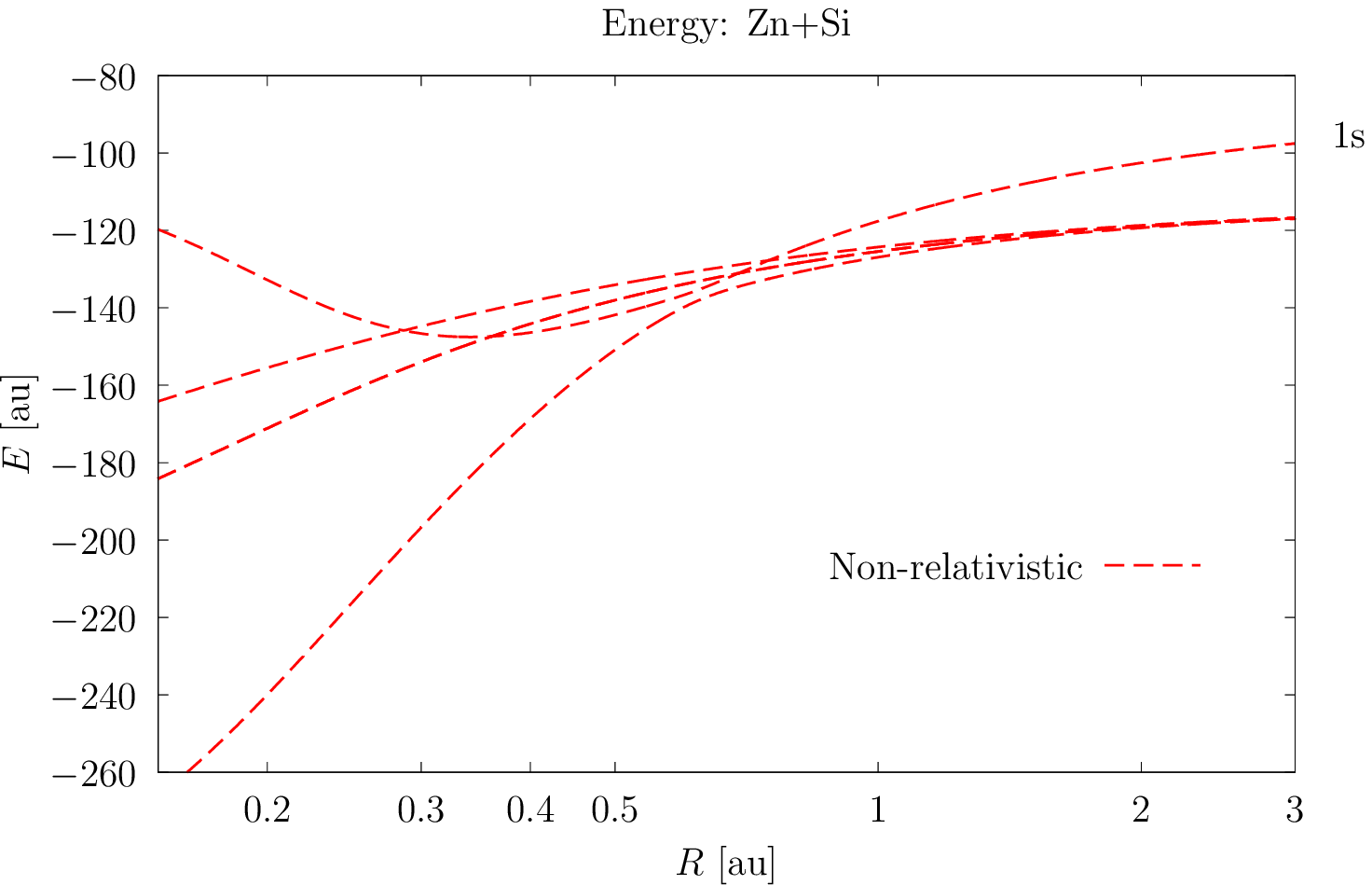}
}\\
%%%%
\subfigure[\quad $Z=15$]{%
\includegraphics[width=0.4\textwidth]{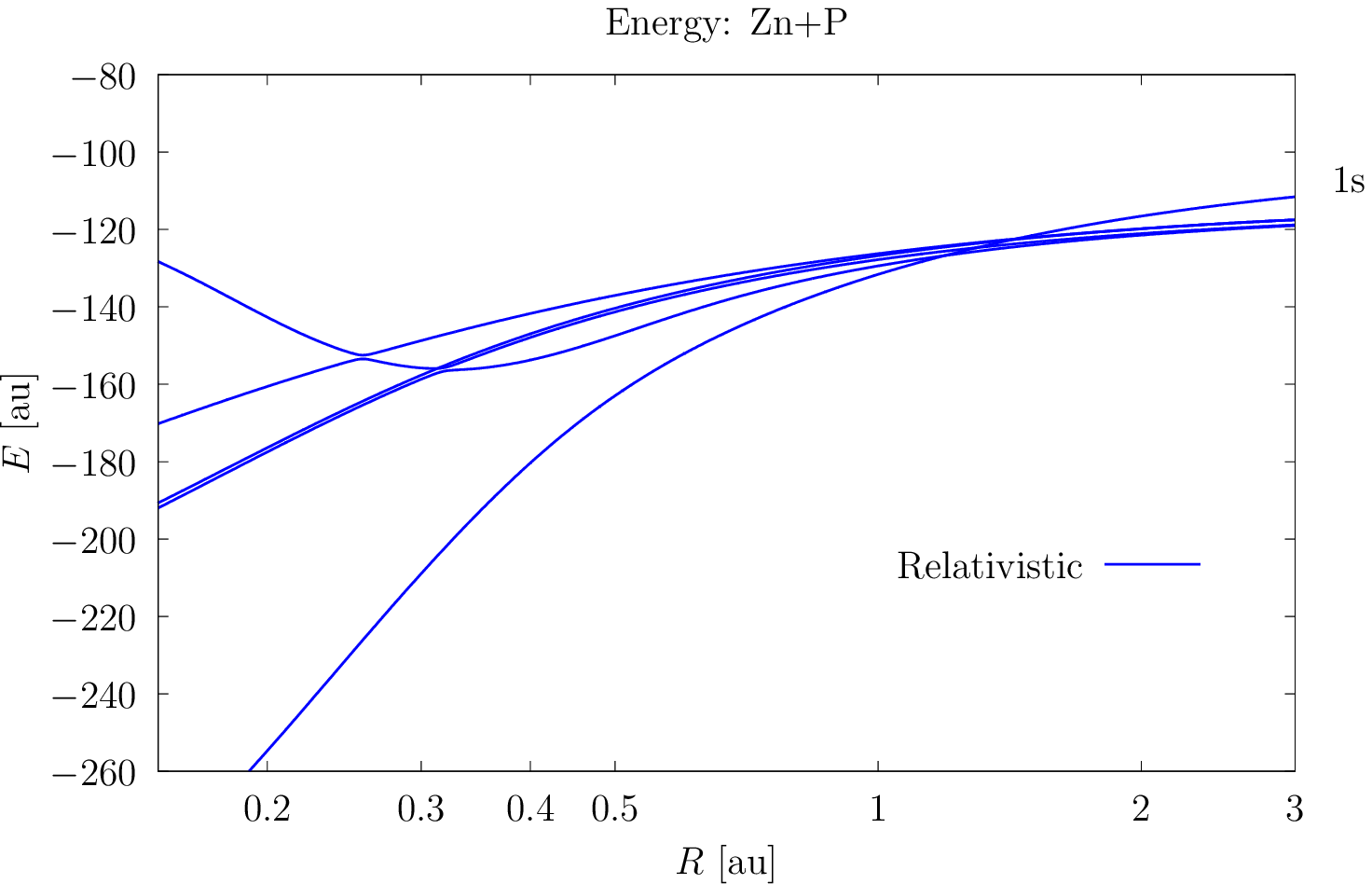}
}
\subfigure[\quad $Z=15$, non-relativistic]{%
\includegraphics[width=0.4\textwidth]{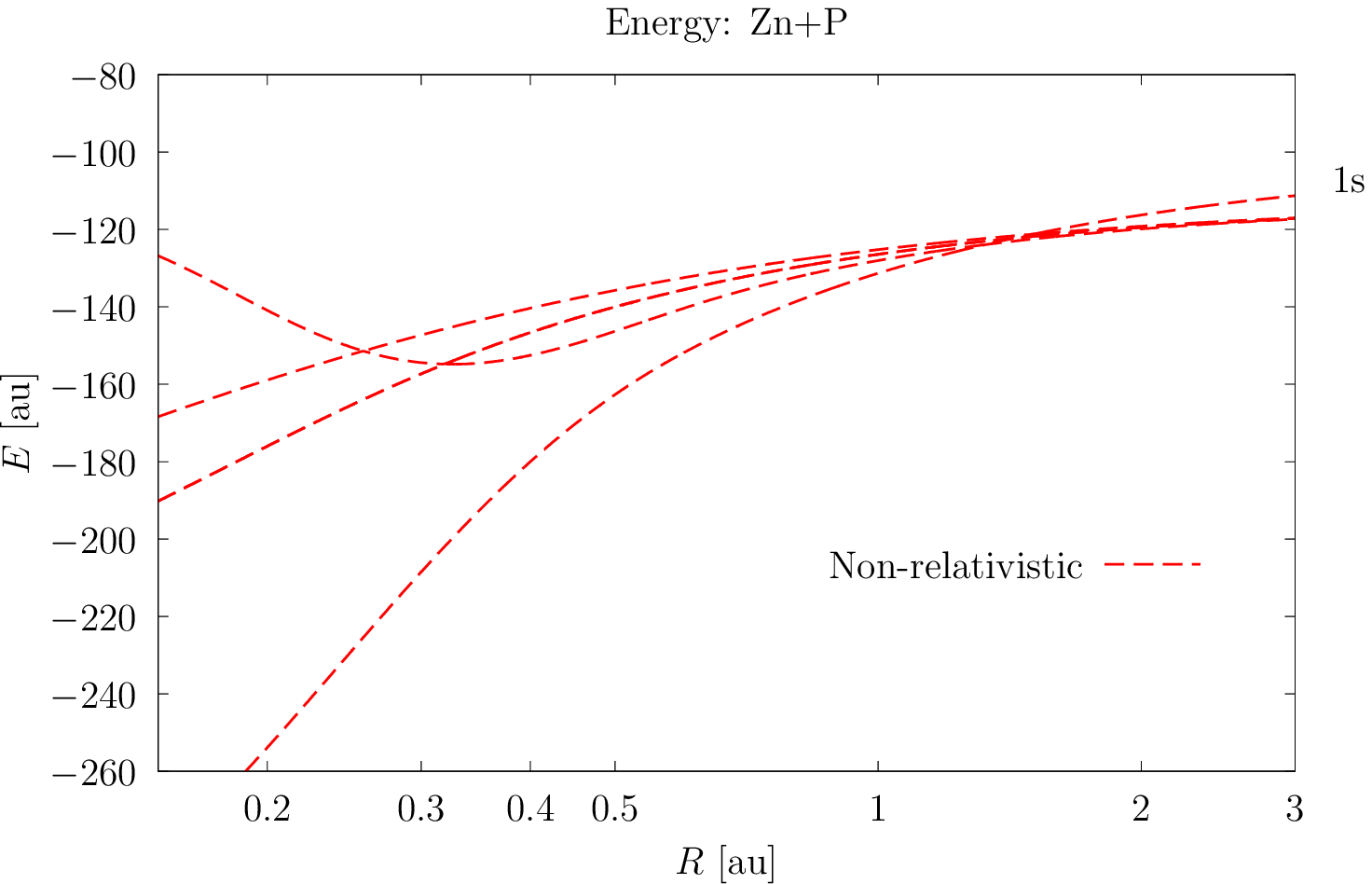}
}
%%%%
%
\end{center}
\caption{\label{fig:energy_Zn}
Correlation diagrams for the Zn$^{30+}$-A$^{(Z-1)+}(1s)$ systems. The energy level (at large internuclear distances) of the $1s$ target state is indicated by "$1s$".}
\end{figure}

Summarizing, extremely huge relativistic effects take place for the EC cross sections of the Th$^{90+}$-Ru$^{43+}(1s)$ collision in the low-energy regime. Also we carried out similar investigation for the three times less nuclear charges of colliding ions. 
The results of EC cross section calculations for the collisions of bare zinc ($Z=30$) nucleus with hydrogenlike targets: magnesium ($Z=12$), aluminium ($Z=13$), silicon ($Z=14$) and phosphorus ($Z=15$) are presented in Fig.~\ref{fig:Zn_EC} as functions of the collision energy. 
%%%%%%%%%%%%%%%%%%%%%%%%%%%%%%%%%%%%%%%%%%%%%%%%%%%%%%%%%%%%%%%%%%%%%  
%%%%%%%%%%%%%%%%%%%%%%%%%%%%%%%%%%%%%%%%%%%%%%%%%%%%%%%%%%%%%%%%%%%%%  
The corresponding correlation diagrams are shown in Fig.~\ref{fig:energy_Zn}. 
%%%%%%%%%%%%%%%%%%%%%%%%%%%%%%%%%%%%%%%%%%%%%%%%%%%%%%%%%%%%%%%%%%%%%  
%%%%%%%%%%%%%%%%%%%%%%%%%%%%%%%%%%%%%%%%%%%%%%%%%%%%%%%%%%%%%%%%%%%%%  
Again one can see resonance and non-resonance behaviors in the low-energy regime. The relativistic effects definitely play a role, but the data analysis becomes not so clear. The correlation diagram structure differences do not look to such an essential. So far one could note that the relativistic effects influence by the strongest way on EC to the $2p_{3/2}$ state. We proceed with the discussion of not too heavy ion-atom collisions in the next subsection.

\subsection{Many-electron systems}  
\label{sec:results:many-electron}  
% \subsection{C$^{4+}$-He collision}  
% \label{sec:results:C}  
%%%%%%%%%%%%%%%%%%%%%%%%%%%%%%%%%%%%%%%%%%%%%%%%%%%%%%%%%%%%%%%%%%%%%%%%%%%%%%%% 
Collision of heavy bare nuclei with hydrogenlike ions is a rather exotic object for experimental study. That is why in the present subsection we consider (quasi)resonance EC processes in many-electron systems.

\begin{figure}
\begin{center}
\subfigure[\quad Total and  $n$-selective SEC cross sections for electron capture to the C$^{3+}$ ion.]{%
\includegraphics[width=0.450\textwidth]{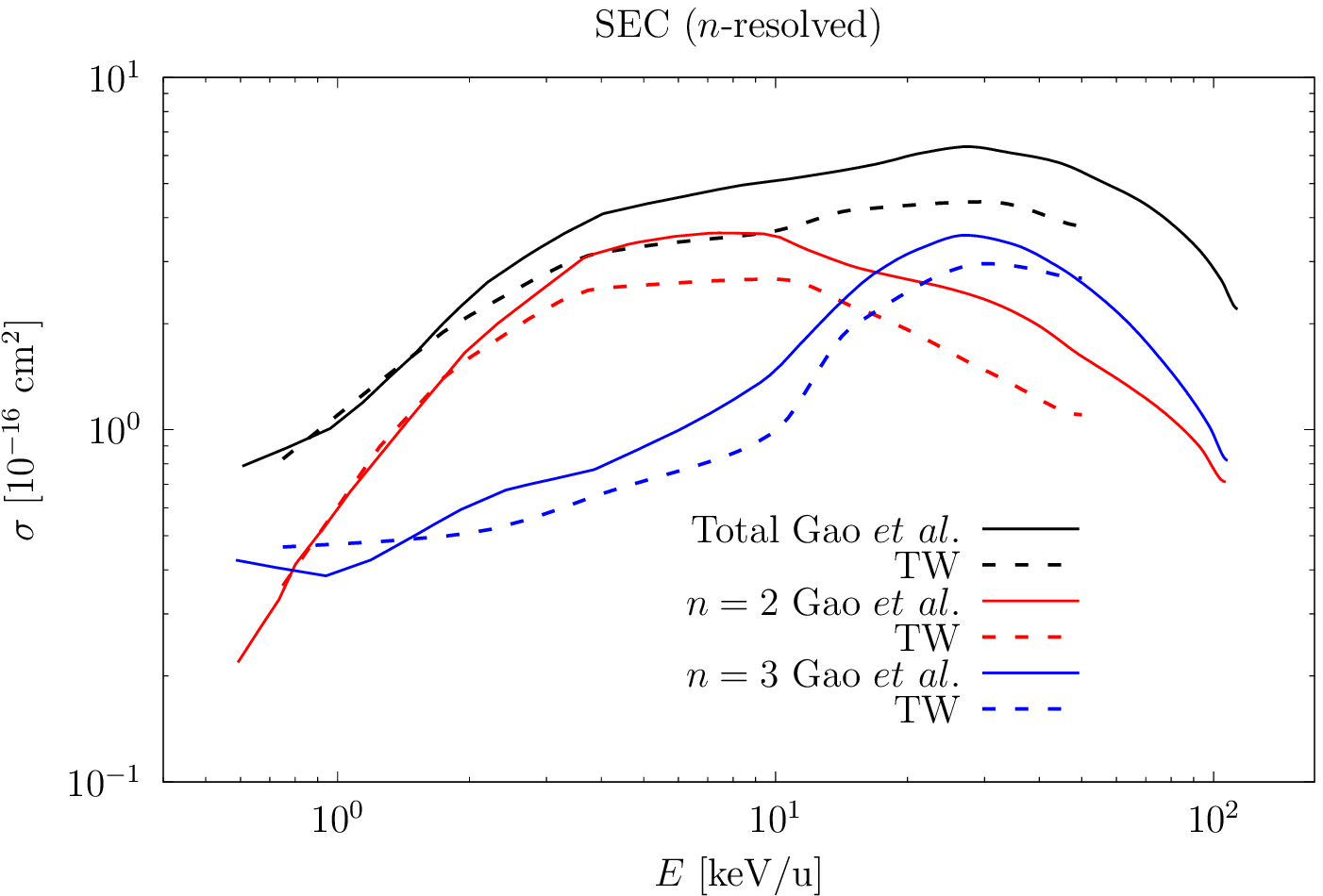}
% \includegraphics[width=0.50\textwidth]{../He_C4+/V_scr4/Graph/DEC_total.eps}
% % \includegraphics[width=0.45\textwidth]{Graph_E3+/1.eps}
}
\subfigure[\quad $2l$-selective SEC cross sections for electron capture to the C$^{3+}$ ion.]{%
\includegraphics[width=0.450\textwidth]{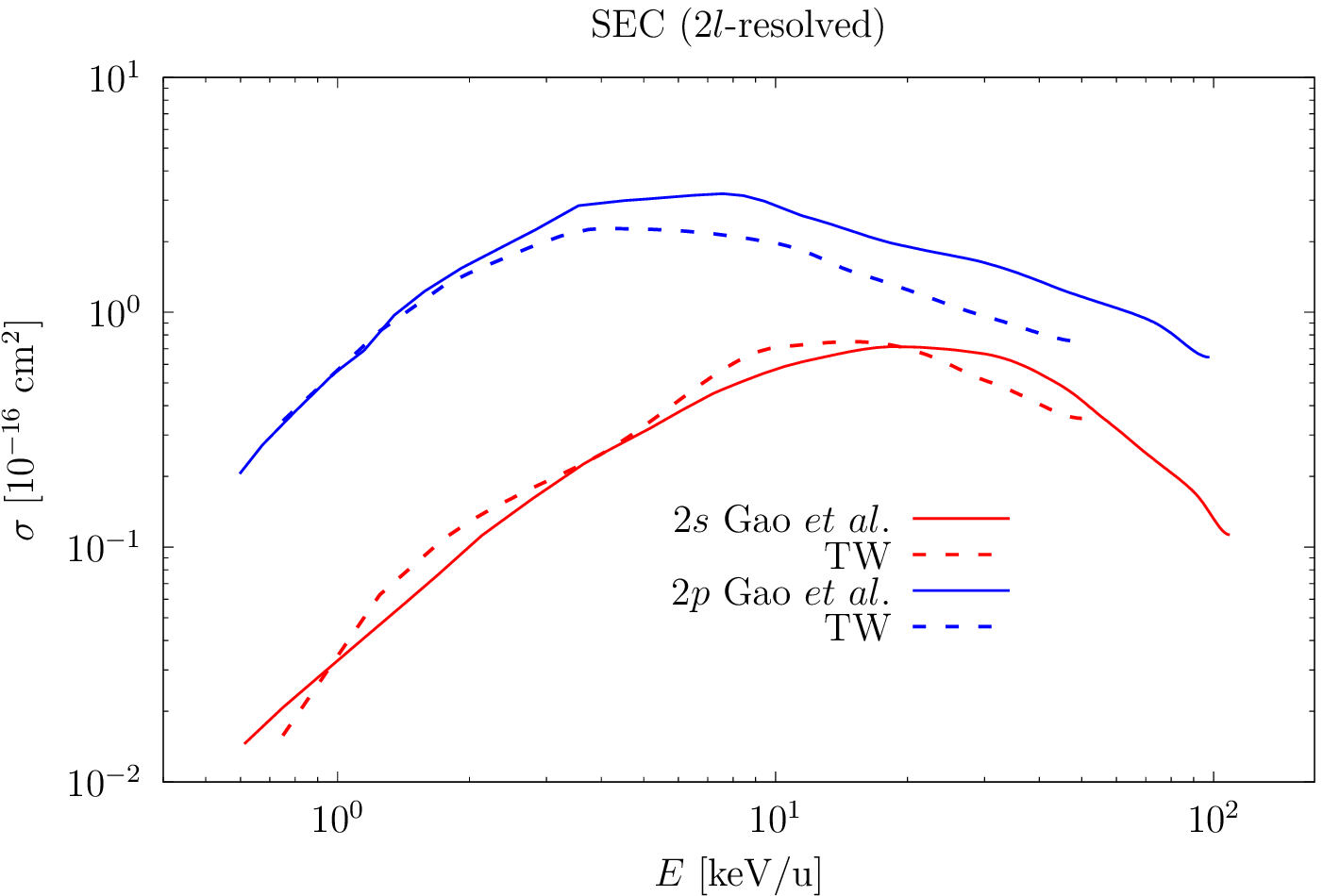}
}\\
%%%%
\subfigure[\quad Total DEC cross sections for electron capture to the C$^{2+}$ ion.]{%
\includegraphics[width=0.450\textwidth]{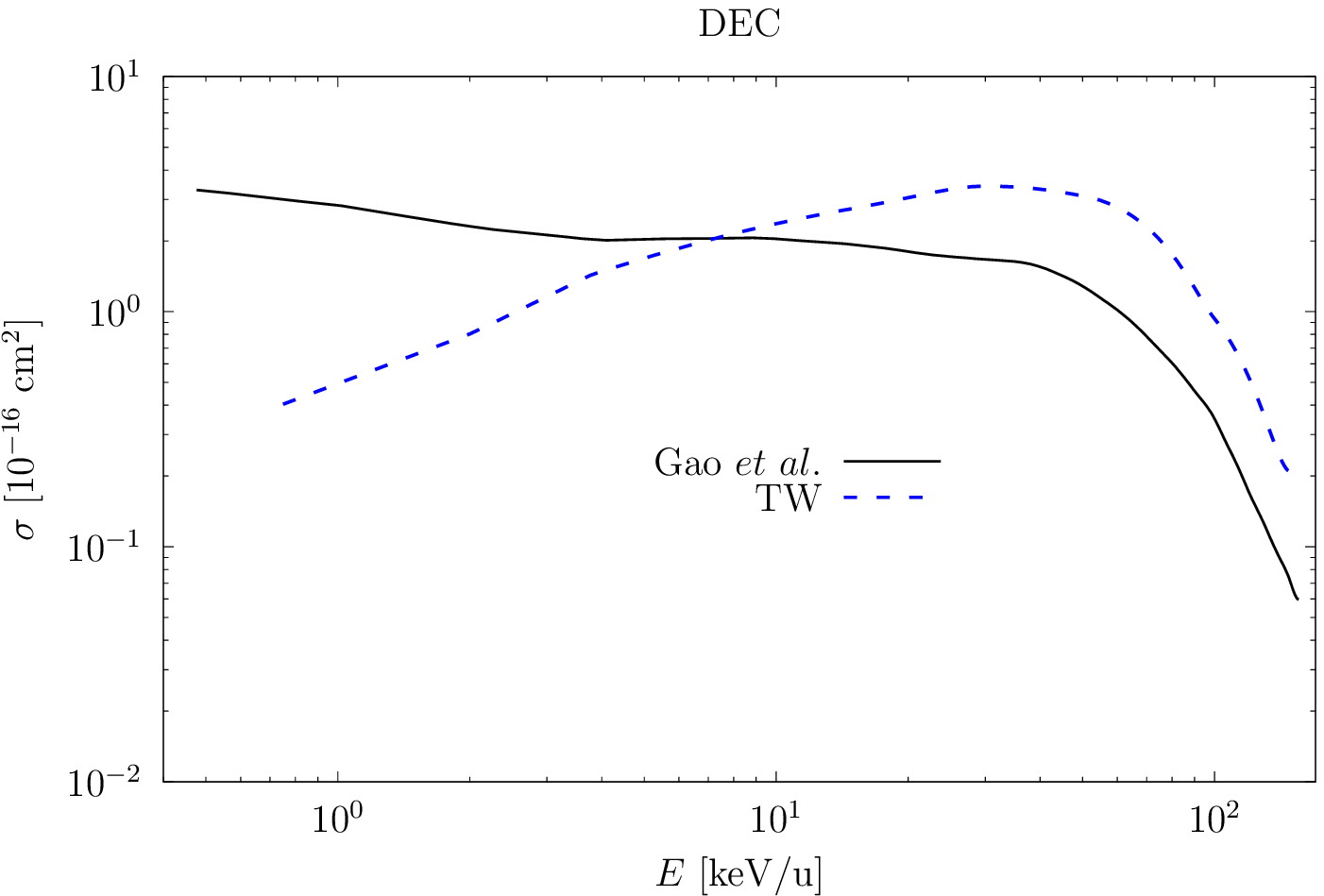}
}
\subfigure[\quad $2l2l^\prime$-selective DEC cross sections for electron capture to the C$^{2+}$ ion.]{%
\includegraphics[width=0.450\textwidth]{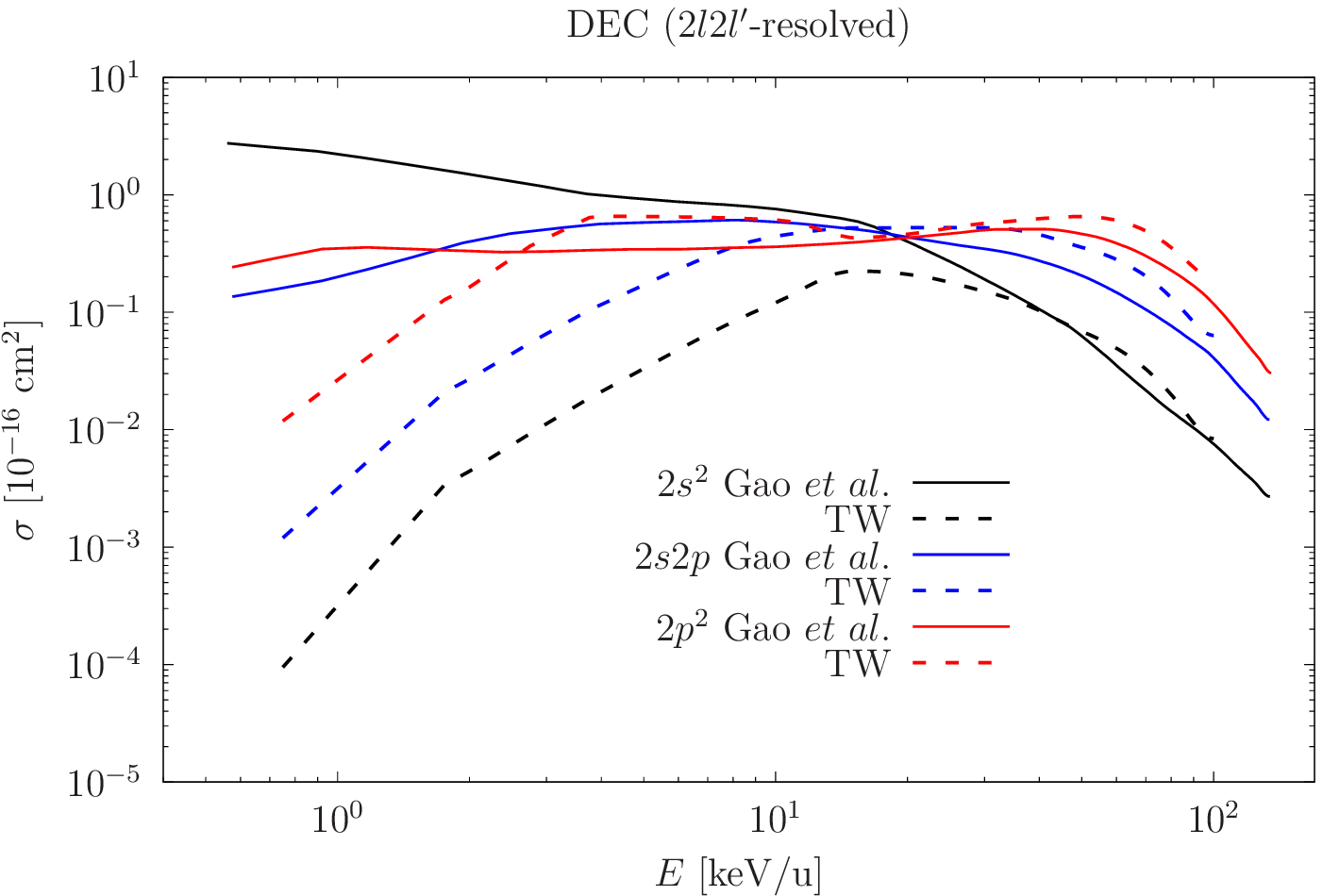}
}

\end{center}
\caption{\label{fig:C}
Cross sections for the C$^{4+}(1s^2)$-He$(1s^2)$ collision as functions of the impact energy.
The theoretical results are from the present calculations (dashed lines marked "TW") and Gao~\textit{et al.}\cite{Gao:PRA:2017} (solid lines).}
\end{figure}

At the first step the method described in Sec.~\ref{sec:Theory} was applied to the non-relativistic C$^{4+}(1s^2)$-He$(1s^2)$ collision system in the range of collision energies $0.8$-$50$ keV/u. 
In Fig.~\ref{fig:C}, we present our results for SEC and DEC cross sections. 
The total and state-selective values are shown. 
The theoretical results obtained by Gao~\textit{et al.}\cite{Gao:PRA:2017} within the advanced non-relativistic theoretical approach are also displayed for comparison. 
It is worth to specially note that our approach is based on the independent particle model, which is not good enough to provide reliable results for highly-correlated systems like helium. 
Nevertheless, the obtained data qualitatively correctly describe the behavior of the cross sections, including state-selective ones. Meanwhile, a considerable deviation for the DEC cross sections at small collision energies exists.  
%%%%%%%%%%%%%%%%%%%%%%%%%%%%%%%%%%%%%%%%%%%%%%%%%%%%%%%%%%%%%%%%%%%%%  
%%%%%
%%%%%%%%%%%%%%%%%%%%%%%%%%%%%%%%%%%%%%%%%%%%%%%%%%%%%%%%%%%%%%%%%%%%%%%%%%%%%%%% 
%%%%%%%%%%%%%%%%%%%%%%%%%%%%%%%%%%%%%%%%%%%%%%%%%%%%%%%%%%%%%%%%%%%%%%%%%%%%%%%% 
% %  
\clearpage

%%%%%%%%%%%%%%%%%%%%%%%%%%%%%%%%%%%
% \subsection{Th$^{90+}$-Ru$^{42+}$ collision}  
% \label{sec:results:Th}  
%%%%%%%%%%%%%%%%%%%%%%%%%%%%%%%%%%%%%%%%%%%%%%%%%%%%%%%%%%%%%%%%%%%%%%%%%%%%%%%% 
In the following, we consider the Th$^{90+}$-Ru$^{42+}(1s^2)$ heavy ion collision, where the resonance condition is fulfilled like in the C$^{4+}(1s^2)$-He$(1s^2)$ one.
In this system the electron-electron interaction is suppressed by the strong electron-nucleus one,  thus making the independent-particle approximation very reasonable. 
% From the other point of view, this system is strongly relativistic and demands using relativistic methods. 
%The similarity coefficients is about $22$. 
As it was demonstrated in Sec.~\ref{sec:results:one-electron}, the EC processes from the $K$ target to the $L$ projectile shells are resonance and are accompanied by the strongest role of the relativistic effects. 
Since the $K$-$L$ transitions are dominant for the system, consideration of the two-electron target ion (Ru$^{42+}(1s^2)$) allows one to obtain the main contributions to the many-electron probabilities and cross sections of EC to the $L$ projectile shell and vacancy creation in the $K$ target shell for the target ions with a lower order of ionization up to the neutral ruthenium atom also.

% These ions are chosen in order to satisfy the resonance energy condition for the $L$ and $K$ shells of thorium and ruthenium, correspondingly.

\begin{figure}
\begin{center}
\subfigure[\quad Total and  $n$-selective SEC cross sections.]{%
\includegraphics[width=0.450\textwidth]{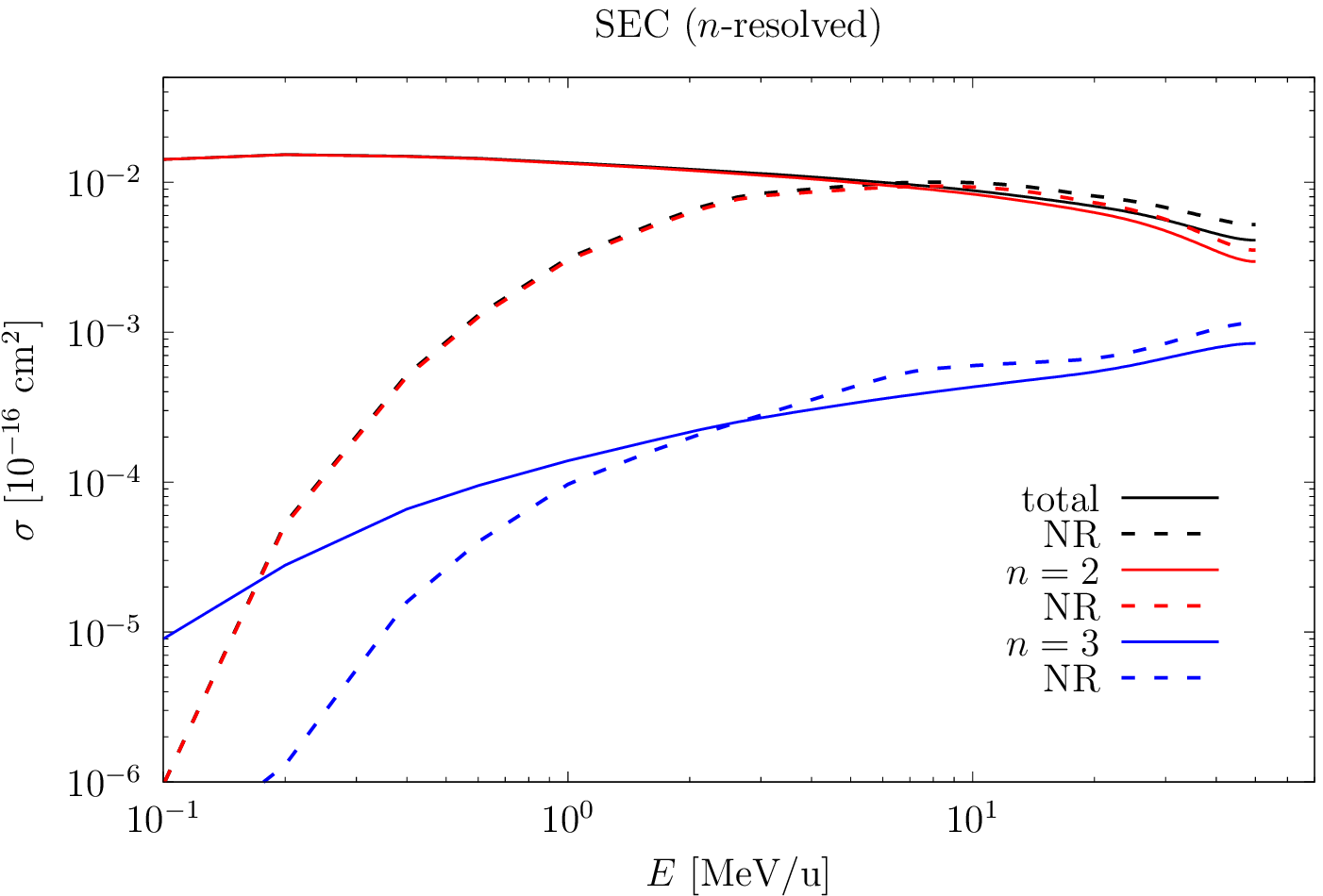}
% \includegraphics[width=0.50\textwidth]{../He_C4+/V_scr4/Graph/DEC_total.eps}
% % \includegraphics[width=0.45\textwidth]{Graph_E3+/1.eps}
}
\subfigure[\quad $2l$-selective SEC cross sections.]{%
\includegraphics[width=0.450\textwidth]{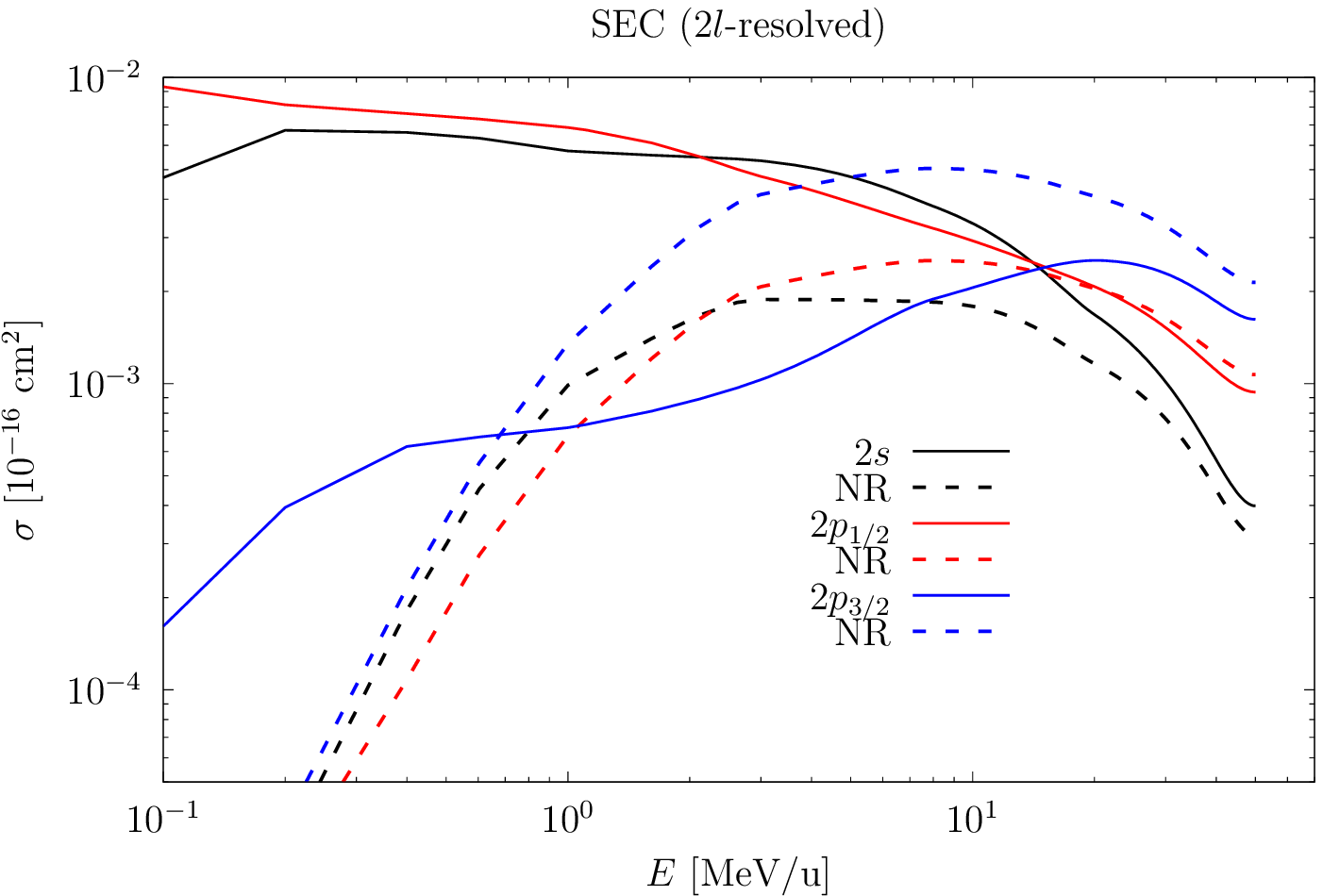}
}
%%%%
%
\end{center}
\caption{\label{fig:Th_SEC}
SEC cross sections for the Th$^{90+}$-Ru$^{42+}(1s^2)$ collision as functions of the impact energy.
The relativistic (solid lines) and non-relativistic results (dashed lines marked "NR")
are presented.}
\end{figure}

\begin{figure}
\begin{center}
\includegraphics[width=0.450\textwidth]{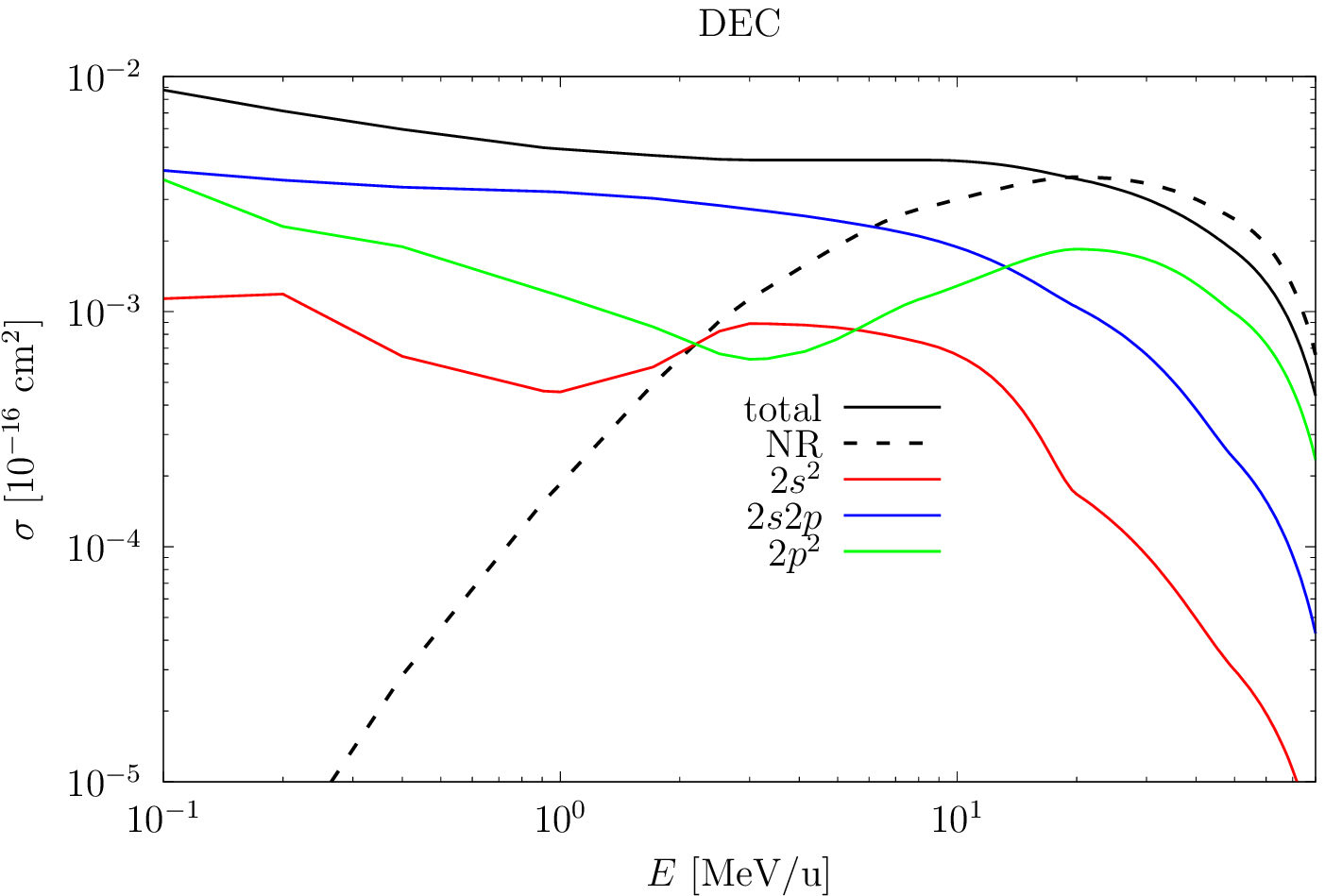}
\end{center}
\caption{\label{fig:Th_DEC1}
Total and state-selective DEC cross sections for the Th$^{90+}$-Ru$^{42+}(1s^2)$ collision as functions of the impact energy. The relativistic (solid lines) and non-relativistic results (dashed lines marked "NR") are presented.}
\end{figure}

\begin{figure}
\begin{center}
\subfigure[\quad $2s2p$-selective DEC cross sections.]{%
\includegraphics[width=0.450\textwidth]{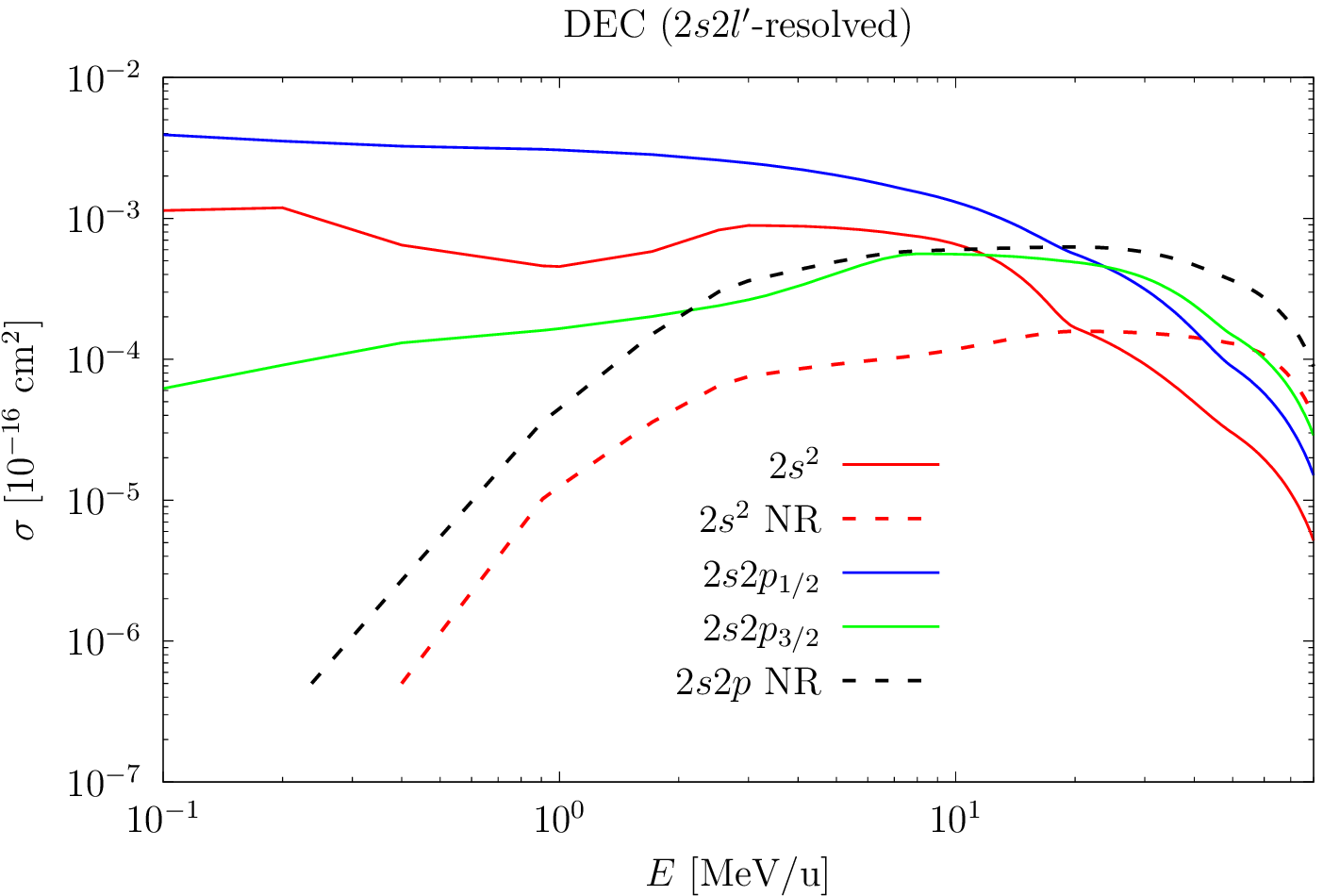}
% \includegraphics[width=0.50\textwidth]{../He_C4+/V_scr4/Graph/DEC_total.eps}
% % \includegraphics[width=0.45\textwidth]{Graph_E3+/1.eps}
}
\subfigure[\quad $2p2p'$-selective DEC cross sections.]{%
\includegraphics[width=0.450\textwidth]{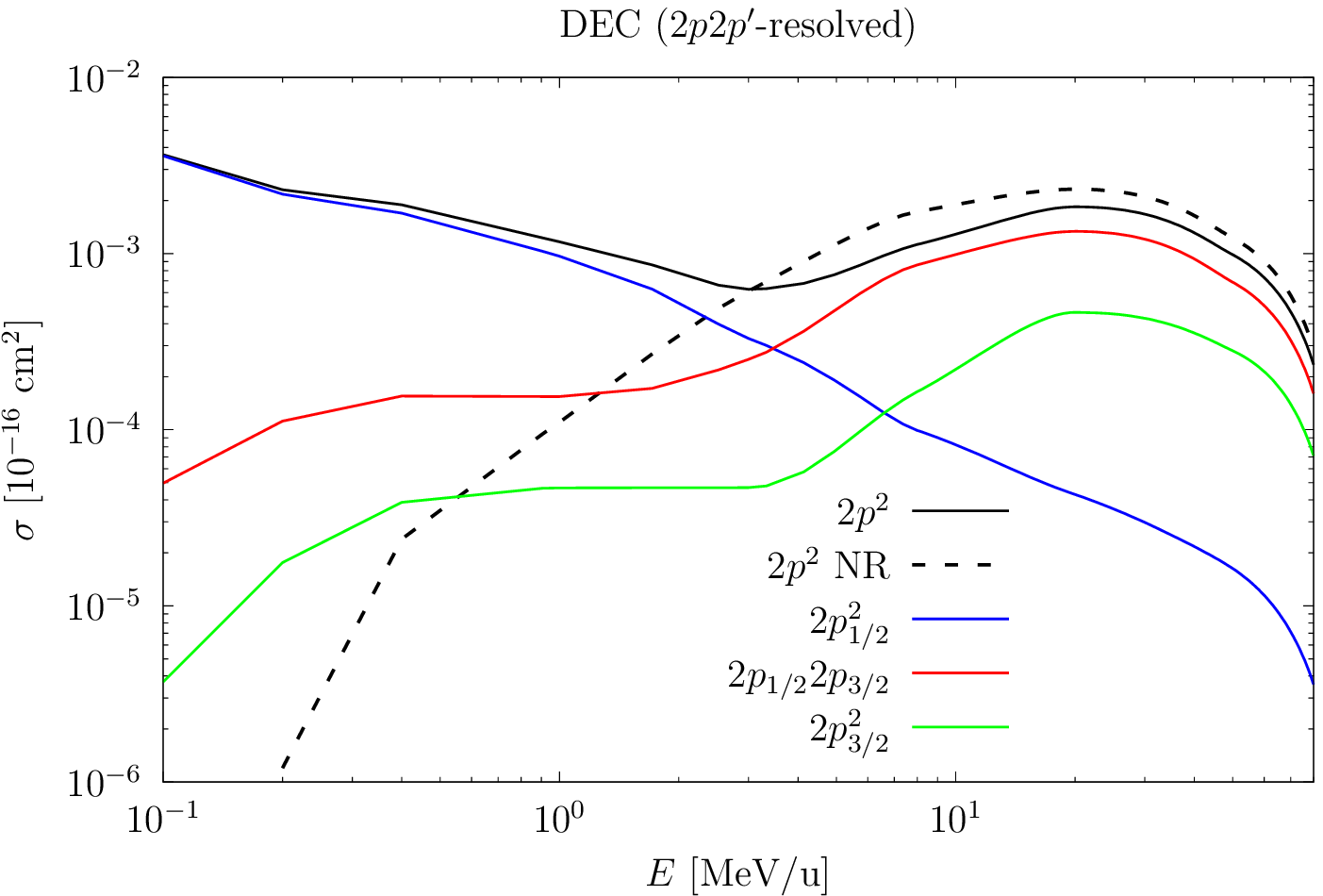}
}
%%%%
%
\end{center}
\caption{\label{fig:Th_DEC2}
State-selective DEC cross sections for the Th$^{90+}$-Ru$^{42+}(1s^2)$ collision as functions of the impact energy.
The relativistic (solid lines) and non-relativistic results (dashed lines marked "NR")
are presented.}
\end{figure}

In figures~\ref{fig:Th_SEC}-\ref{fig:Th_DEC2},  the results of the SEC and DEC cross sections for the 
Th$^{90+}$-Ru$^{42+}(1s^2)$ collision in the energy region $0.5$-$50$~MeV/u are presented. The data obtained in the non-relativistic limit are also displayed in the figures for comparison. 
As in the one-electron case, we stress again a very important role of the relativistic effects, which for low-energy collision becomes crucial. 
%%%%%%%%%%%%%%%%%%%%%%%%%%%%%%%%%%%
%%%%%
%
%%%%%%%%%%%%%%%%%%%%%%%%%%%%%%%%%%%
%%%%%
%%%%%%%%%%%%%%%%%%%%%%%%%%%%%%%%%%%
%%%%%%%%%%%%%%%%%%%%%%%%%%%%%%%%%%%
%%%%%
%%%%%%%%%%%%%%%%%%%%%%%%%%%%%%%%%%%%%%%%%%%%%%%%%%%%%%%%%%%%%%%%%%%%%  

\begin{figure}
\begin{center}
\subfigure[]{%
\includegraphics[width=0.45\textwidth]{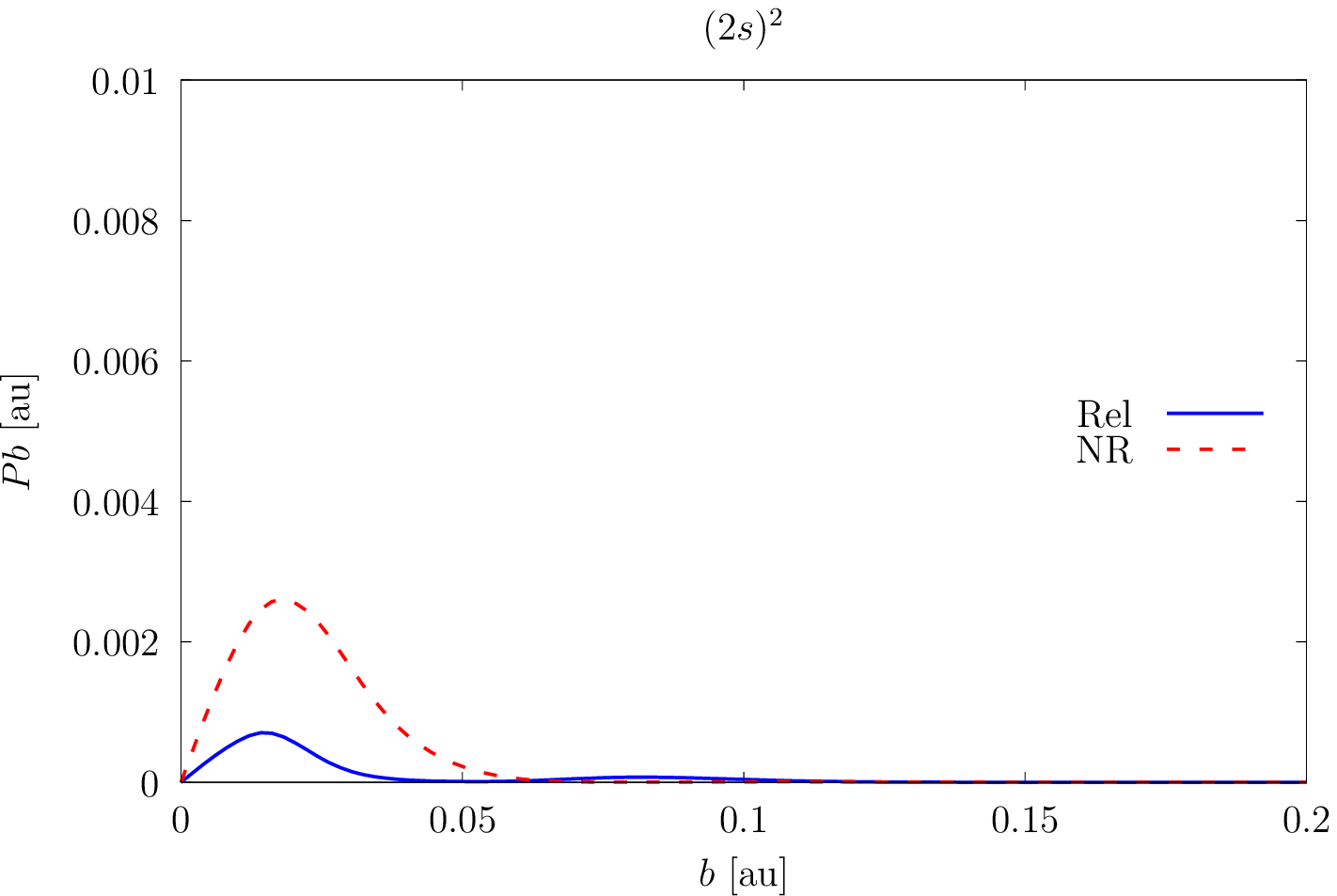}
}
\subfigure[]{%
\includegraphics[width=0.45\textwidth]{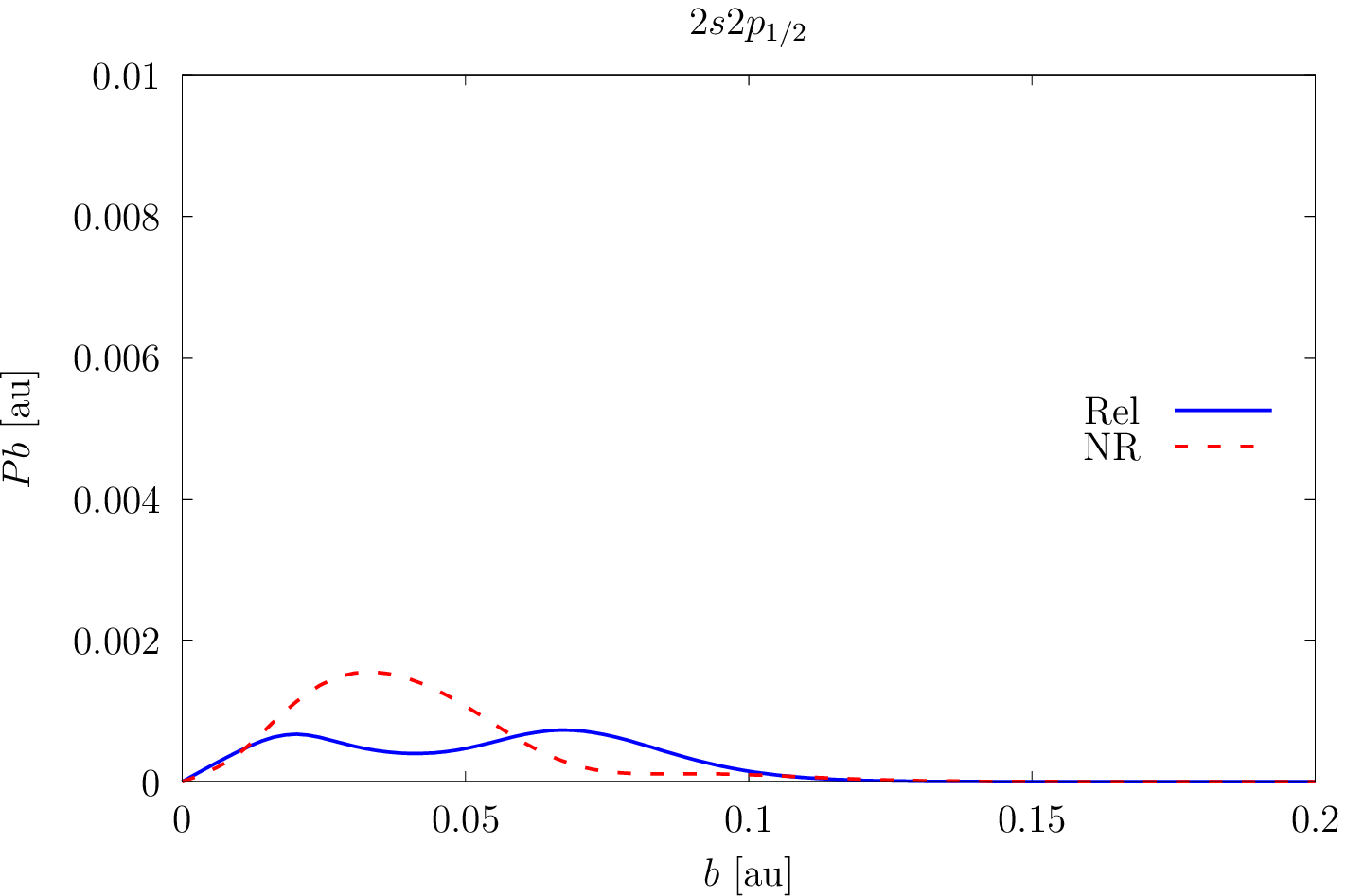}
}\\
\subfigure[]{%
\includegraphics[width=0.45\textwidth]{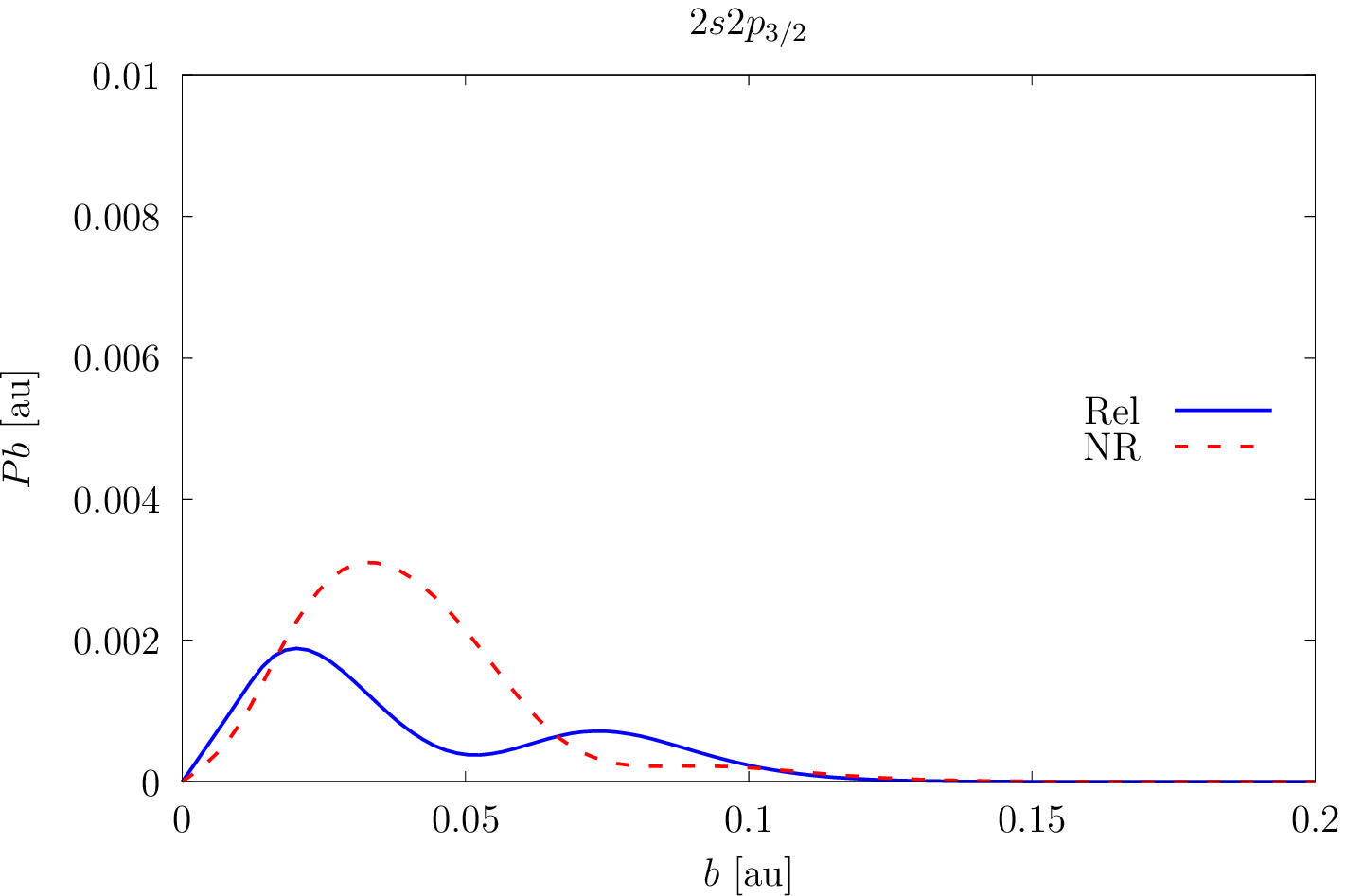}
}
%%%%
\subfigure[]{%
\includegraphics[width=0.450\textwidth]{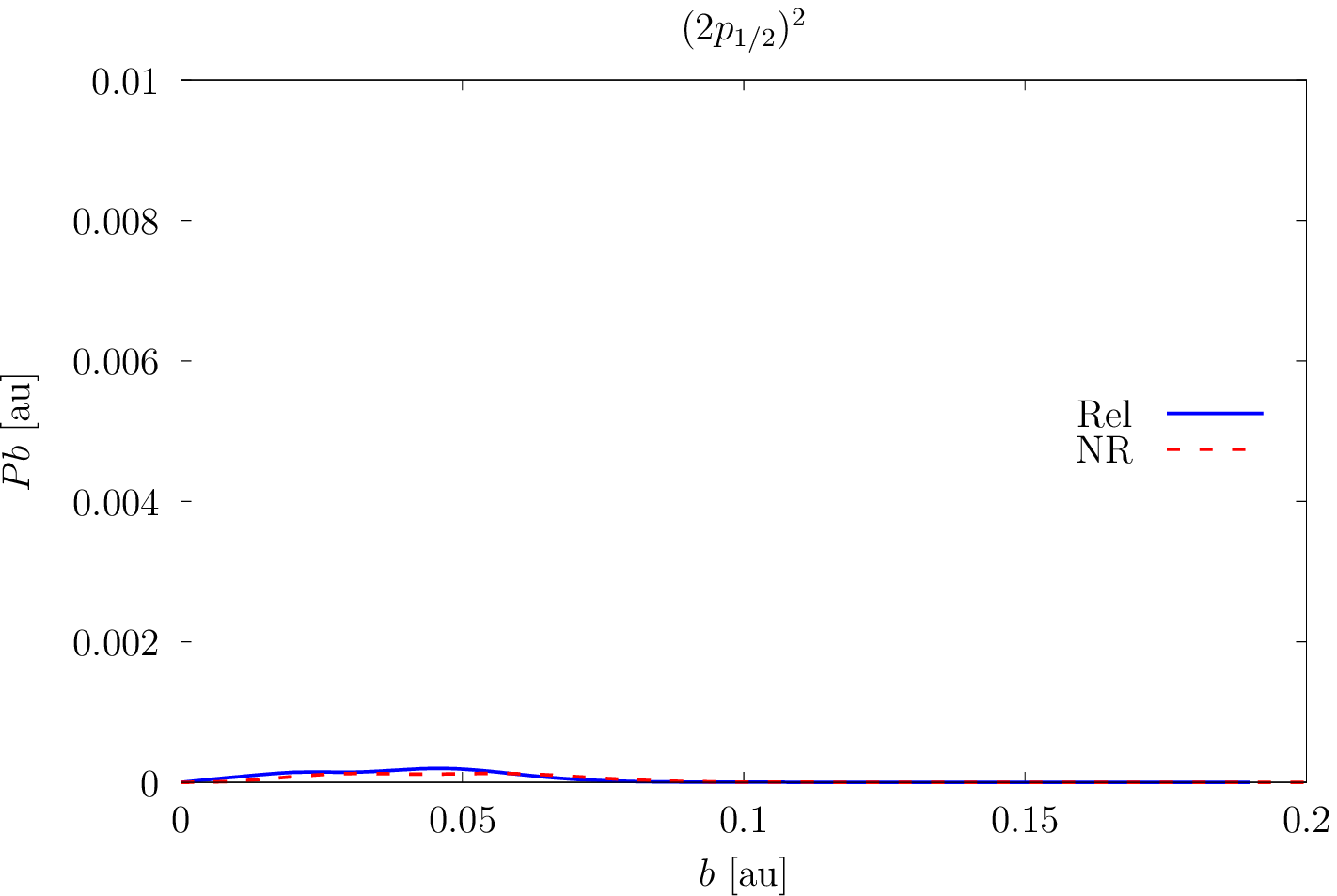}
}\\
\subfigure[]{%
\includegraphics[width=0.450\textwidth]{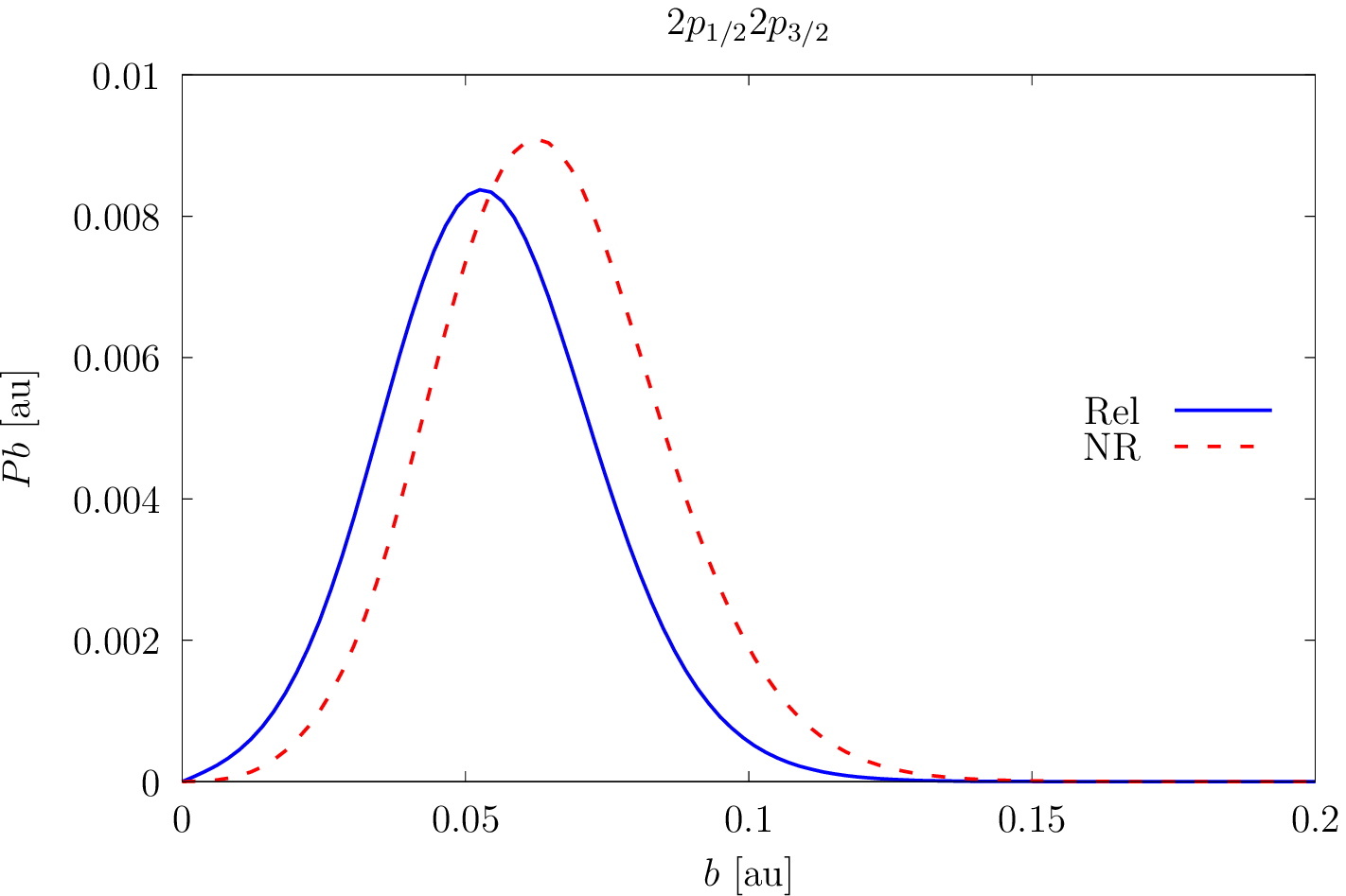}
}
\subfigure[]{%
\includegraphics[width=0.450\textwidth]{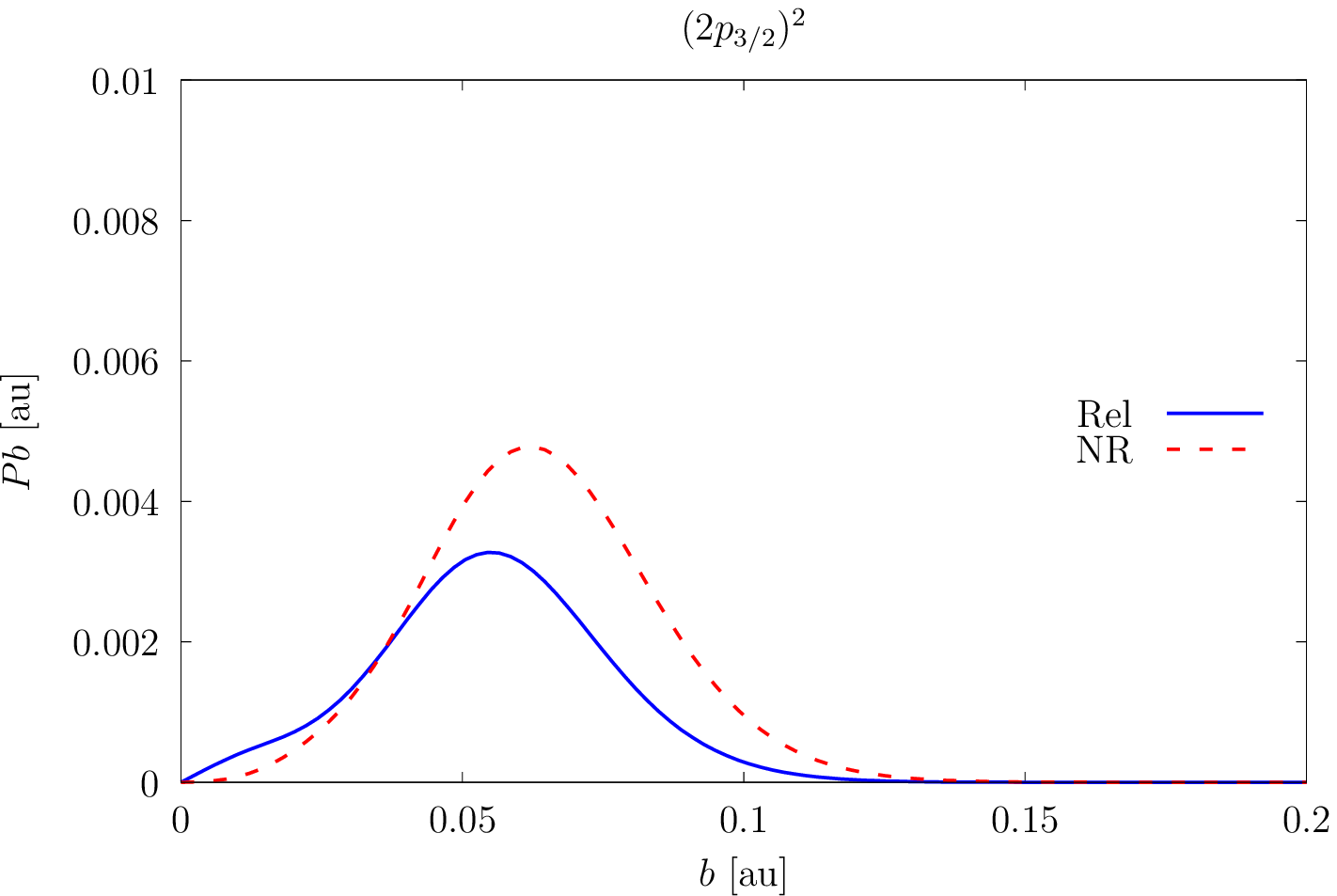}
}
%%%%
%
\end{center}
\caption{\label{fig:Th_DEC_50}
State-selective DEC probabilities weighted by the impact parameter for the Th$^{90+}$-Ru$^{42+}(1s^2)$ @ $50$~MeV/u collision as functions of the impact parameter.
The relativistic (solid lines marked "Rel") and non-relativistic results (dashed lines marked "NR") are presented.}
\end{figure}

\begin{figure}
\begin{center}
\subfigure[]{%
\includegraphics[width=0.45\textwidth]{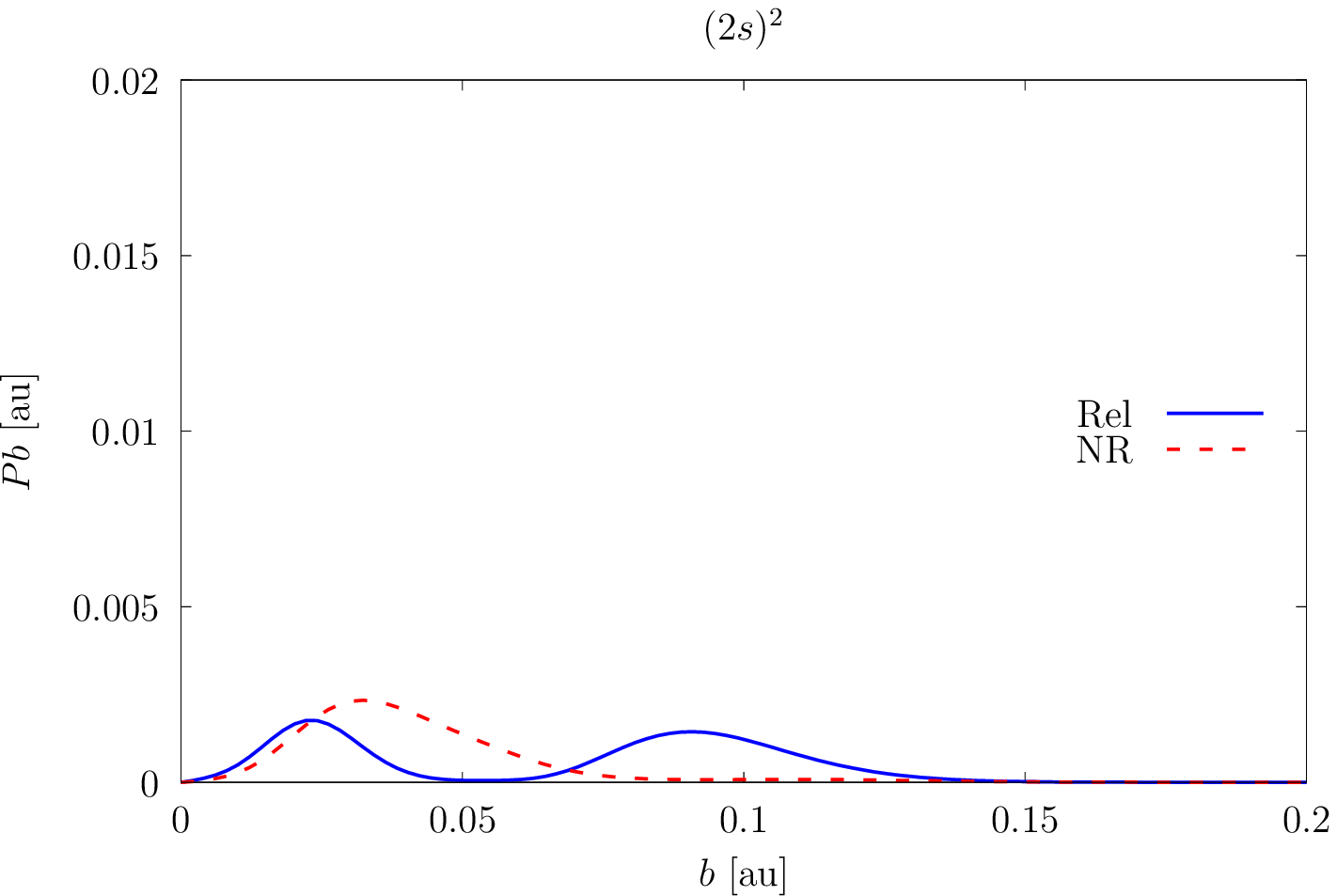}
}
\subfigure[]{%
\includegraphics[width=0.45\textwidth]{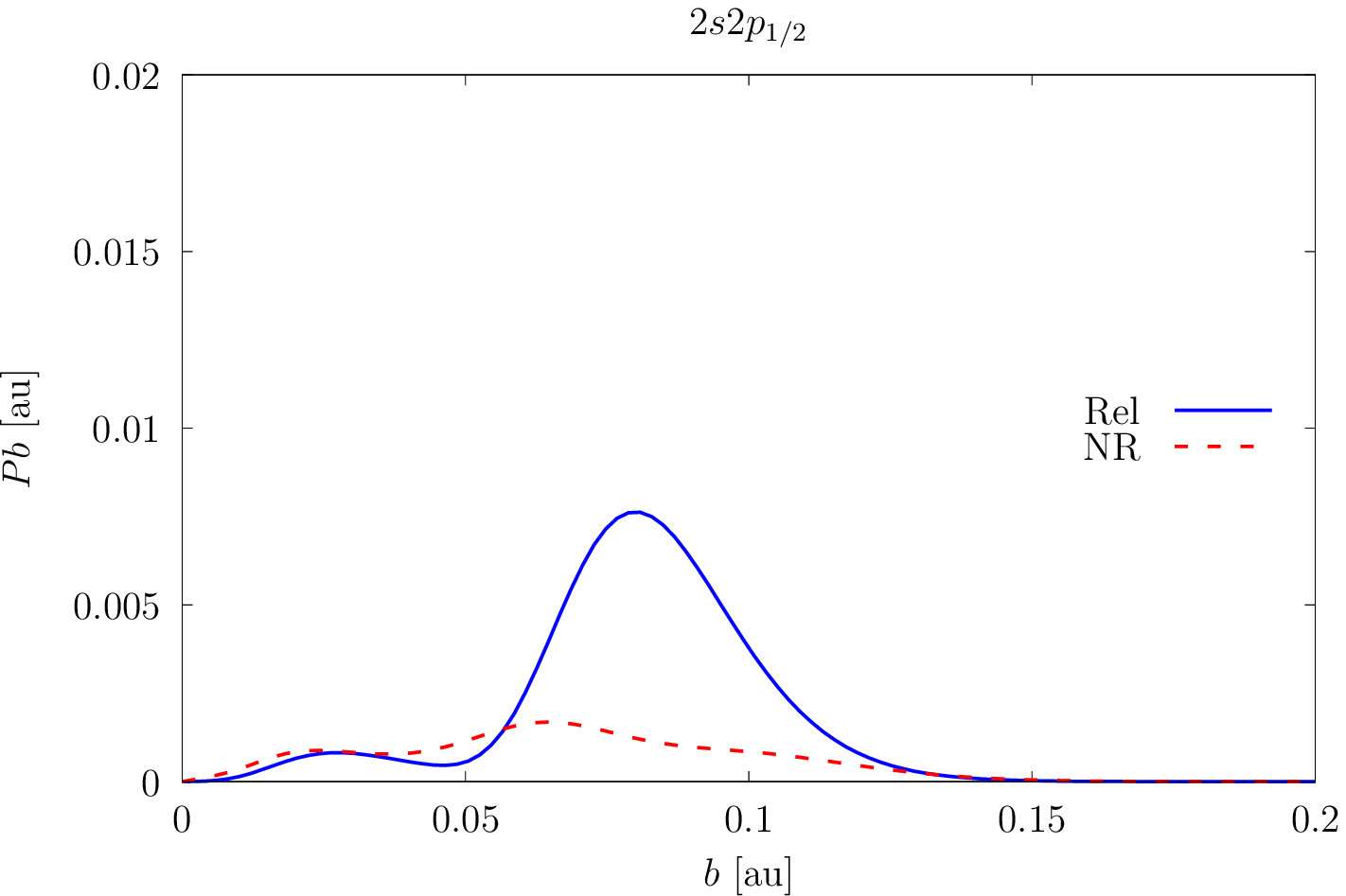}
}\\
\subfigure[]{%
\includegraphics[width=0.45\textwidth]{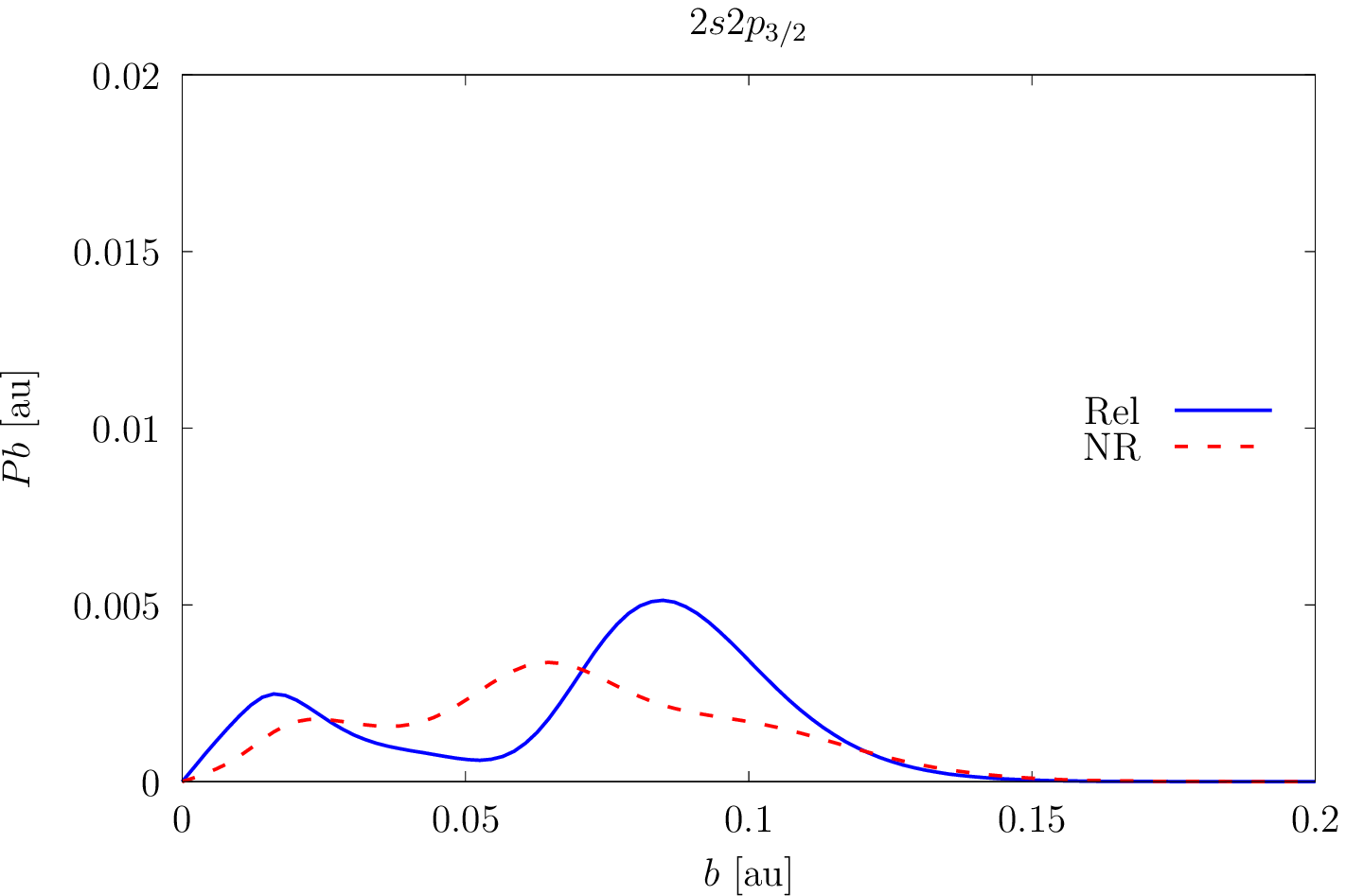}
}
%%%%
\subfigure[]{%
\includegraphics[width=0.450\textwidth]{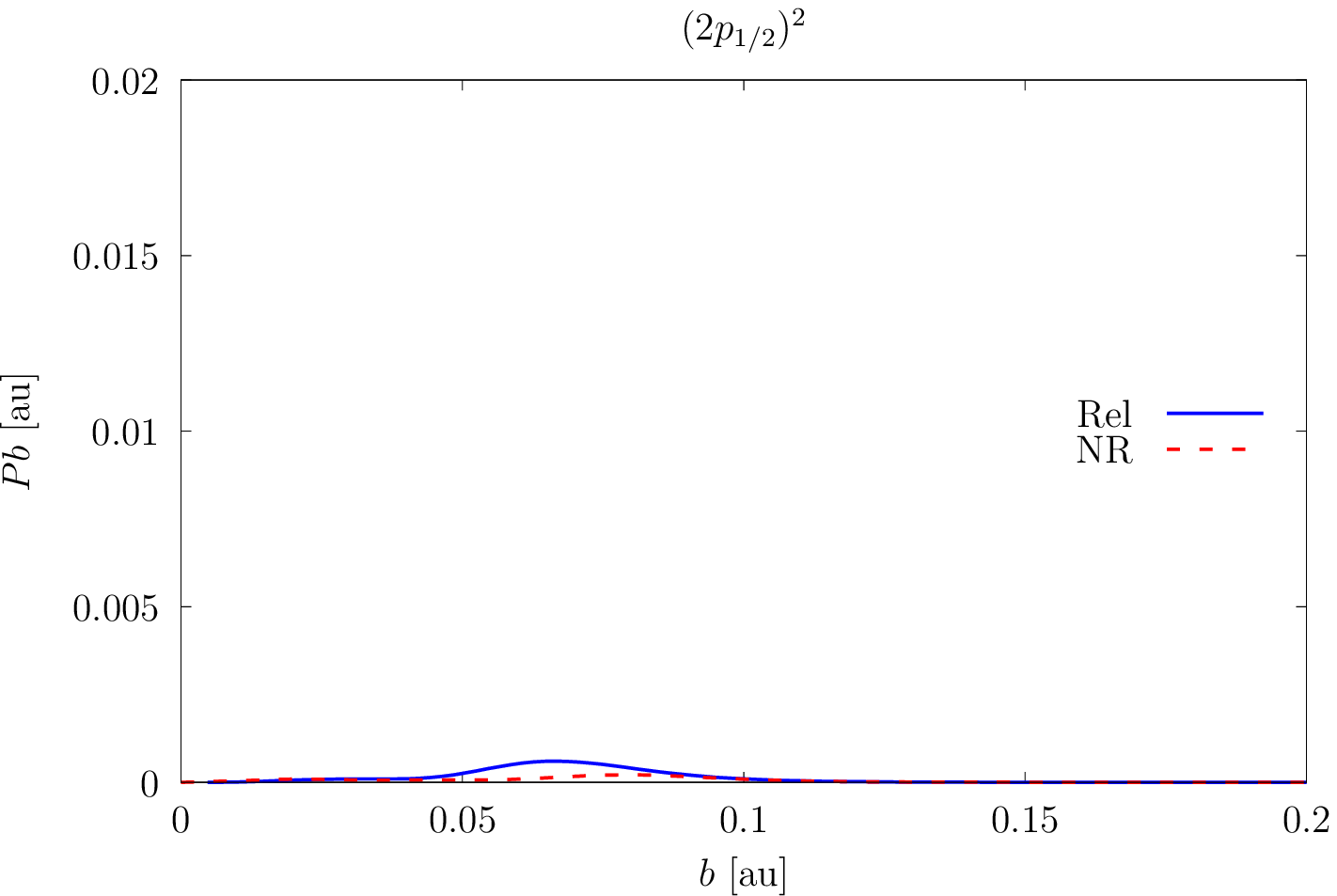}
}\\
\subfigure[]{%
\includegraphics[width=0.450\textwidth]{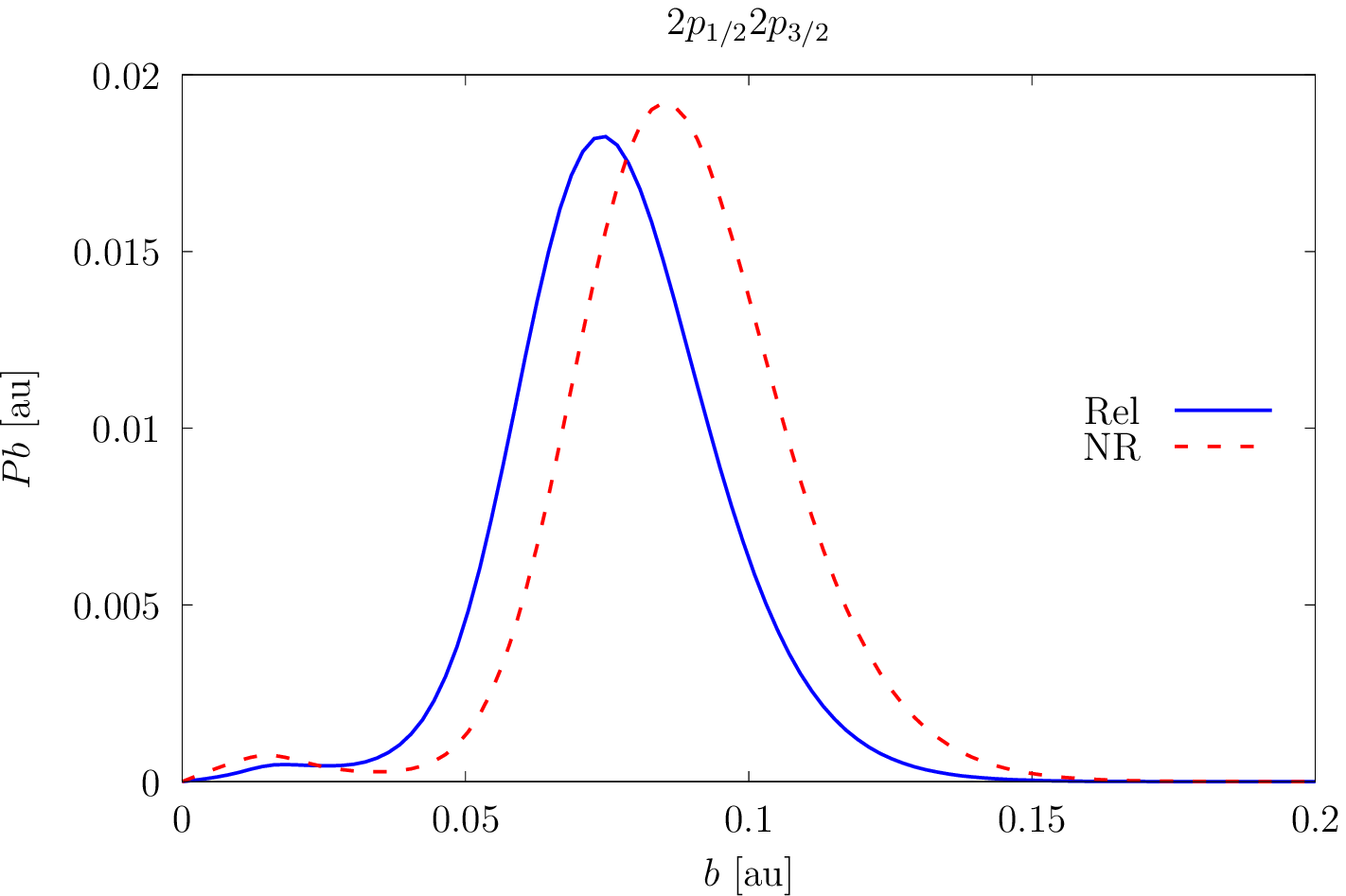}
}
\subfigure[]{%
\includegraphics[width=0.450\textwidth]{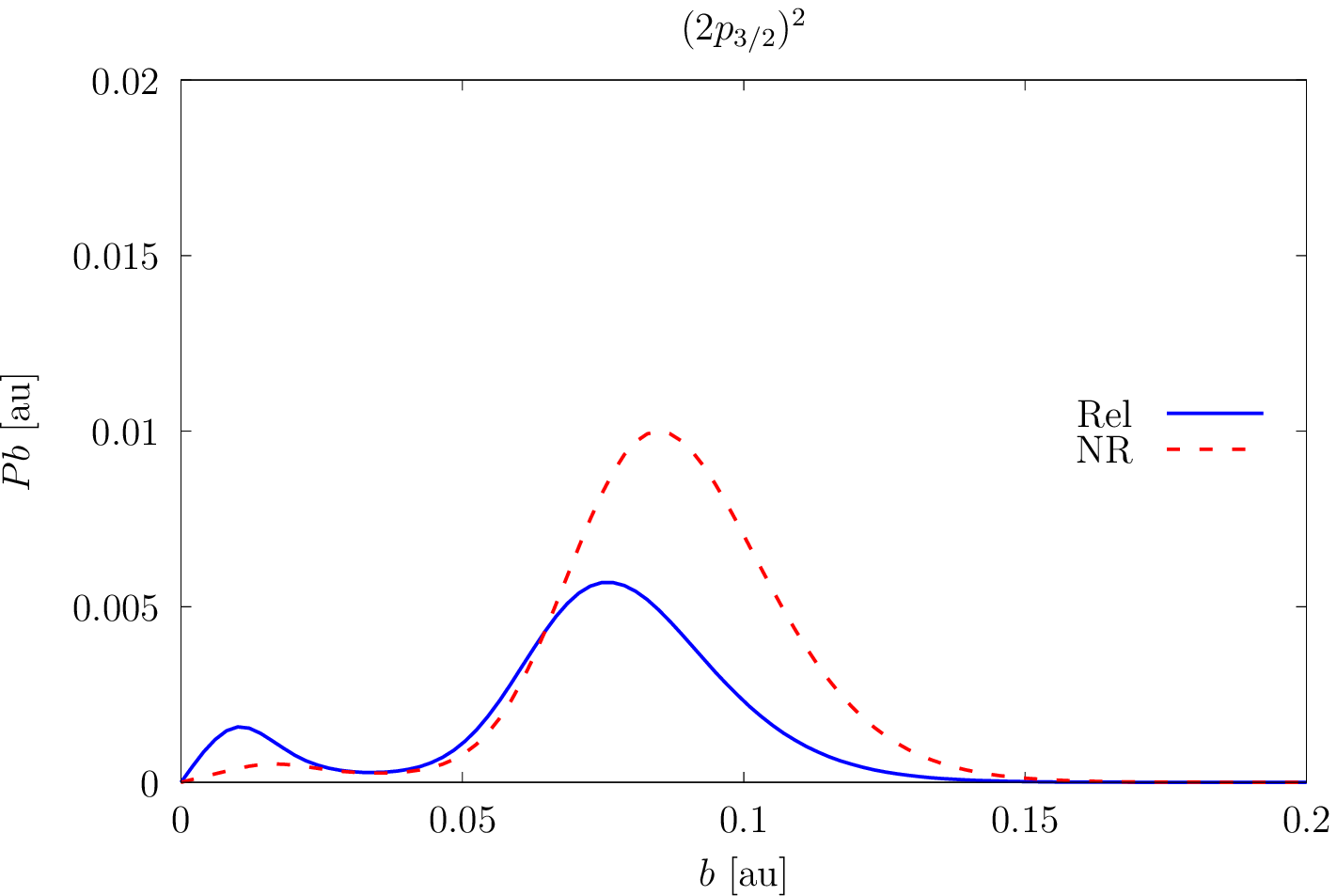}
}
%%%%
%
\end{center}
\caption{\label{fig:Th_DEC_20}
State-selective DEC probabilities weighted by the impact parameter for the Th$^{90+}$-Ru$^{42+}(1s^2)$ @ $20$~MeV/u collision as functions of the impact parameter.
The relativistic (solid lines marked "Rel") and non-relativistic results (dashed lines marked "NR") are presented.
}
\end{figure}

\begin{figure}
\begin{center}
\subfigure[]{%
\includegraphics[width=0.45\textwidth]{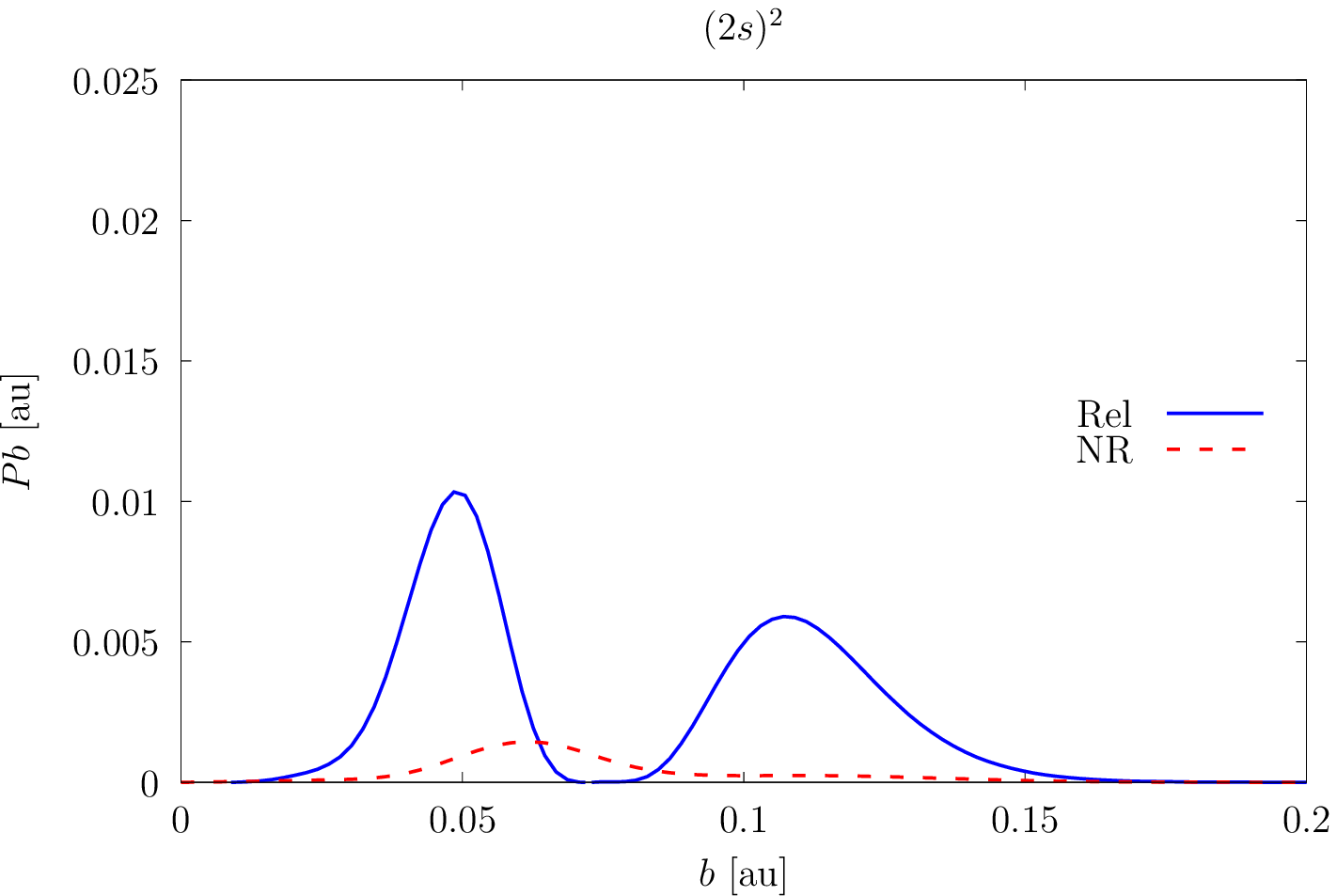}
}
\subfigure[]{%
\includegraphics[width=0.45\textwidth]{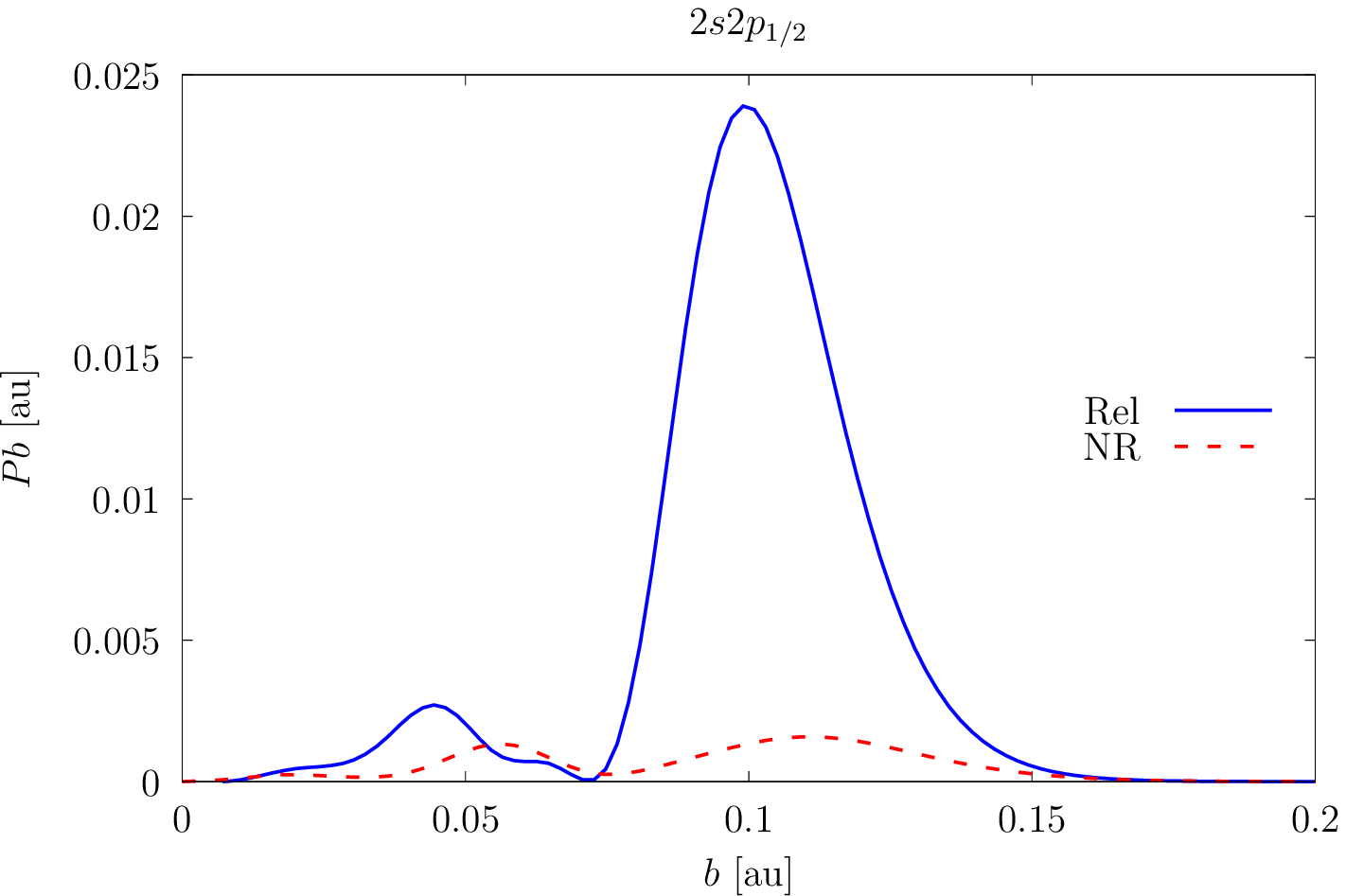}
}\\
\subfigure[]{%
\includegraphics[width=0.45\textwidth]{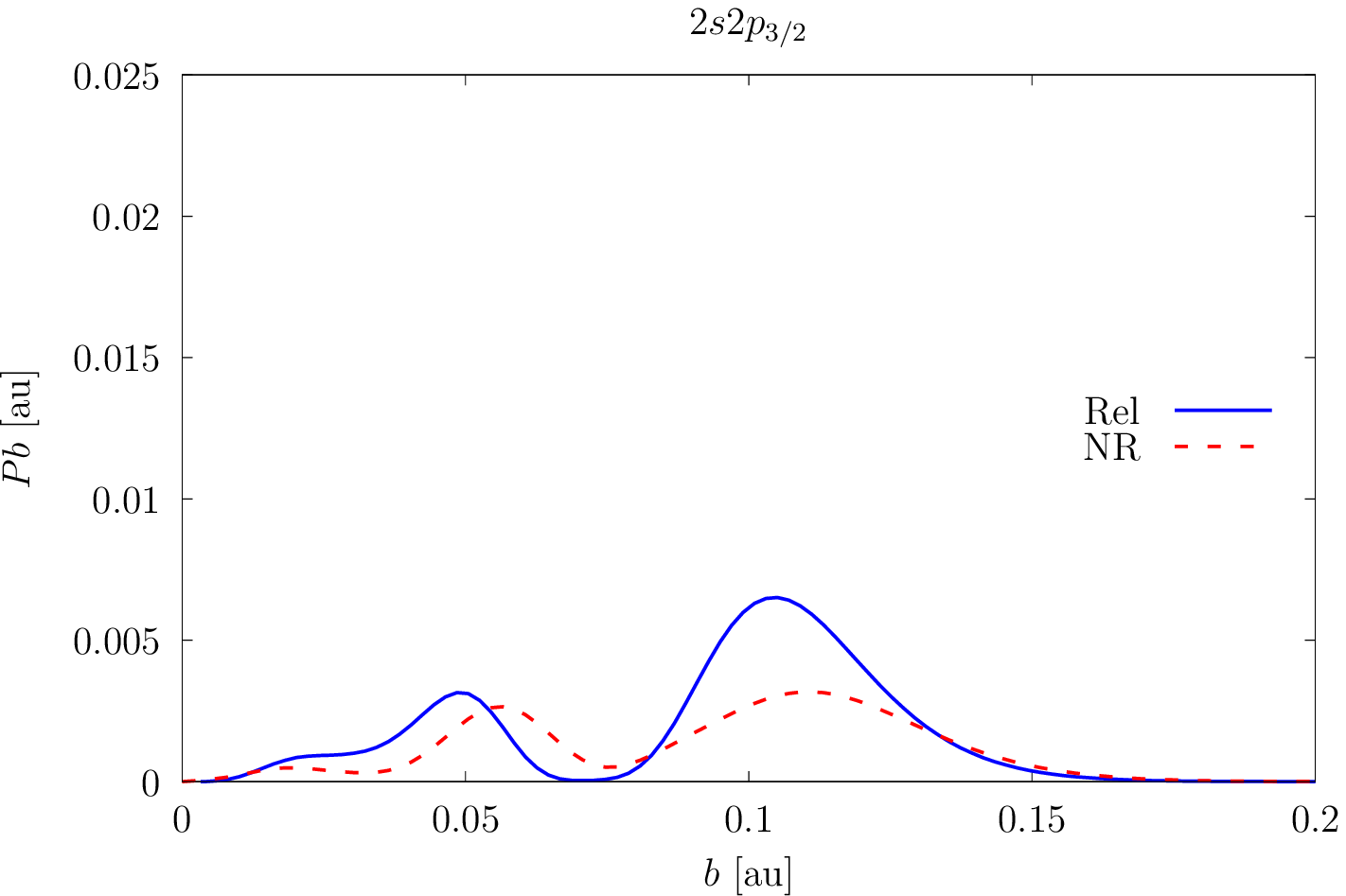}
}
%%%%
\subfigure[]{%
\includegraphics[width=0.450\textwidth]{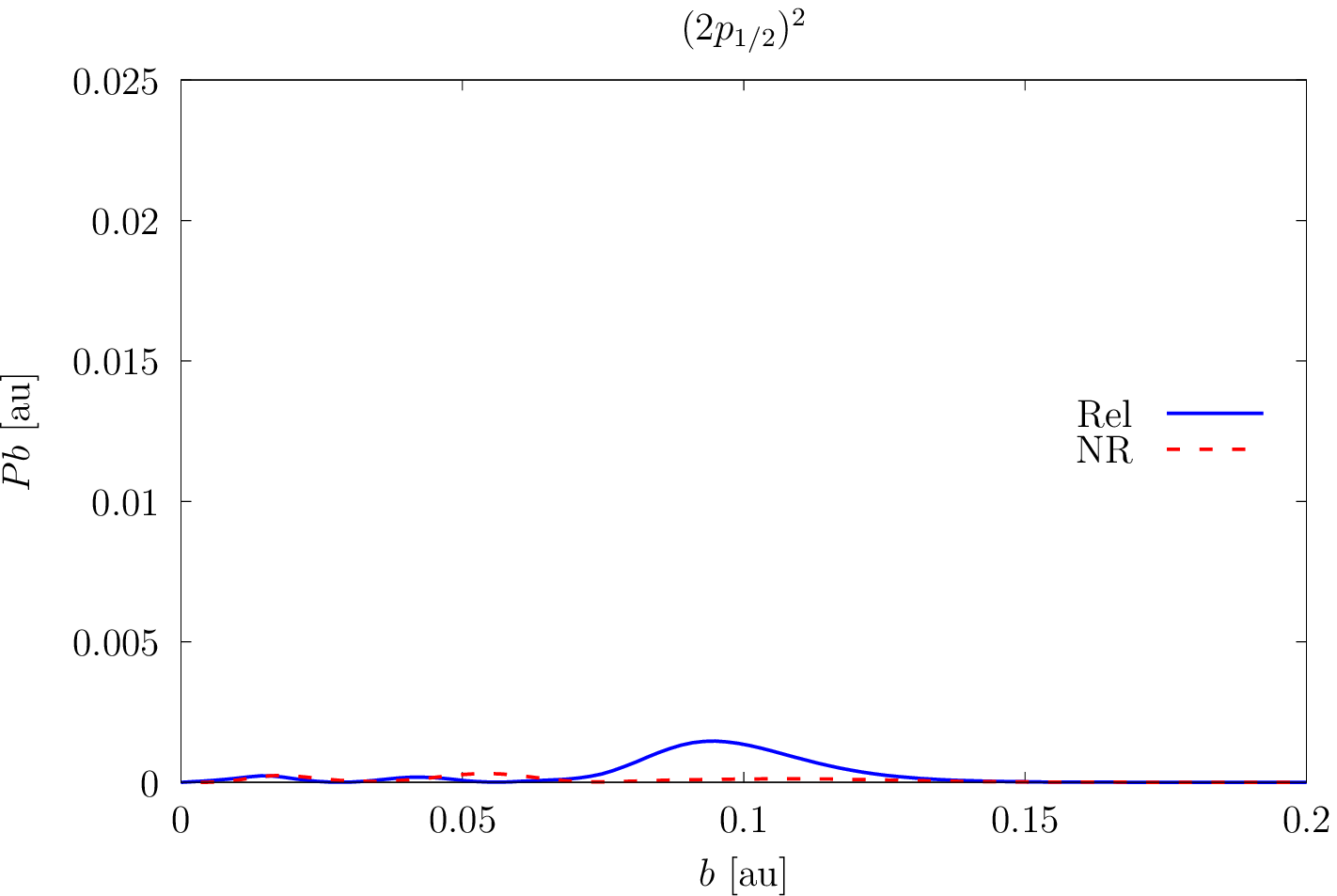}
}\\
\subfigure[]{%
\includegraphics[width=0.450\textwidth]{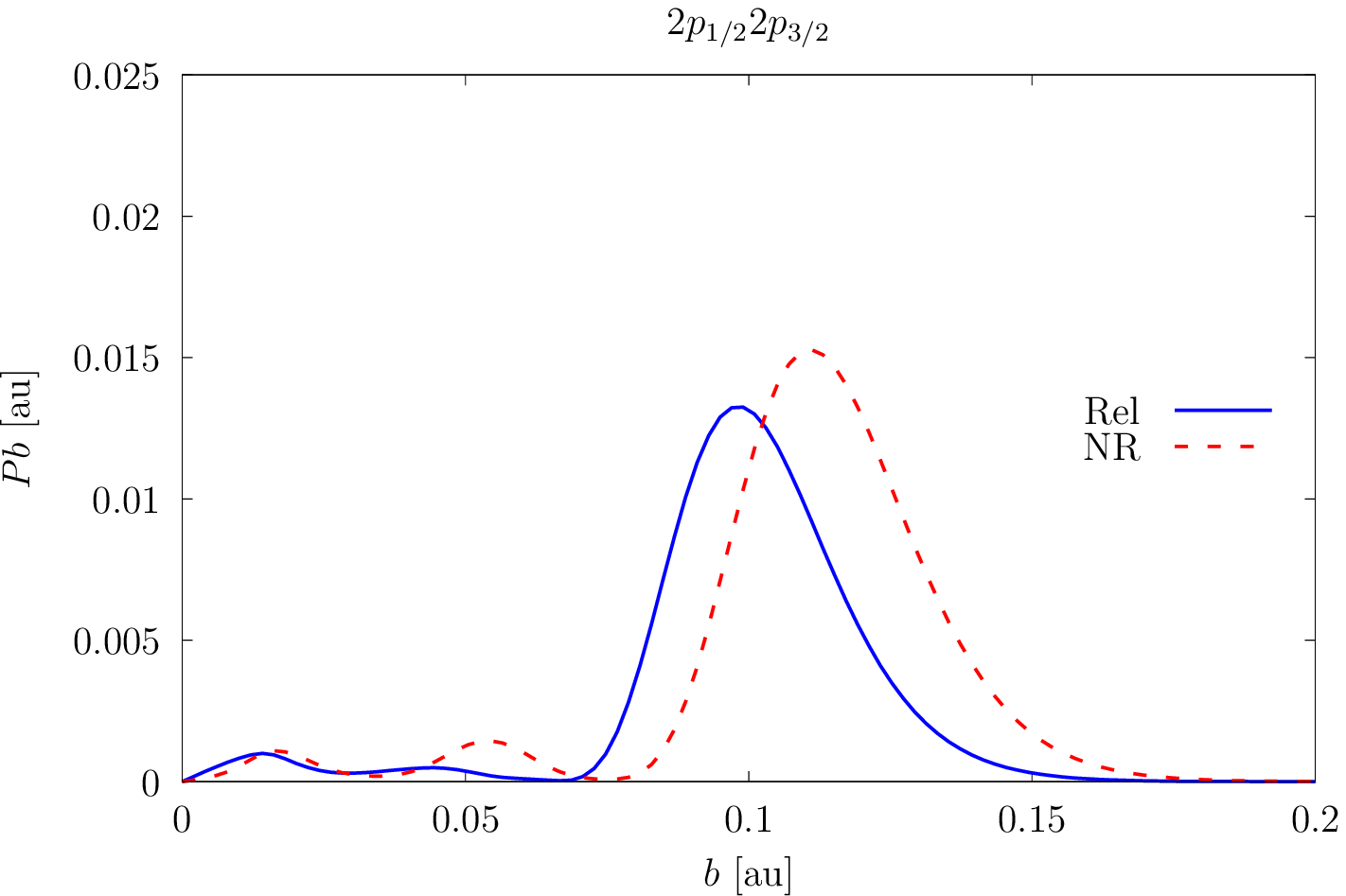}
}
\subfigure[]{%
\includegraphics[width=0.450\textwidth]{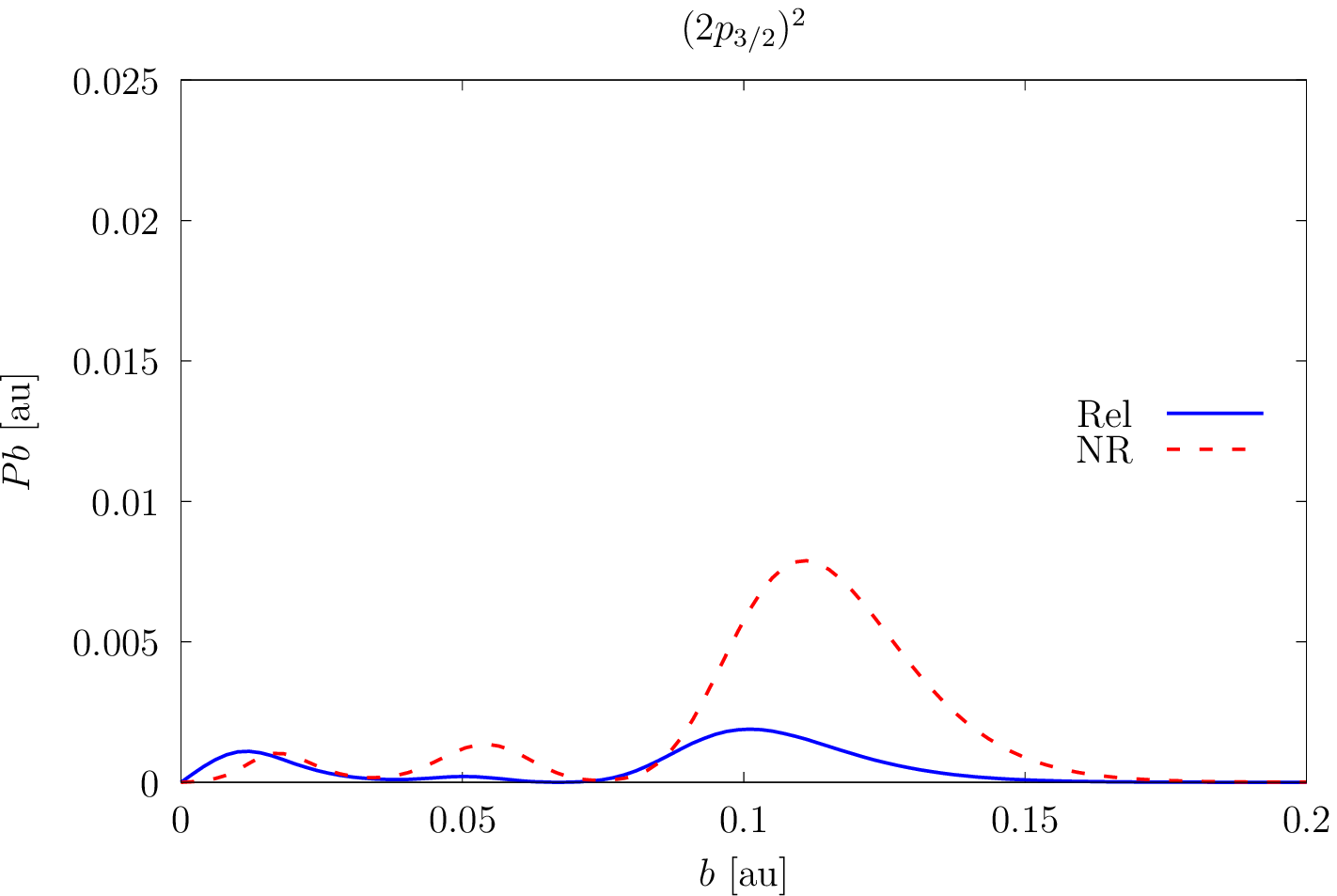}
}
%%%%
%
\end{center}
\caption{\label{fig:Th_DEC_8}
State-selective DEC probabilities weighted by the impact parameter for the Th$^{90+}$-Ru$^{42+}(1s^2)$ @ $8$~MeV/u collision as functions of the impact parameter.
The relativistic (solid lines marked "Rel") and non-relativistic results (dashed lines marked "NR") are presented.
}
\end{figure}

\begin{figure}
\begin{center}
\subfigure[]{%
\includegraphics[width=0.45\textwidth]{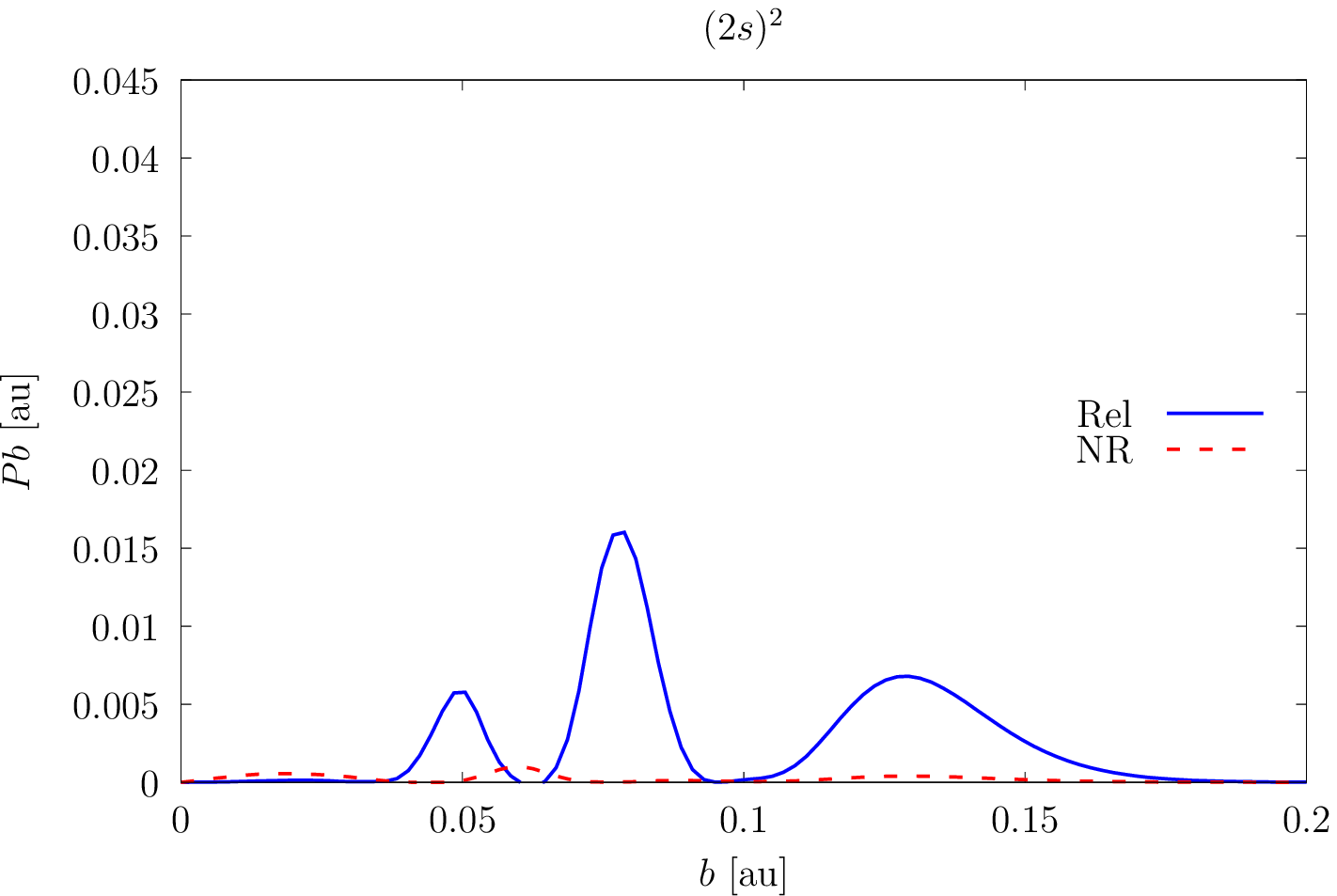}
}
\subfigure[]{%
\includegraphics[width=0.45\textwidth]{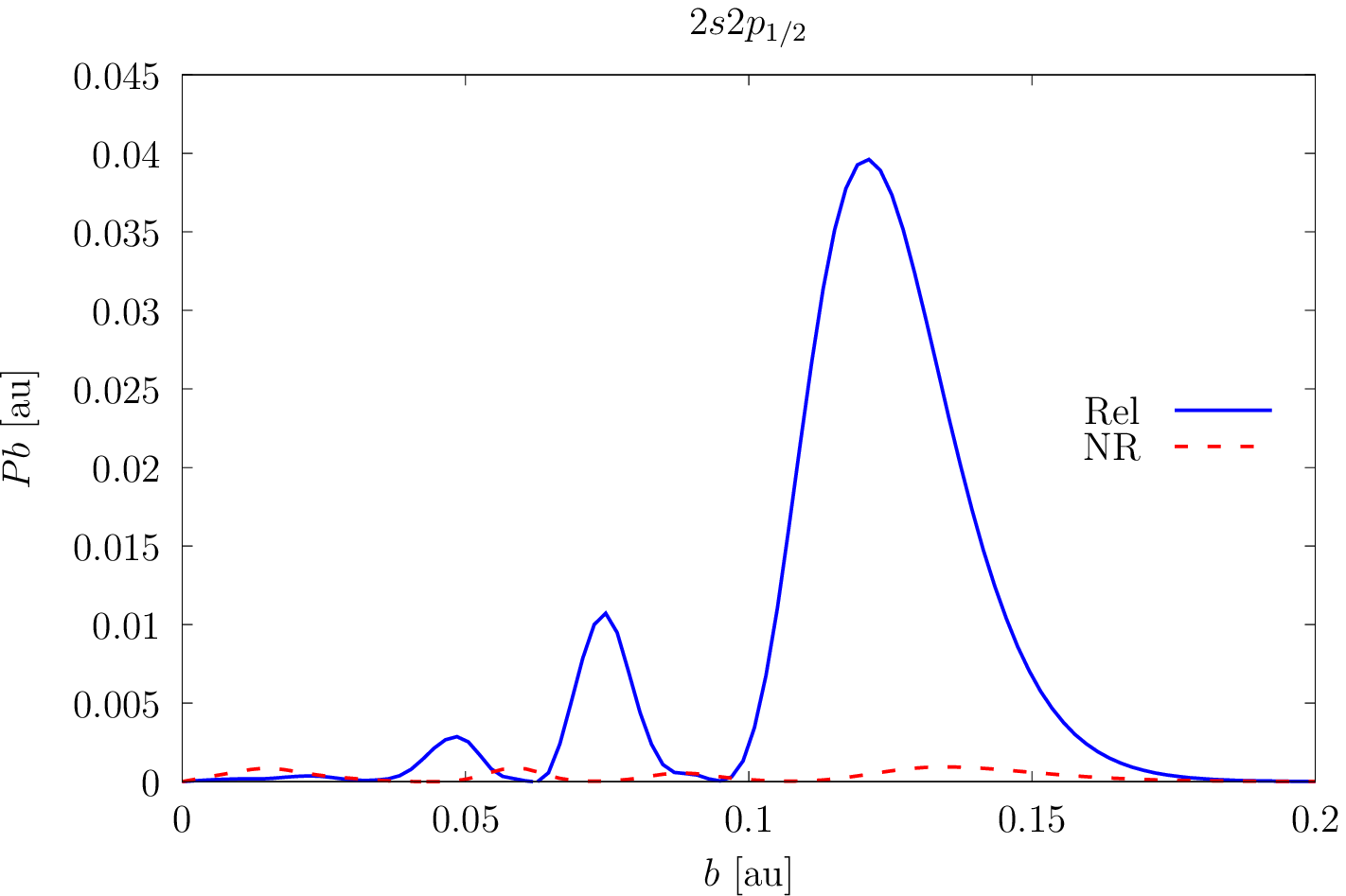}
}\\
\subfigure[]{%
\includegraphics[width=0.45\textwidth]{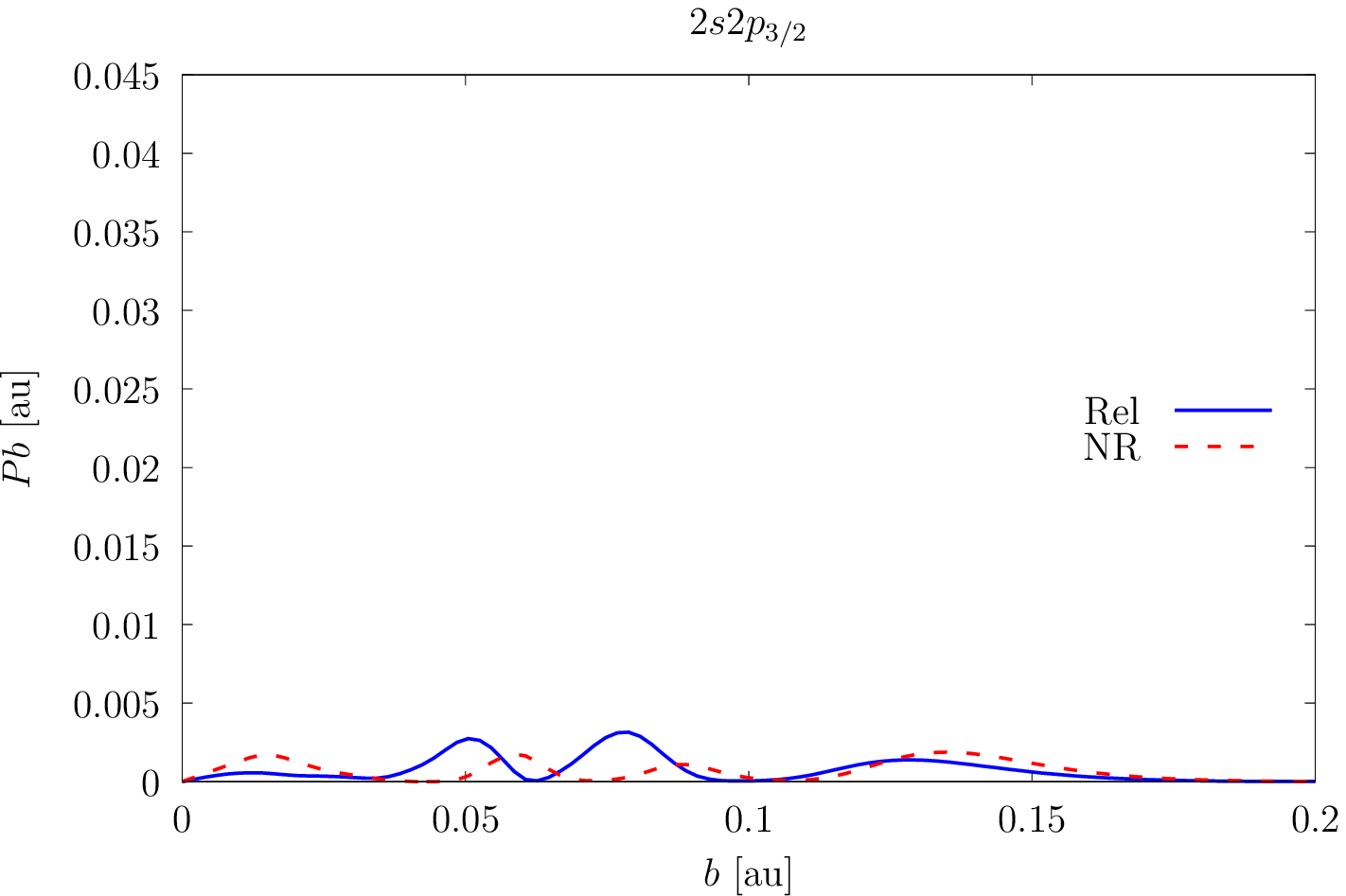}
}
%%%%
\subfigure[]{%
\includegraphics[width=0.450\textwidth]{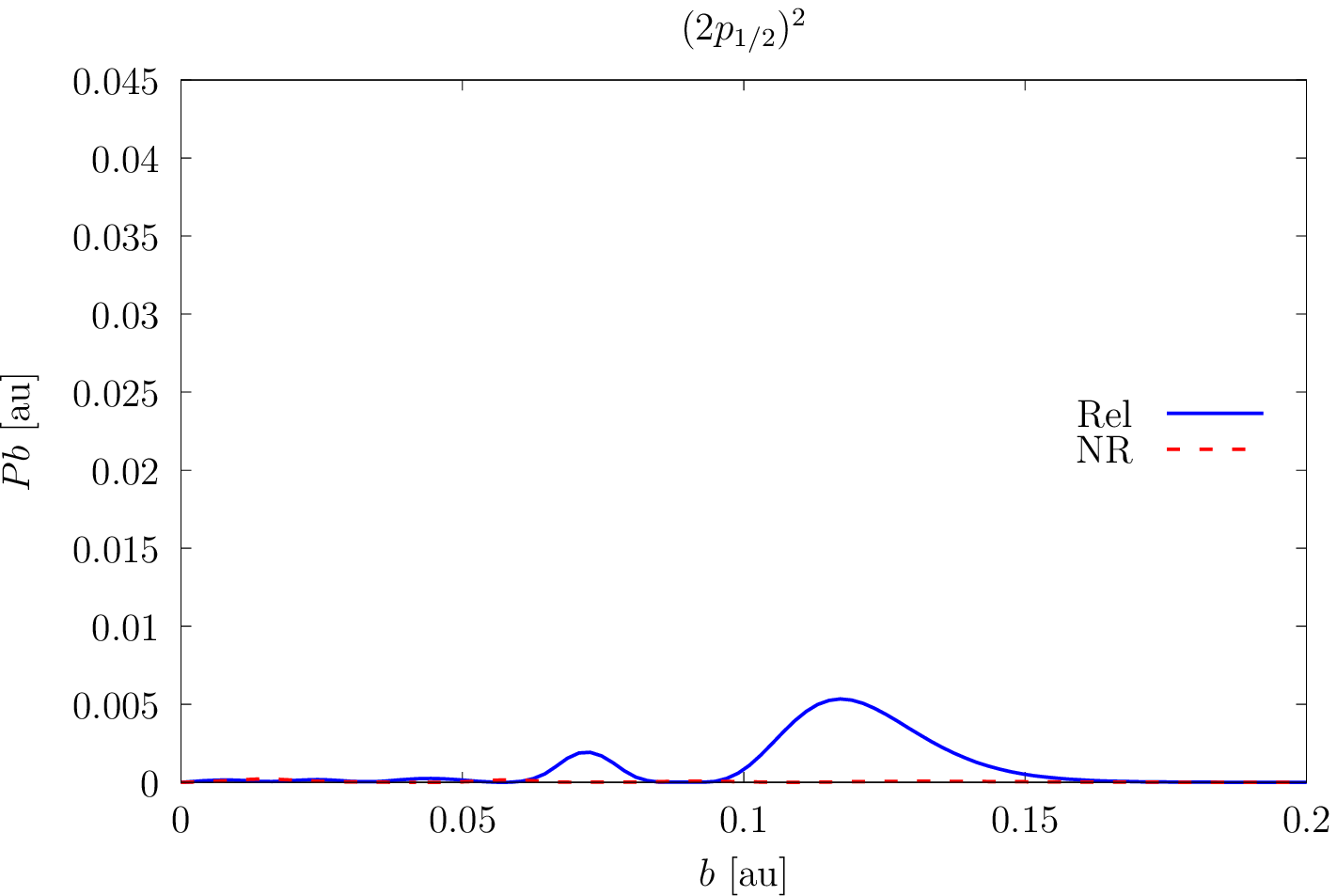}
}\\
\subfigure[]{%
\includegraphics[width=0.450\textwidth]{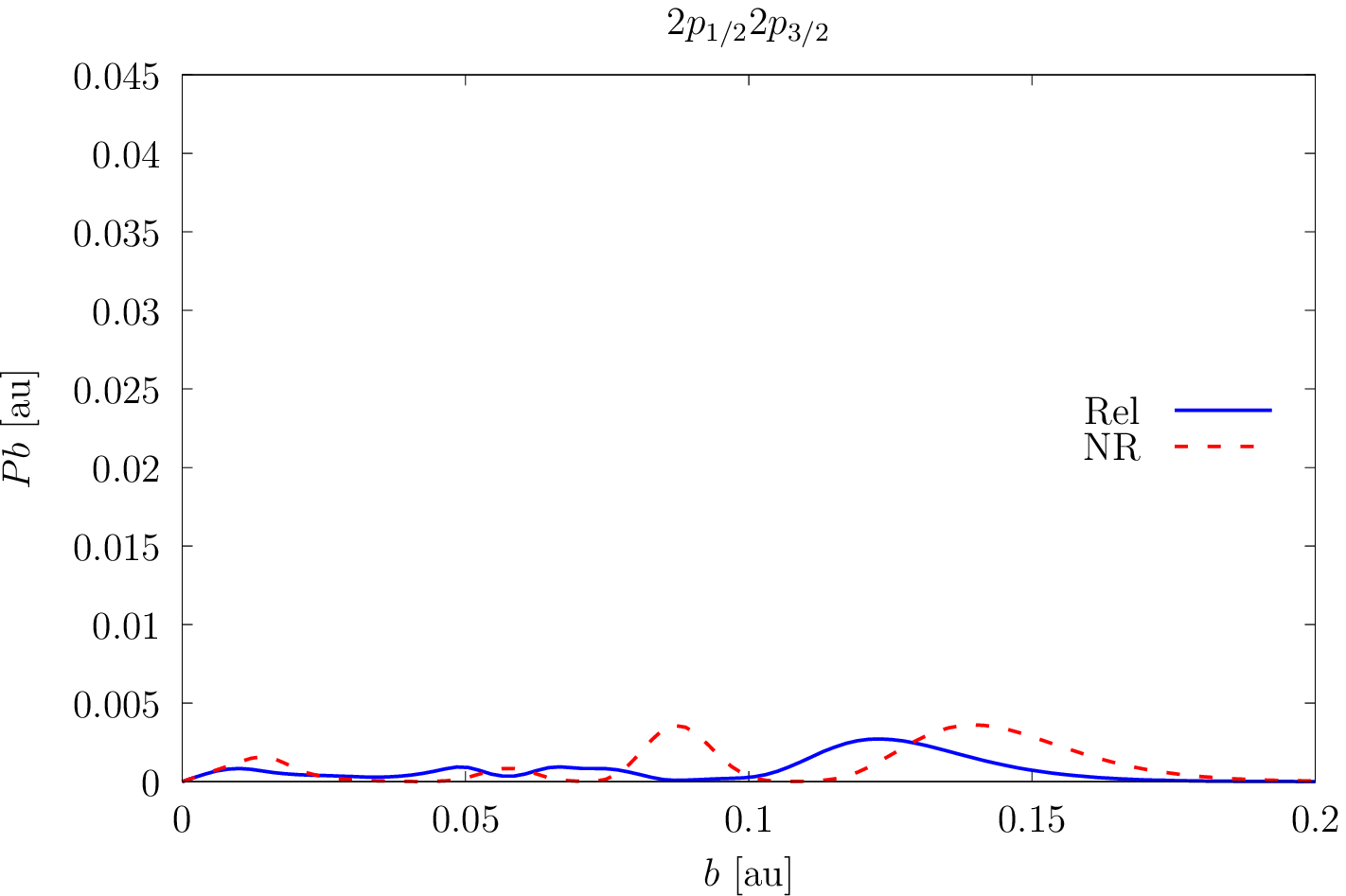}
}
\subfigure[]{%
\includegraphics[width=0.450\textwidth]{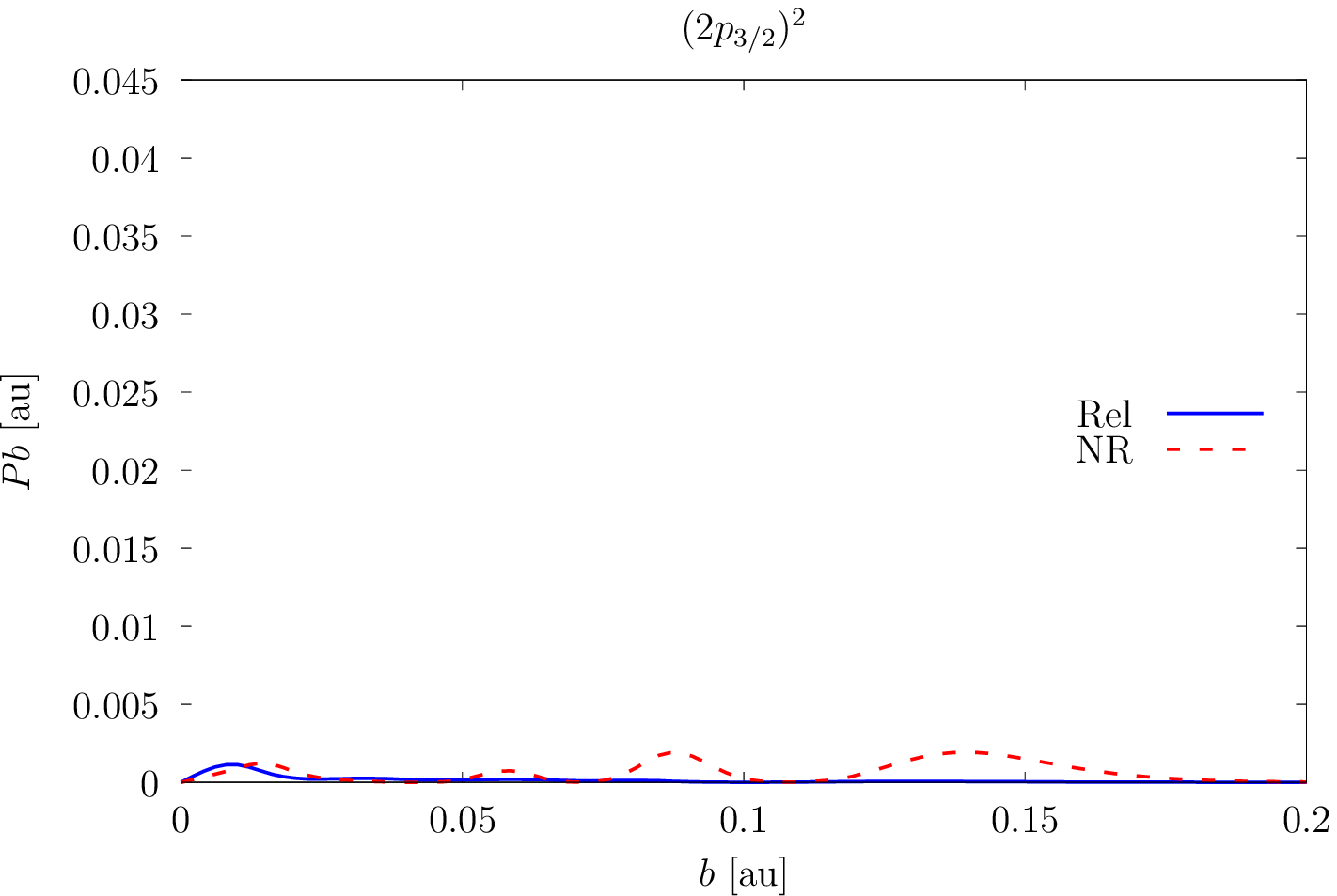}
}
%%%%
%
\end{center}
\caption{\label{fig:Th_DEC_3}
State-selective DEC probabilities weighted by the impact parameter for the Th$^{90+}$-Ru$^{42+}(1s^2)$ @ $3$~MeV/u collision as functions of the impact parameter.
The relativistic (solid lines marked "Rel") and non-relativistic results (dashed lines marked "NR") are presented.}
\end{figure}

Figures~\ref{fig:Th_DEC_50}-\ref{fig:Th_DEC_3} show the results of calculations of various DEC processes as functions of the impact parameter for $50$, $20$, $8$ and $3$~MeV/u collision energies, correspondingly. The capture probabilities are essential for the impact parameter less than $0.2$~a.u., which is approximately a fourfold size of the $1s$ orbital of ruthenium and the $2s$, $2p$ orbitals of thorium. It can be seen that the probability of the processes increases (in average) with decreasing the collision energy, but the shape of the curves becomes more complex (more minima and maxima appear). As expected, the relativistic effects are very important as for magnitudes of the probabilities as for shapes of the curves. The relativistic effects become especially strong at small collision energies.

%%%%%%%%%%%%%%%%%%%%%%%%%%%%%%%%%%%%%%%%%%%%%%%%%%%%%%%%%%%%%%%%%%%%%  
%%%%%%%%%%%%%%%%%%%%%%%%%%%%%%%%%%%%%%%%%%%%%%%%%%%%%%%%%%%%%%%%%%%%%  
%%%%%
%%%%%%%%%%%%%%%%%%%%%%%%%%%%%%%%%%%%%%%%%%%%%%%%%%%%%%%%%%%%%%%%%%%%%  
%%%%%
%%%%%%%%%%%%%%%%%%%%%%%%%%%%%%%%%%%%%%%%%%%%%%%%%%%%%%%%%%%%%%%%%%%%%  
%%%%%
%%%%%%%%%%%%%%%%%%%%%%%%%%%%%%%%%%%%%%%%%%%%%%%%%%%%%%%%%%%%%%%%%%%%%  

\begin{figure}
\begin{center}
\subfigure[]{%
\includegraphics[width=0.45\textwidth]{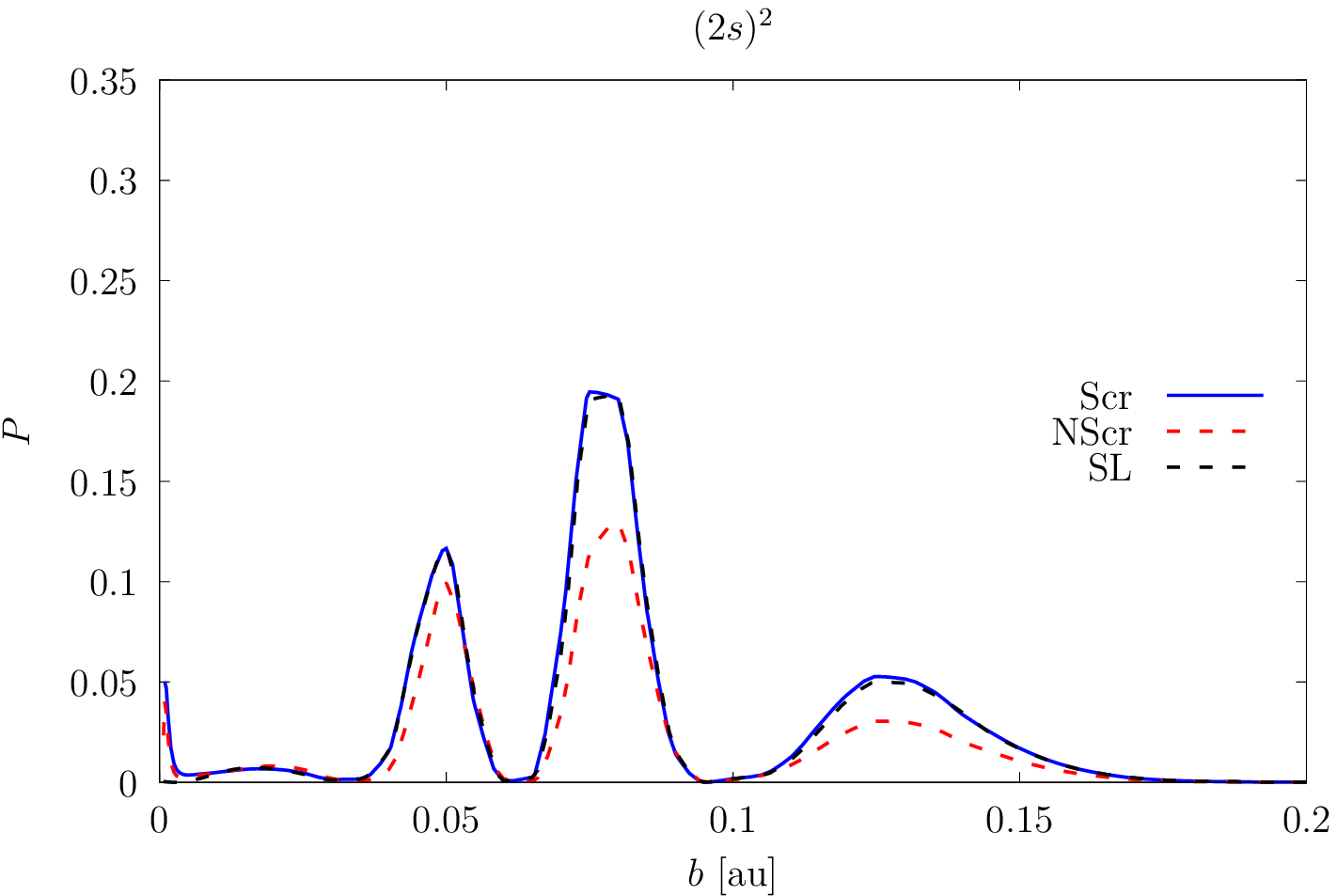}
}
\subfigure[]{%
\includegraphics[width=0.45\textwidth]{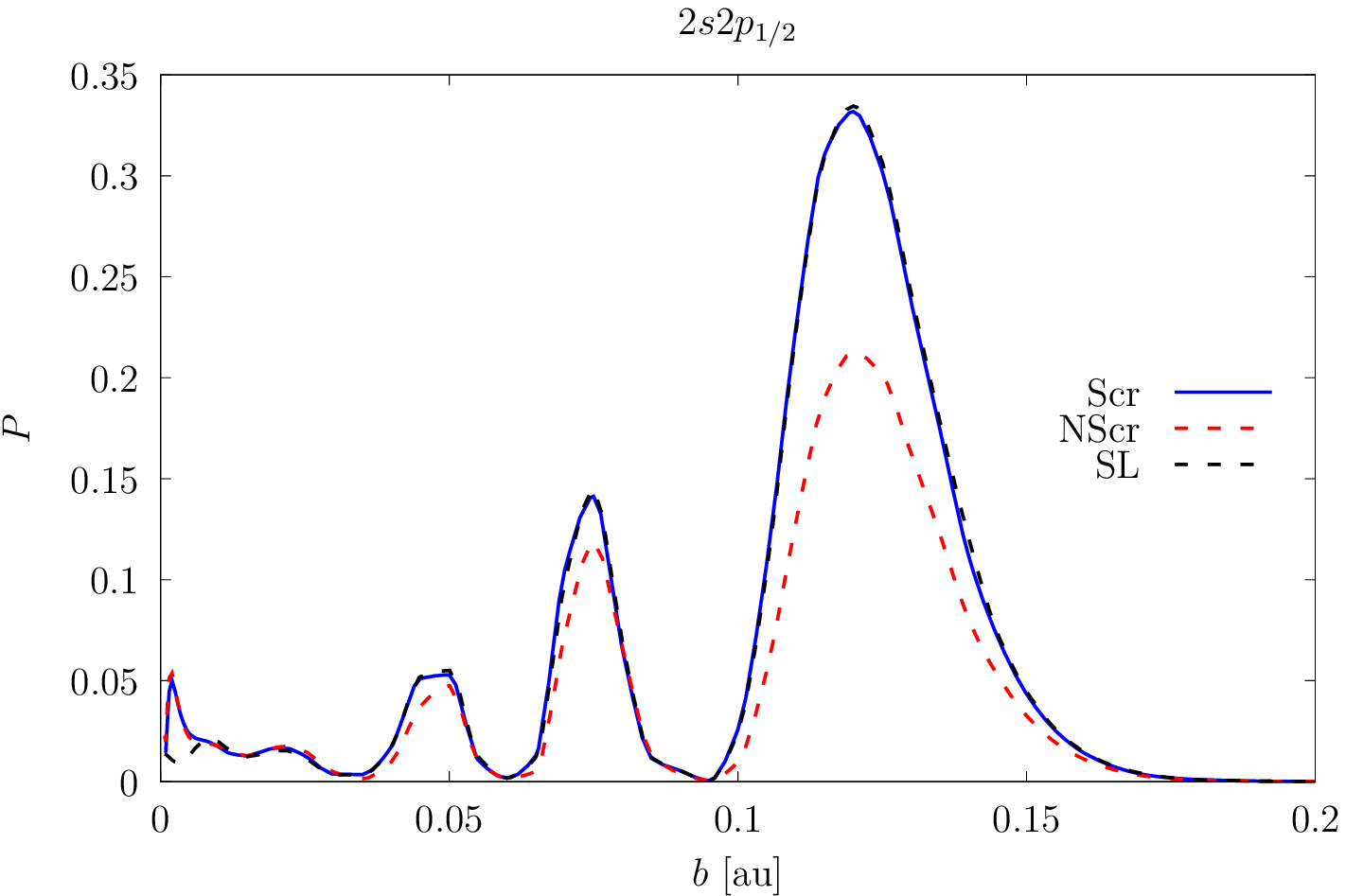}
}\\
\subfigure[]{%
\includegraphics[width=0.45\textwidth]{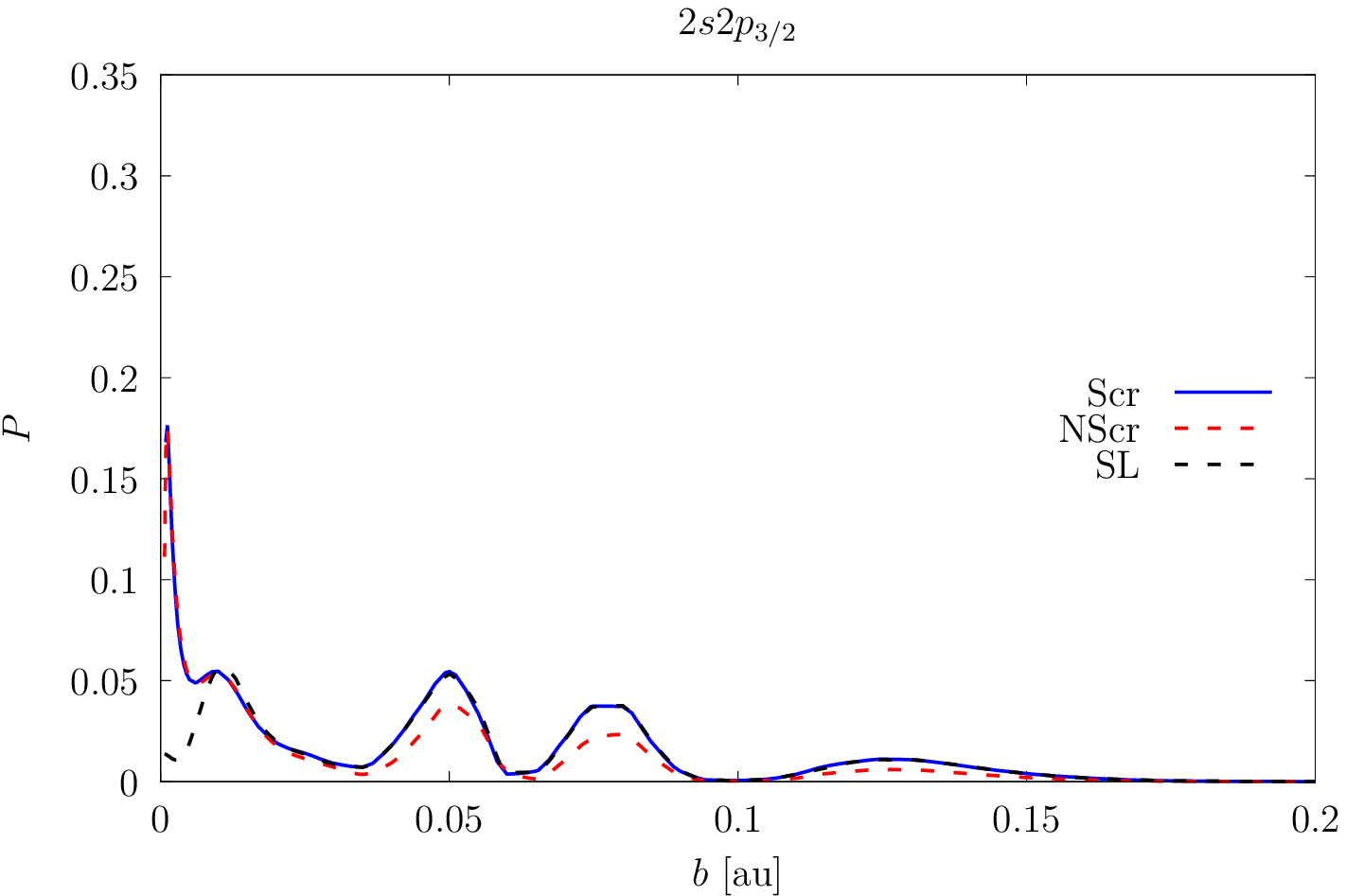}
}
%%%%
\subfigure[]{%
\includegraphics[width=0.450\textwidth]{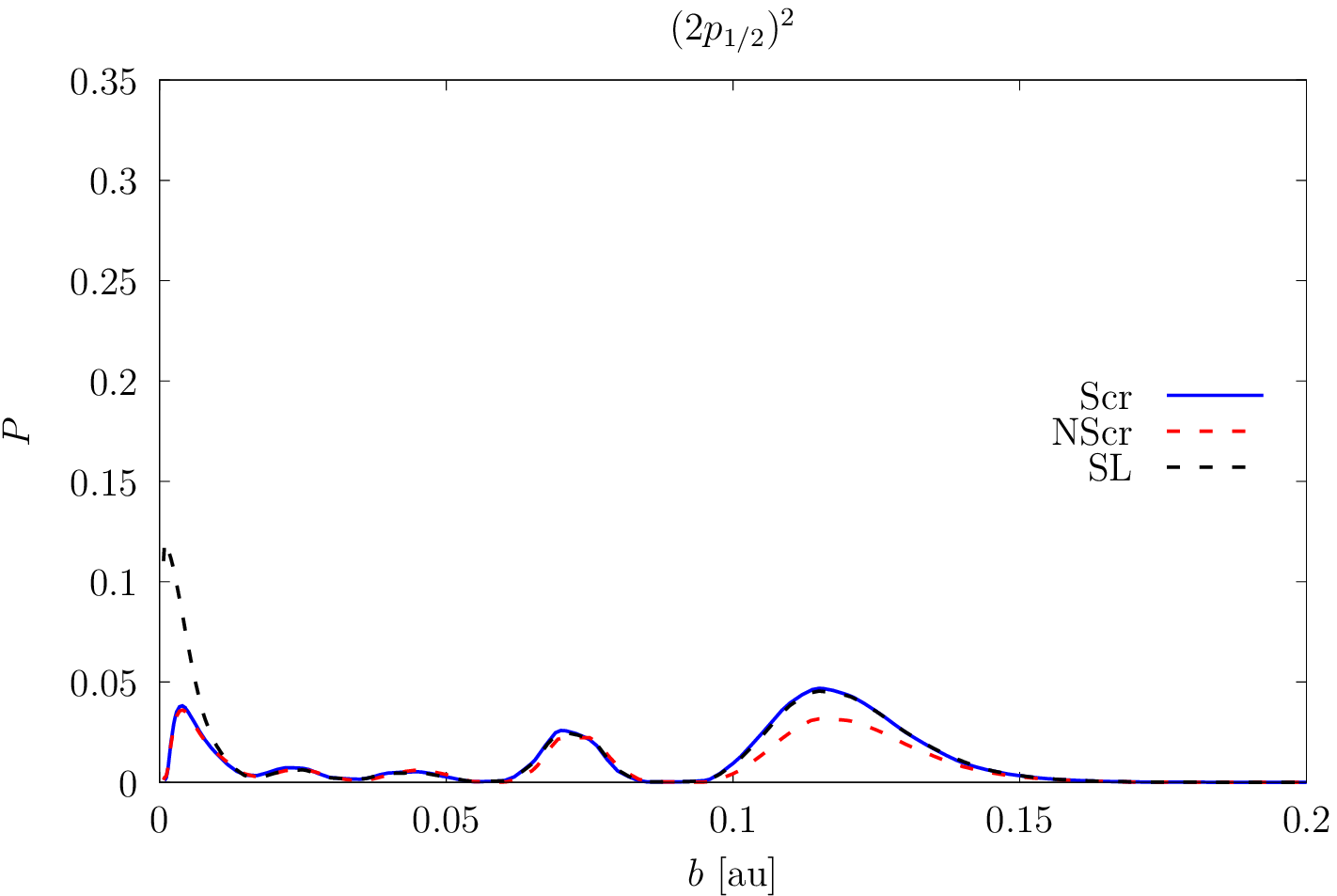}
}\\
\subfigure[]{%
\includegraphics[width=0.450\textwidth]{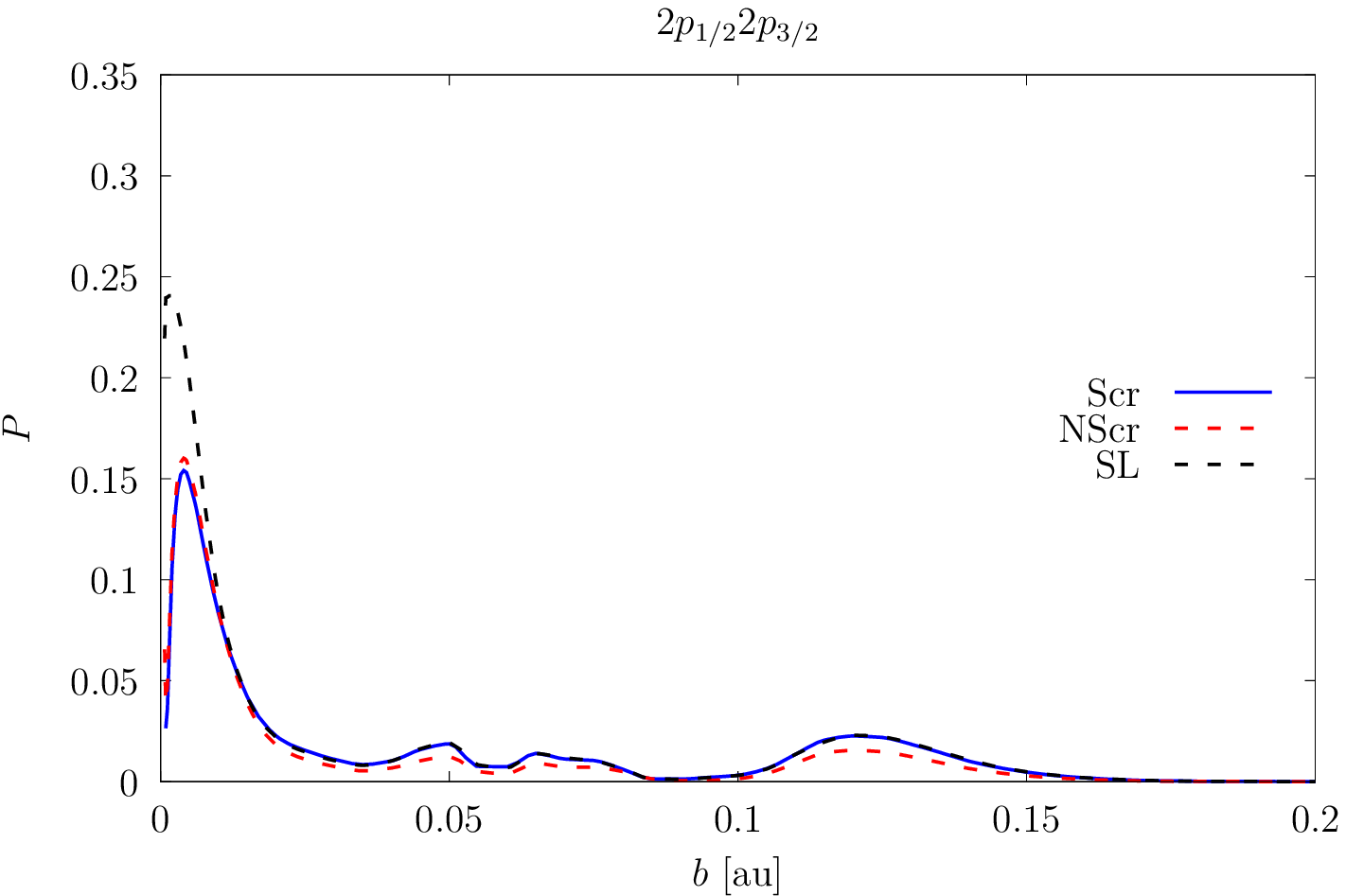}
}
\subfigure[]{%
\includegraphics[width=0.450\textwidth]{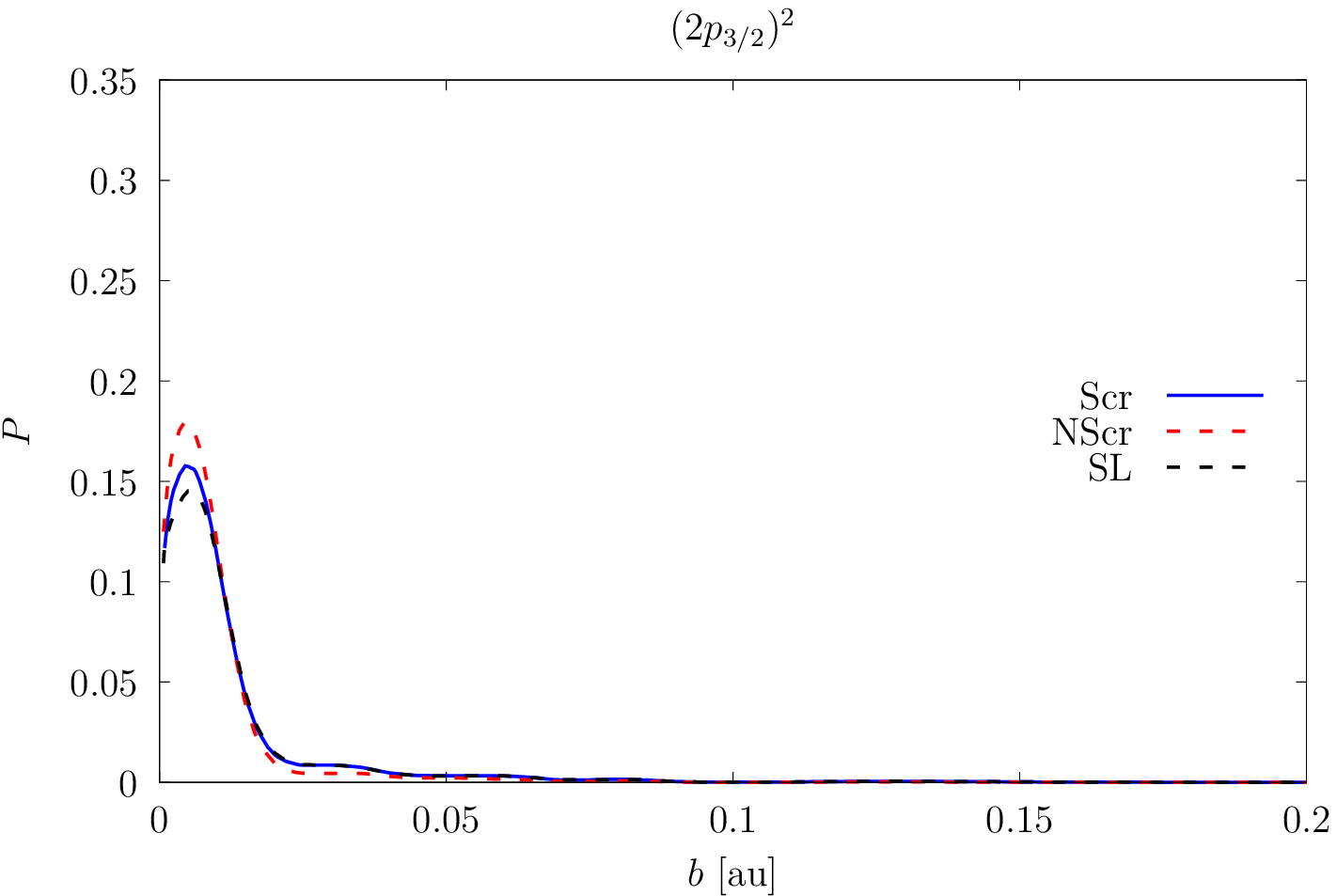}
}
%%%%
%
\end{center}
\caption{\label{fig:Th_DEC_3_test}
State-selective DEC probabilities for the Th$^{90+}$-Ru$^{42+}(1s^2)$ @ $3$~MeV/u collision as functions of the impact parameter.
The results of calculations with a screening potential and the Rutherford trajectory (solid lines marked "Scr"), without screening potential (red dashed lines marked "NScr"), and with a screening potential and the straight-line trajectory (black dashed lines marked "SL") are presented.}
\end{figure}

We also studied the role of a screening potential, which describes approximately the electron-electron interaction,  and the influence of taking into account the Rutherford type of the nuclear trajectories compared to the straight-line ones. The results of the calculations obtained with and without screening potential, and for the straight-line trajectories are presented in Fig.~\ref{fig:Th_DEC_3_test} for the $3$~MeV/u collision energy.
As one can see from the figure, the screening potential influences only the magnitudes of probabilities, but not the shape of the curves. 
The account for the Rutherford trajectories compared to the straight-line ones is noticeable only for small impact parameters (less than $0.01$~au) and, hence, has no effect on the total cross section of the processes.   

%%%%%%%%%%%%%%%%%%%%%%%%%%%%%%%%%%%%%%%%%%%%%%%%%%%%%%%%%%%%%%%%%%%%%  
%%%%%%%%

\begin{figure}
\begin{center}
\subfigure[\quad $Z=36$]{%
\includegraphics[width=0.45\textwidth]{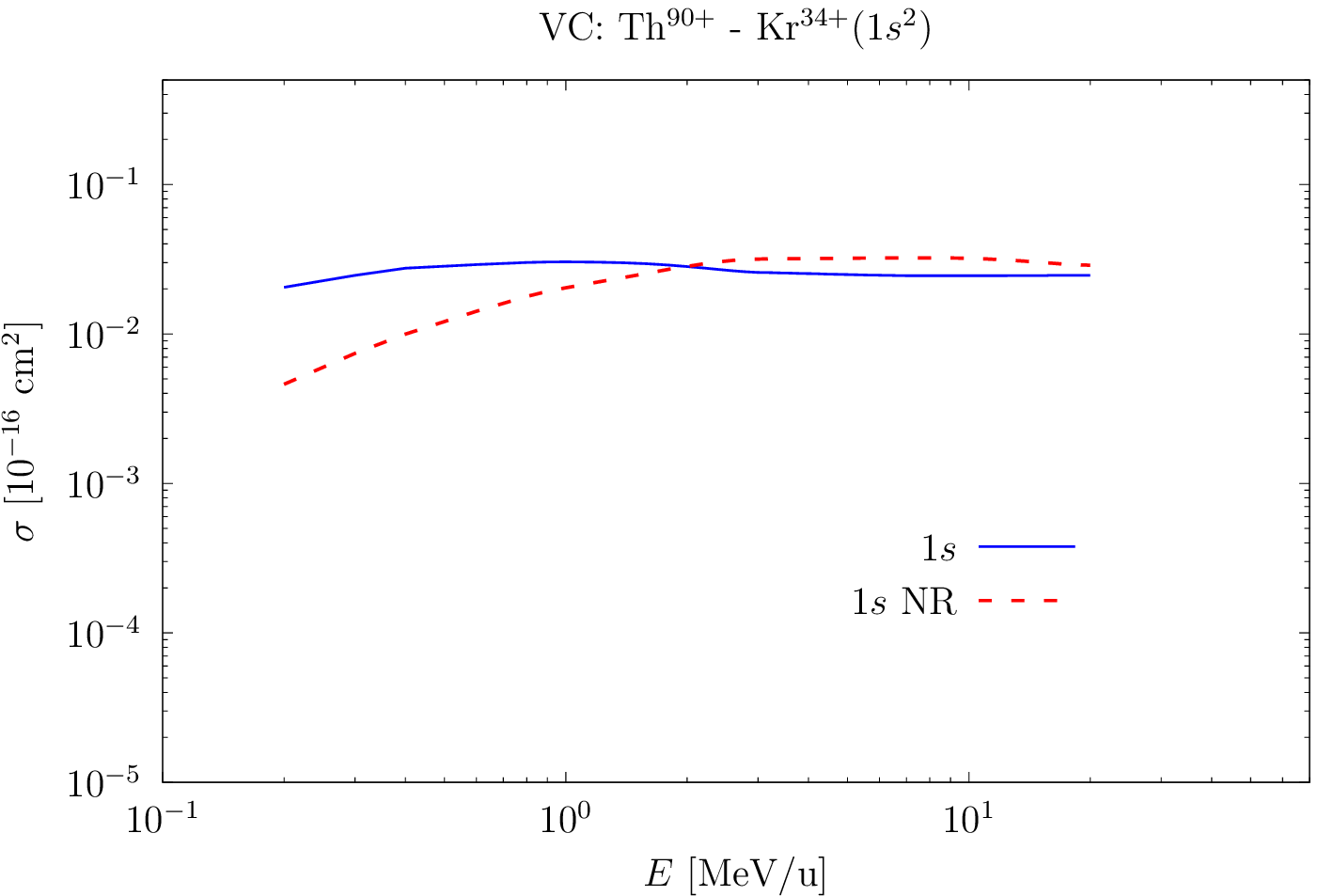}
}
\subfigure[\quad $Z=40$]{%
\includegraphics[width=0.45\textwidth]{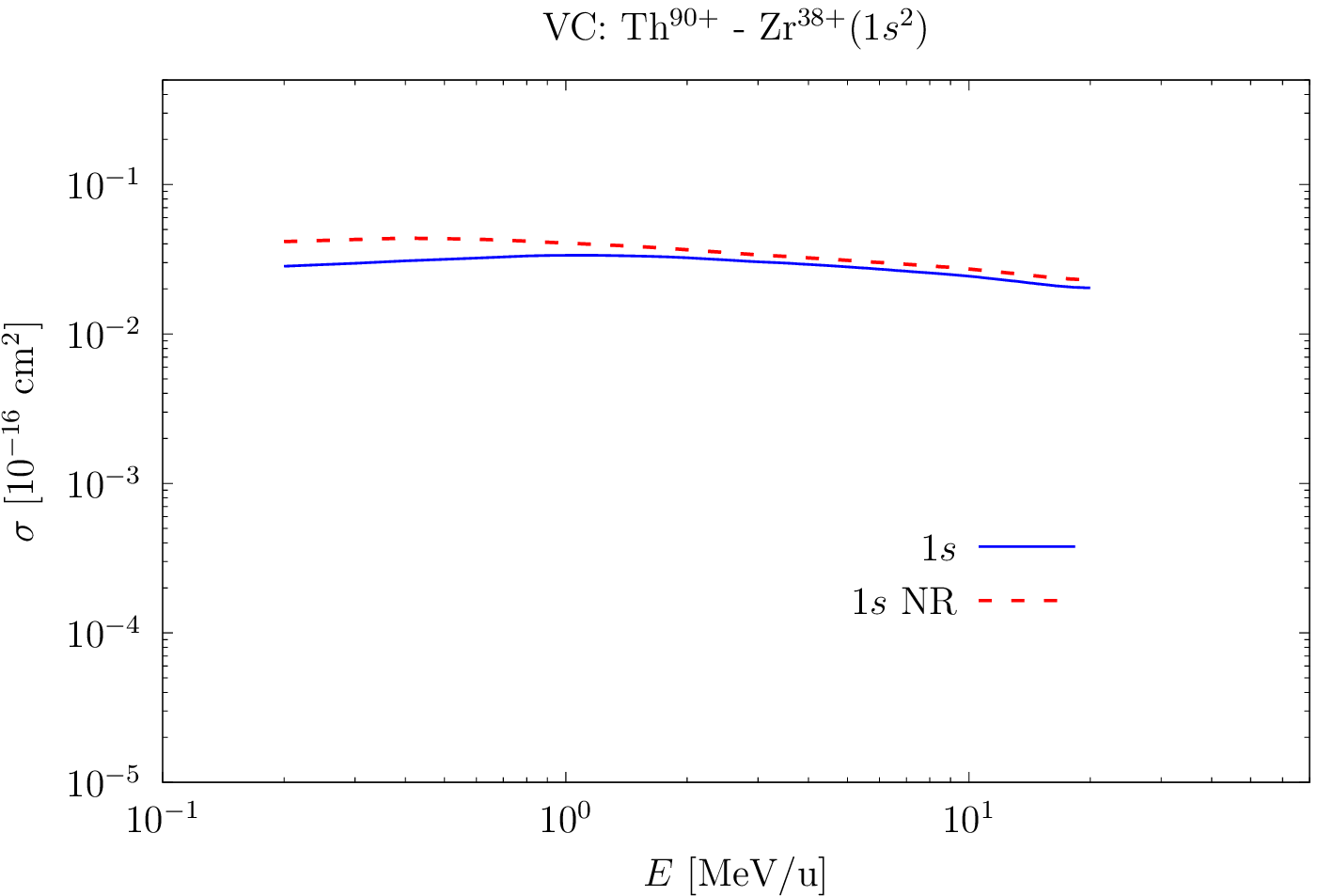}
}\\
\subfigure[\quad $Z=44$]{%
\includegraphics[width=0.45\textwidth]{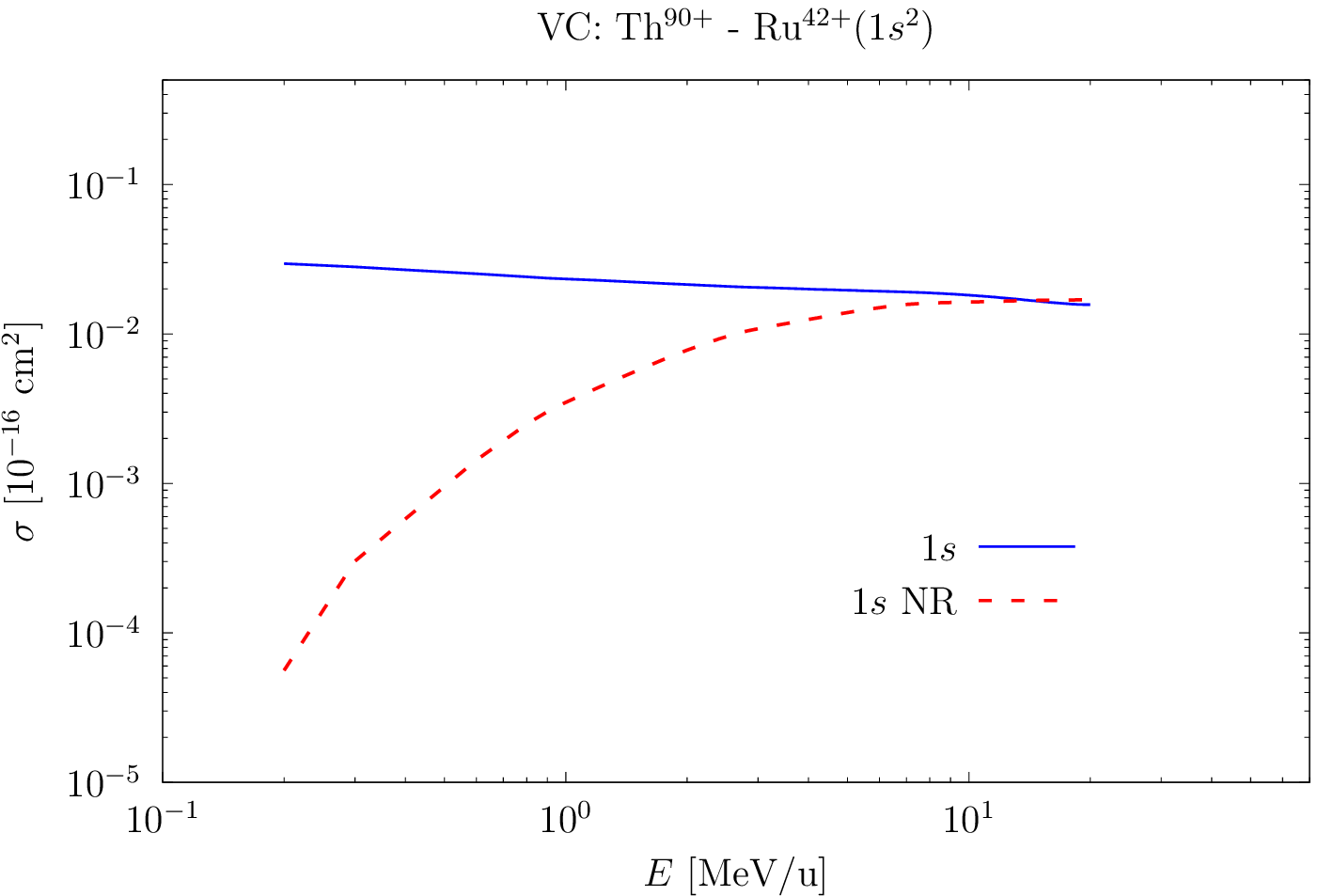}
}
%%%%
\subfigure[\quad $Z=47$]{%
\includegraphics[width=0.45\textwidth]{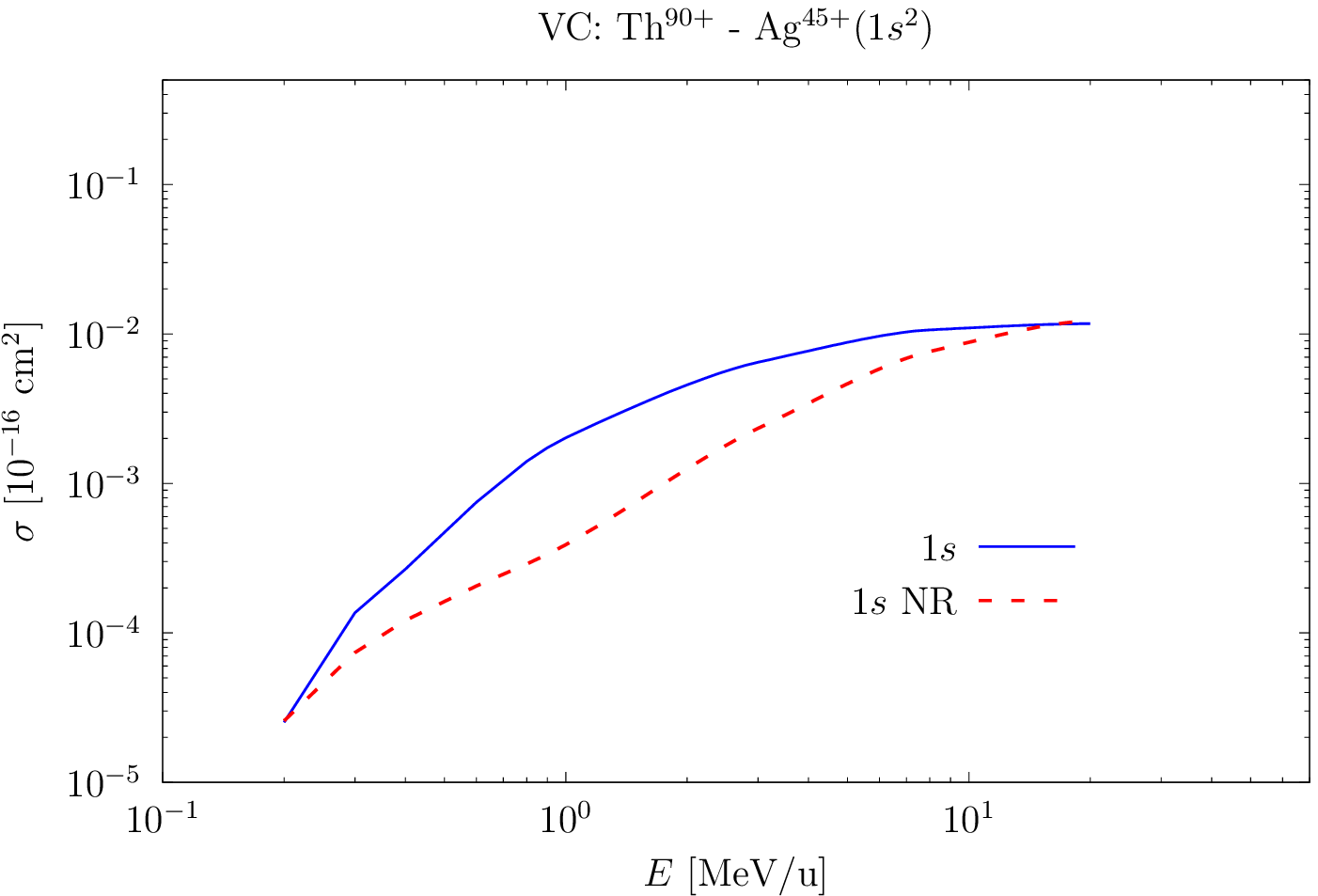}
}
%%%%
%
\end{center}
\caption{\label{fig:Th_vac}
Target K-shell vacancy creation (VC) cross sections for the Th$^{90+}$-A$^{(Z-2)+}(1s^2)$ collisions as functions of the impact energy.
The relativistic (solid blue lines) and non-relativistic results (dashed red lines marked "NR") are presented.}
\end{figure}

\begin{figure}
\begin{center}
\subfigure[\quad $Z=12$]{%
\includegraphics[width=0.45\textwidth]{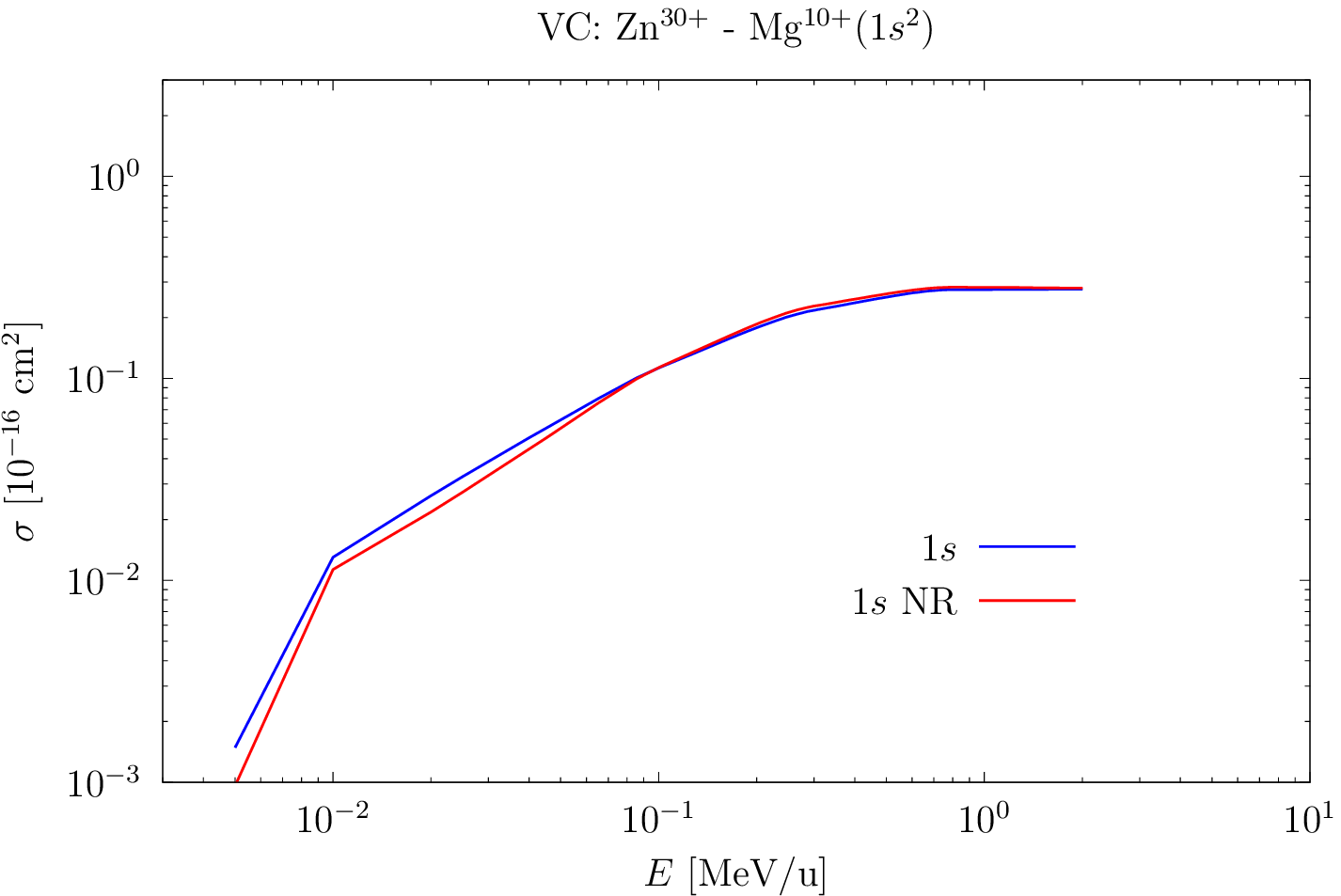}
}
\subfigure[\quad $Z=13$]{%
\includegraphics[width=0.45\textwidth]{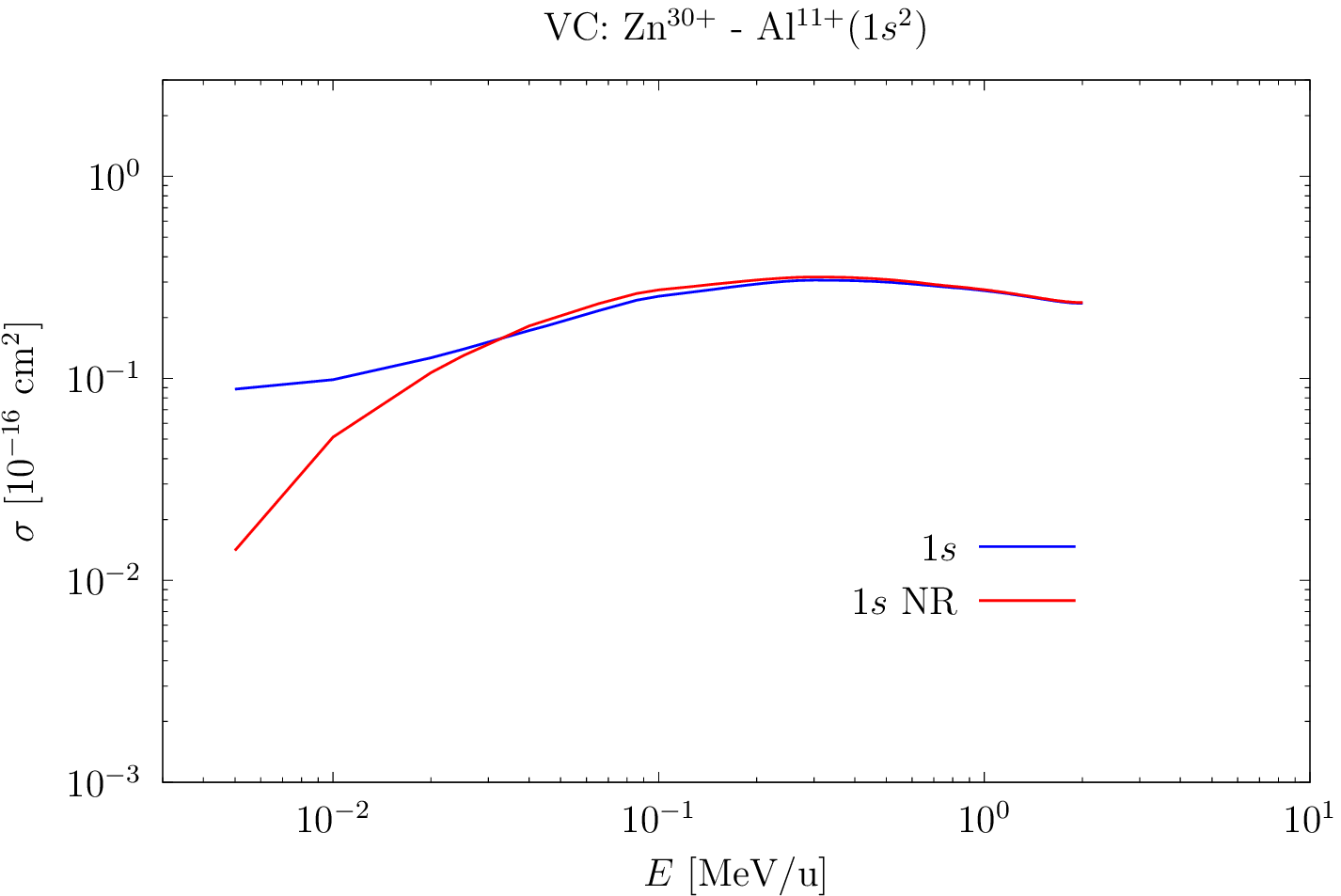}
}\\
\subfigure[\quad $Z=14$]{%
\includegraphics[width=0.45\textwidth]{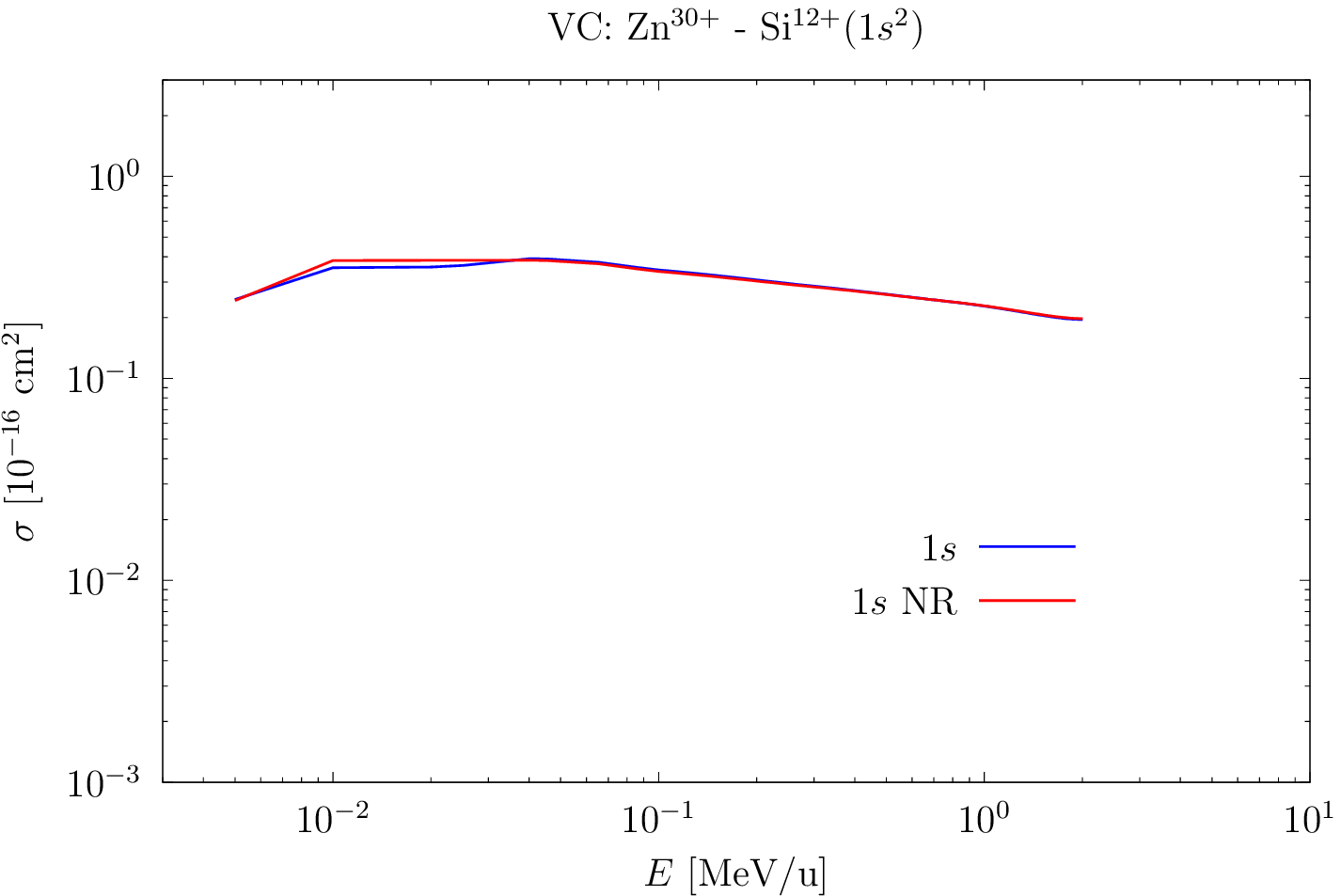}
}
%%%%
\subfigure[\quad $Z=15$]{%
\includegraphics[width=0.45\textwidth]{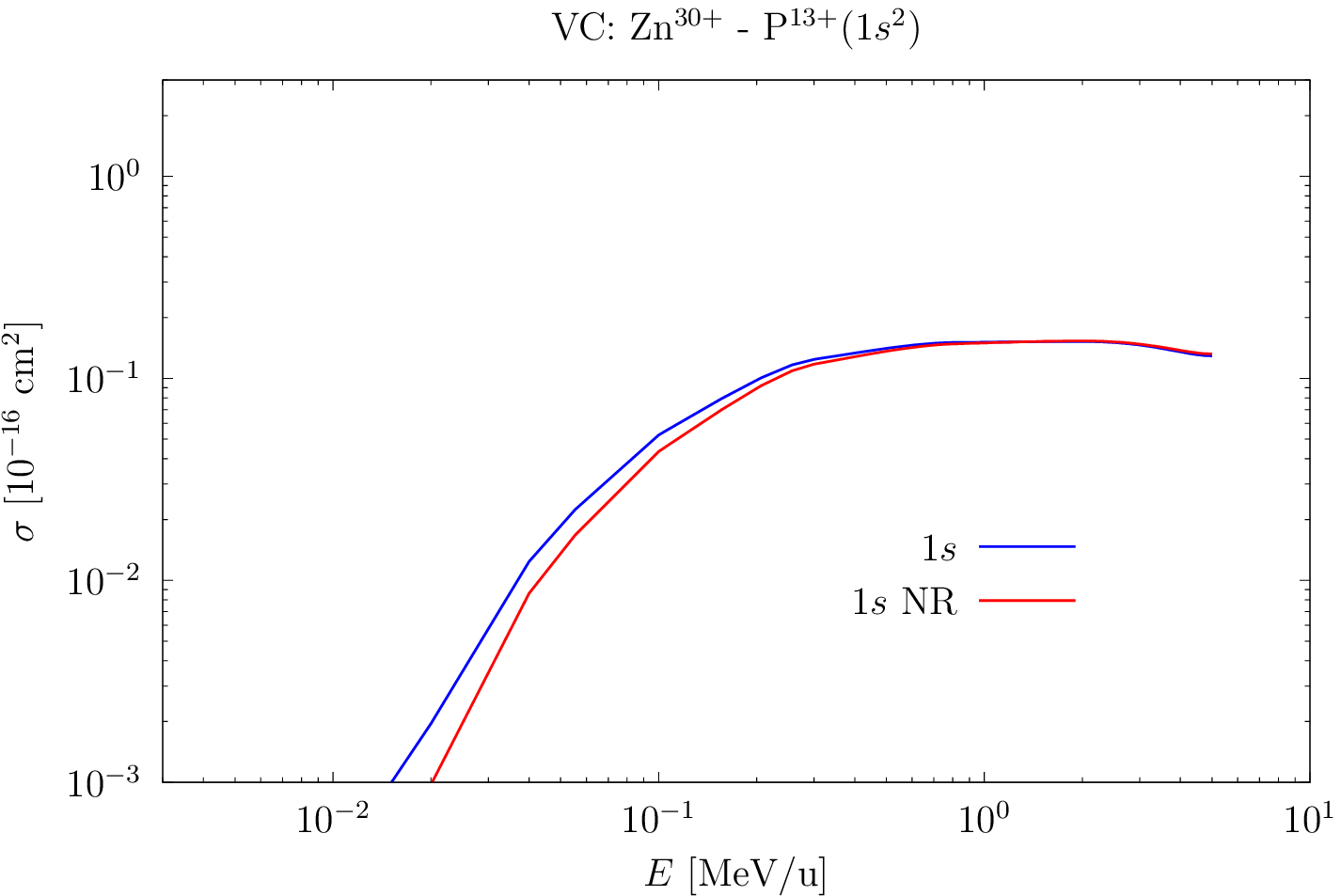}
}
%%%%
%
\end{center}
\caption{\label{fig:Zn_vac}
Target K-shell vacancy creation (VC) cross sections for the 
Zn$^{30+}$-A$^{(Z-2)+}(1s^2)$ collisions as functions of the impact energy.
The relativistic (solid blue lines) and non-relativistic results (dashed red lines marked "NR") are presented.}
\end{figure}

Experimental study of the inner-shell EC in heavy ion-ion collision traditionally based on analysis of the characteristic x-ray spectra of the ions after collision. Investigation of the low-energy collisions with neutral atoms is especially challenging since it demands taking into account EC to the excited states and deexcitation cascades strongly contributing to the spectra. Meanwhile the target $K$-shell vacancy creation and the corresponding $K$ radiation become well defined for neutral or low-ionized target atoms. That is why we have evaluated the target $K$-shell vacancy creation cross sections for collisions of bare nuclei with two electron ions. 
The obtained data should be very reliable for the collision with neutral atoms also, because of the capture to the projectile $L$ shell is the main transition channel for the target $K$-shell electrons and their excitations do not play any meaningful role (at least at the resonance conditions).
The results of the target $K$-shell vacancy creation (VC) cross section calculations for the collisions of bare thorium ($Z=90$) nucleus with heliumlike targets: krypton ($Z=36$), zirconium ($Z=40$), ruthenium ($Z=44$) and silver ($Z=47$) are presented in Fig.~\ref{fig:Th_vac} as functions of the collision energy. 
%%%%%%%%%%%%%%%%%%%%%%%%%%%%%%%%%%%%%%%%%%%%%%%%%%%%%%%%%%%%%%%%%%%%%  
%%%%%%%%%%%%%%%%%%%%%%%%%%%%%%%%%%%%%%%%%%%%%%%%%%%%%%%%%%%%%%%%%%%%%  
As one could expect, there are totally different behavior of the curves for the different targets as well as for relativistic and non-relativistic calculations. Again we note a crucial role of the relativistic effects for the collisions with ruthenium in the low-energy regime. The results of calculations for collisions of not so heavy ions:
bare zinc ($Z=30$) nucleus with heliumlike targets: magnesium ($Z=12$), aluminium ($Z=13$), silicon ($Z=14$) and phosphorus ($Z=15$) are presented in Fig.~\ref{fig:Zn_vac} as functions of the collision energy. The strongest relativistic effect could be observed for collision with aluminium. Obviously, this is due to EC to the $2p_{3/2}$ state (see Fig.~\ref{fig:Zn_EC}(b)), but the relativistic value is only around two times larger than the non-relativistic one for the collision energy about $10$~keV/u.

%%%%%%%%%%%%%%%%%%%%%%%%%%%%%%%%%%%%%%%%%%%%%%%%%%%%%%%%%%%%%%%%%%%%%  
%%%%%%%%%%%%%%%%%%%%%%%%%%%%%%%%%%%%%%%%%%%%%%%%%%%%%%%%%%%%%%%%%%%%%  

%%%%%%%%%%%%%%%%%%%%%%%%%%%%%%%%%%%%%%%%%%%%%%%%%%%%%%%%%%%%%%%%%%%%%%%%%%%%  
\section{Conclusion}  
\label{sec:Conclusion}  
%%%%%%%%%%%%%%%%%%%%%%%%%%%%%%%%%%%%%%%%%%%%%%%%%%%%%%%%%%%%%%%%%%%%%%%%%%%%%%%% 
In the present paper, the (quasi)resonance EC transitions in the non-symmetrical $(Z_{\rm P} \simeq 2Z_{\rm T})$ ion-atom collisions have been investigated. A special attention has been paid to study a role of the relativistic effects. Systematic calculations of the EC of the target $K$-shell electrons to the $L$ subshells of the projectile have been carried out for the collisions of a bare thorium nucleus $(Z_{\rm P}=90)$ with hydrogenlike ions $(Z_{\rm T}=36$-$47)$ and a bare zinc nucleus $(Z_{\rm P}=30)$ with hydrogenlike ions $(Z_{\rm T}=12$-$15)$. Strong relativistic effects, which are crucial for the case of Th$^{90+}$-Ru$^{43+}(1s)$ collisions in the low-energy regime are found. 
The one- and two-electron capture processes occurring in course of the collisions 
of the two electron system Th$^{90+}$-Ru$^{42+}(1s^2)$
have been studied in details. 
The total and state-selective SEC and DEC cross sections have been evaluated in the wide range of collision energies $0.5$-$50$~MeV/u.
Impact parameter dependencies of various DEC processes have been presented. 
The analysis shows that the probabilities of EC to the $L$ projectile shell and VC in the $K$ target shell obtained for the collision of bare nuclei with two-electron ions are relevant with a good accuracy also for collisions with target ions of a lower ionization order up to neutral atoms. Because of an experimental investigation of the target $K$-shell VC seems much more probable than EC, the corresponding cross sections for collisions of bare thorium ($Z=90$) nucleus with heliumlike targets: krypton ($Z=36$), zirconium ($Z=40$), ruthenium ($Z=44$) and silver ($Z=47$) and bare zinc ($Z=30$) nucleus with heliumlike targets: magnesium ($Z=12$), aluminium ($Z=13$), silicon ($Z=14$) and phosphorus ($Z=15$) have been obtained. Again, the role of relativistic effects for the thorium-ruthenium collisions is crucial in the low-energy regime. 
The calculations have been performed within the framework of the independent particle model using the coupled-channel approach with the atomic Dirac-Fock-Sturm  
orbitals.
% Large basis set is employed to reach controlled reasonable convergence.
% We also investigated steadiness of the obtained results on used screening potential for describing interelectron interaction and type of collided ions trajectories. 

Our study demonstrates a very significant role of the relativistic  
effects for the EC processes, which becomes crucial in the thorium-ruthenium collisions in the low-energy regime. Thus,   
investigations of heavy highly charged ion-atom collisions seem very   
promising for tests of relativistic and QED effects in  
scattering processes.

%%%%%%%%%%%%%%%%%%%%%%%%%%%%%%%%%%%%%%%%%%%%%%%%%%%%%%%%%%%%%%%%%%%%%%%%%%%%%%%% 
%  
% \clearpage  
%%%%%%%%%%%%%%%%%%%%%%%%%%%%%%%%%%%%%%%%%%%%%%%%%%%%%%%%%%%%%%%%%%%%%%%%%%%%%%%% 
\section{Acknowledgments}  
This work was supported by 
RFBR (Grants No.~18-32-20063),
SPbSU-DFG (Grants No.~11.65.41.2017 and No.~STO 346/5-1). 
Y.S.K. acknowledges support from the Chinese Academy of the Sciences President’s International Fellowship Initiative (PIFI) under Grant No.~2018VMC0010 and SPbSU (COLLAB~2019: No.~41160833).
The work of A.I.B. was also supported by the Ministry of Education and Science of the Russian Federation and SPbSU (TRAIN2019: No.~41159402).
V.M.S. acknowledges support by CAS President International Fellowship Initiative (PIFI), SPbSU (COLLAB~2019: No.~37722582) and the Foundation for the Advancement of Theoretical Physics and Mathematics “BASIS”.
The research was carried out using computational resources provided by the Resource Center “Computer Center of SPbSU”.

%%%%%%%%%%%%%%%%%%%%%%%%%%%%%%%%%%%%%%%%%%%%%%%%%%%%%%%%%%%%%%%%%%%%%%%%%%%%  
\clearpage  
%%%%%%%%%%%%%%%%%%%%%%%%%%%%%%%%%%%%%%%%%%%%%%%%%%%%%%%%%%%%%%%%%%%%%%%%%%%%  
% \begin{thebibliography}{99}  
%%%%%%%%%%%%%%%%%%%%%%%%%%%%%%%%%%%%%%%%%%%%%%%%%%%%%%%%%%%%%%%%%%%%%%%%%%%%  
%  
% \bibliography{/home/yury/Yandex.Disk/Task/2center/Report/2center}
% \bibliography{../../../Report/2center}
\bibliography{collision2019}
% \bibliography{2center}
%%%%%%%%%%%%%%%%%%%%%%%%%%%%%%%%%%%%%%%%%%%%%%%%%%%%%%%%%%%%%%%%%%%%%%%%%%%%%  
% \end{thebibliography}  
%  
%%%%%%%%%%%%%%%%%%%%%%%%%%%%%%%%%%%%%%%%%%%%%%%%%%%%%%%%%%%%%%%%%%%%%%%%%%%%%  
\end {document}